\documentclass[a4paper,11pt]{article}
\pdfoutput=1 

\usepackage{jheppub} 

\usepackage[T1]{fontenc} 
\usepackage{fontawesome}
\usepackage[table]{xcolor}
\usepackage[caption=false]{subfig}
\usepackage{epsfig}
\usepackage[T1]{fontenc} 
\usepackage{tikz-feynman}
\usepackage{hepnames}
\usepackage{adjustbox}
\usepackage{multirow}
\usepackage{slashed}
\usepackage{lipsum}
\usepackage{chngcntr}
\usepackage{bigints}
\usepackage[normalem]{ulem}
\usepackage{amsmath}
\usepackage{textcomp}
\usepackage{cleveref}
\usepackage{lscape}
\usepackage[utf8x]{inputenc}
\usepackage{float}
\usepackage{dirtytalk}
\usepackage{colortbl}
\usepackage{booktabs}
\crefname{figure}{fig\,.}{figs\,.} 
\crefname{equation}{eq\,.}{eqs\,.} 
\newcommand{\stkout}[1]{\ifmmode\text{\sout{\ensuremath{#1}}}\else\sout{#1}\fi}
\definecolor{calpolypomonagreen}{rgb}{0.12, 0.3, 0.17}
\newcommand{\bea}{\begin{eqnarray}}
\newcommand{\eea}{\end{eqnarray}}

\newcommand{\nn}{\nonumber}
\usepackage{xcolor}
\title{ Exploring Constraints on Simplified Dark Matter Model Through Flavour and Electroweak Observables}

\author[a]{Lipika Kolay,}
\author[a]{Soumitra Nandi.}

\affiliation[a]{Department of Physics, Indian Institute of Technology Guwahati,\\North Guwahati, Assam-781039, India,}

\emailAdd{klipika@iitg.ac.in}
\emailAdd{soumitra.nandi@iitg.ac.in}
\abstract{This study focuses on a combined analysis of various available inputs to constrain the parameter spaces of a simplified dark matter (SDM) model featuring a spin-0 mediator and fermionic dark matter (DM). The spin-0 mediator interacts with standard model (SM) fermions, SM gauge bosons, and DM. We constrain the parameter spaces of different relevant couplings, DM mass, and the mediator mass, using the data from flavour-changing charged and neutral current processes, CKM matrices, $W$ and $Z$-pole observables, DM relic density, direct and indirect detection bounds. We have calculated bounds on the couplings from both separate and simultaneous analyses of the mentioned processes. We identify correlated parameter spaces for all the relevant parameters which include the couplings and the masses. For the DM and mediator masses, we have scanned the region between 100 GeV and 1000 GeV. Using our results, we have obtained bounds on the couplings of possible higher dimensional operators from which we can formulate our SDM. } 

\begin{document}
	\maketitle
	\flushbottom
%

\section{Introduction}

The standard model (SM) of particle physics fails to explain some key aspects of nature. However, it is a successful theory that describes many phenomena of nature at the fundamental level. The SM can not provide a candidate for dark matter (DM) nor accommodate the observed baryon asymmetry. Also, the model could not provide a mechanism for neutrino mass generation. Different extensions of the SM introduce new degrees of freedom beyond the SM with new interactions, which could account for the observed deficiencies of the SM. In particular, many new physics models exist beyond the SM that provide a suitable DM candidate. However, in the context of experimental searches, it is convenient to study the signatures of this DM candidate in a model-independent way employing the effective field theory approach (EFT). In the EFT approaches with DM (DMEFT), one writes down a set of non-renormalizable operators that parametrize the interaction of the DM particle with SM fields in terms of one effective scale $\Lambda$ and of the DM mass. Given the limitations of the applicability of the DMEFT at the center-of-mass energies at the LHC, the recent searches at the ATLAS and CMS rely on the simplified DM models (SDM), which can describe the full kinematics of DM production at the LHC correctly \cite{Bai:2013iqa,Schmeier:2013kda,Buckley:2014fba,Abdallah:2014hon,Abdallah:2015ter,Berlin:2015wwa,Baek:2015lna,Englert:2016joy,Albert:2016osu,DeSimone:2016fbz,Arcadi:2017kky,Bauer:2017fsw,Arcadi:2017wqi,Abercrombie:2015wmb,LHCDarkMatterWorkingGroup:2018ufk,Arina:2018zcq,Arcadi:2019lka,Arcadi:2020gge,Arcadi:2020jqf}. The SDM have a moderately increased number of parameters, with the most crucial state mediating the interaction of the DM particle with the SM.      
The simplest way to devise a dark matter model is by considering a scalar, fermionic, or vector field obeying the SM gauge symmetries whose stability can be ensured by an additional discrete $\mathbb{Z}_2$ symmetry under which the DM is odd, but all other SM particles are even. However, to annihilate the SM particles and give rise to the correct relic abundance, there has to be a mediator between the dark and the visible sectors. The interactions of the mediator with the visible sector may include a non-zero vertex with the SM quark fields, among others, such that the DM can scatter off fixed target nuclei and be detected from any hint of nuclear recoil. In this paper, we consider an SDM model with fermionic DM and a spin-0 mediator; for example, see the refs. \cite{Dolan:2014ska,Harris:2014hga,Kahlhoefer:2015bea,Backovic:2015soa,Buckley:2015ctj,Englert:2016joy,Buchmueller:2017uqu,Kahlhoefer:2017umn,Morgante:2018tiq,LHCDarkMatterWorkingGroup:2018ufk,Arina:2018zcq}.  The spin-0 mediator interacts with the SM fermions, gauge bosons, and fermionic DM. Such interactions impact many other vital observables related to flavour-changing charged current (FCCC) and flavour-changing neutral current (FCNC) processes, the $ W $ and $ Z $ -pole observables, etc. The low-energy FCCC and FCNC observables will hence be helpful in constraining the new physics couplings, the mediator mass, and the correlations among them. In addition, since we are considering the interaction of the mediator with the SM gauge bosons, the electroweak precision test like $W$ and $Z$-pole observables might play an essential role in constraining the relevant couplings.  

To constrain the interaction strength of the spin-0 mediator with the SM fermions, the gauge bosons, and to the DM particle, we have taken the following inputs in our analysis:
\begin{itemize}
	\item The available inputs on the exclusive semileptonic and rare FCNC decays via the following transitions: $b\to s (d) \ell^+\ell^-$, $s\to d \ell^+ \ell^-$, $c\to u\ell^+\ell^-$, $b\to s \nu\bar{\nu}$ and $s\to d\nu\bar{\nu}$.
	
	\item  We have included the inputs on the neutral meson mixing amplitudes, like $ B_{d}^{0}-\bar{B}_{d}^{0} $,  $B_{s}^{0}-\bar{B}_{s}^{0} $ and $ K_0 - \bar{K}_0 $, which are also FCNC processes in the SM.
	
	\item The inputs on the low energy FCCC processes include the extracted values of the CKM elements, the branching fractions of a few leptonic decays of $B^-$, $D^-$, $D_s^-$, $K^-$ and hadronic decays of $\tau$ lepton. In addition, we include the available constraints on the $t\to b W$ effective couplings to which our SDM will contribute.
	
	\item Among the $W$ and $Z$-pole observables, we consider the inputs on the electroweak precision observables from $S, T, U$ parameters, the inputs on the mass of the $W$ boson, the ratios of $Z$ bosons decay rates to lepton and quarks and the several polarization asymmetries.
	
	\item DM relic density and the direct detection cross sections.
\end{itemize}
We have scanned the allowed parameter spaces, including all the inputs at a time. Also, we have separately studied the impact of the FCNC, FCCC, and electroweak precision observables, as well as the data on the relic and direct detection cross-section. Usually, in different phenomenological analyses, different benchmark points are used without paying attention to the probable correlations between those parameters coming from the combined data. This is the first analysis that provides correlated solution points for a combined data set.

	\section{Theory Framework}
 	\subsection{Working Model}
 	\subsubsection{Simplified model}\label{Working_Model}
 	Effective field theories are widely used to explain the low energy phenomenon, where the EFT operators are non-renormalizable \cite{Rothstein:2003mp, Manohar:2018aog}. In some cases, interpreting the collider measurements using EFT might be arguable. On the contrary, simplified models have attained more applicability in collider searches. Also, they are sufficient to explain DM phenomenology without going through the detailed study of the UV complete models, which might be a cumbersome job. Here, we are interested in simplified models, which are a competent alternative to EFT operators and contain a DM and a mediator only. Different kinds of simplified models have been studied in different contexts  \cite{Matsumoto:2018acr,Kim:2008pp,Arina:2016cqj,Haisch:2015ioa,Biswas:2021pic,Li:2016uph,Li:2018qip,Berlin:2014tja}.
 	
 	We have considered an extension of the SM by a singlet Dirac fermion dark matter $ \chi $, which communicates with the SM via a spin-0 singlet $S$. We impose a discrete $ \mathbb{Z}_2 $ symmetry under which $ \chi $ is odd ($ \chi \rightarrow -\chi $) and all other particles are even. 
 	The relevant Lagrangian can be written as \cite{Matsumoto:2018acr,Haisch:2015ioa,Englert:2016joy} : 	
 	\begin{equation} \label{eq:model}
 	\begin{split}
 	\mathcal{L} &=\mathcal{L}_{SM} + \frac{1}{2} \bar{\chi} (i \slashed{\partial} - m_{\chi})\chi  + \frac{1}{2} \partial_{\mu}S \partial^ {\mu} S -\mathcal{L}_{\bar{\chi}\chi S} -\mathcal{L}_{\bar{\psi}\psi S}- V(S,H) + \mathcal{L}_{VVS}. 
 	\end{split}
 	\end{equation}
 	The interactions of $S$ with the DM ($\chi$) and the SM fermions are written as 
 	\begin{align}\label{eq:fermionS}
 	\mathcal{L}_{\bar{\chi}\chi S} &= \bar{\chi}(c_{s\chi} + i c_{p\chi} \gamma_5)\chi S, \nonumber \\ 
 	\mathcal{L}_{\bar{\psi}\psi S} &=\bar{\psi} (g_{s\psi} + i g_{p\psi}\gamma_{5}) \psi S.
 	\end{align}
 	In principle, the mediator $S$ could mix with the SM Higgs. However, we have kept those mixing parameters to be small. Also, to avoid the tree level FCNC, we have written the couplings $g_{s(p)\psi}$ similar to those of the Higgs Yukawa's and defined as given below: 	
 	\begin{equation}\label{eq:fermion_coupling}
 	\begin{split}
 	\mathcal{L}_{\bar{\psi}\psi S} & =  \bar{\psi} (g_{s\psi} + i g_{p\psi}\gamma_{5}) \psi S \\ 
 	& = \bar{\psi} \biggl ( \sqrt{2}m_{\psi}\frac{g_s}{v} + i\sqrt{2}m_{\psi}\frac{g_p}{v} \gamma_{5} \biggr )\psi S  \\ 
 	& = m_{\psi} \bar{\psi} (c_s + i c_p \gamma_{5})\psi S.
 	\end{split}
 	\end{equation}
 	where, $ c_s = \frac{\sqrt{2}g_s}{v}, c_p = \frac{\sqrt{2}g_p}{v}  $, $ v $ being the VEV of SM Higgs. Also, taking the couplings in this manner naturally gives us the mass dependent non-universal couplings between the spin-0 mediator and SM fermion pairs. 
 	
 Also, the interactions of the spin-$0$ boson $S$ with the SM gauge bosons are defined by  
 	\begin{equation}\label{eq:gaugeint}
 	\mathcal{L}_{VVS} = - c_W' ~W^+_{\mu}W^{\mu-}S   - c_Z' ~Z_{\mu}Z^{\mu} S.
 	\end{equation}
 Here the gauge bosons-mediator couplings are defined as (SM type) 
 	\begin{equation}\label{eq:gauge_coupling}
 	c_V' = 2 m_V^2 c_G,
 	\end{equation}
 	with $ c_G = \frac{g_{V}}{v} $, so that the NP couplings will be in the same scale as the previous $ S\bar{f}f $ couplings. 
 	
 	Finally, the scalar potential $V(S,H)$ is defined as  
 	\begin{equation}
 	V(S,H) = \frac{\mu_{S}^2}{2}  S^2 + \frac{\lambda_3}{3!} S^3  + \frac{\lambda_4}{4!} S^4 +  \lambda_1 S H^{\dagger} H + \lambda_2 S^2  H^{\dagger} H + \mu_h^2 H^{\dagger} H + \frac{1}{2} \lambda_H (H^{\dagger}H)^2.
 	\end{equation}
 	Here  $ H $  is SM Higgs doublet given by: 
 	\begin{equation}
 	H =  
 	\begin{pmatrix}
 	0\\
 	\frac{v+h}{\sqrt{2}}
 	\end{pmatrix}.
 	\end{equation}
 Note that we are working within a simplified model which is not UV complete. In eq. \eqref{eq:model} the SM-mediator interaction terms are renormalizable but not invariant under the SM gauge group. The calculations in this framework will be valid up to the scale $\Lambda$, beyond which one may need to consider the presence of new particles or interactions. The kind of interactions defined in eq.~\eqref{eq:fermion_coupling} can be obtained in an UV complete model, like two-higgs-doublet-model (2HDM) with an extra pseudo-scalar \cite{LHCDarkMatterWorkingGroup:2018ufk,Ipek:2014gua,No:2015xqa,Liu:2015oaa, Goncalves:2016iyg, Bauer:2017ota,Kahlhoefer:2017umn,Arcadi:2020gge,Arcadi:2022lpp}. Also, these kinds of interactions could be obtained from a higher dimension operator, which will be invariant under the SM gauge group and suppressed by the powers in $ \Lambda $. In the following subsection, we will discuss one such possibility.  	 		
\subsubsection{Possible Higher Dimensional Operators}\label{higher_dim_model}
In the simplified model discussed above, the interactions of the SM fermions and the gauge bosons with the spin-0 mediator can be obtained in 2HDM with extensions. Here, we will discuss the possibility of generating these interactions from higher dimensional SM gauge invariant operators. A nice discussion in this direction could be seen in the ref. \cite{Englert:2016joy}. In our case, we can write down a Lagrangian consisting of parts involving dimension-5 gauge invariant operators, for example  
	\begin{equation}\label{eq:lagdim5}
	  	\mathcal{L}_{dim5} = \mathcal{L}_{ferm} + \mathcal{L}_{gauge}\,,
	 \end{equation} 
		with 
		\begin{equation}
		\mathcal{L}_{ferm} = -\frac{C}{\Lambda} [\bar{\psi}_L i \gamma_{5} H \psi_{R} P ] - y_f [\bar{\psi}_L H \psi_{R}] + h.c.
		\end{equation}
		and 
\begin{equation}\label{gauge_higherDim}
\mathcal{L}_{gauge}=\frac{C'}{\Lambda} P |D_{\mu}H|^2.
\end{equation}
Here, $H$ is the SM Higgs doublet field as defined in the last subsection, and the field $P$ is defined as
\begin{equation}
P = u + S_1,
\end{equation}
with $u$ as the vacuum expectation value associated with the $P$ field. We have shown in the appendix-\ref{sec:apphigherdimopr} that after expanding the above Lagrangian and giving a chiral rotation to the fermionic fields, we get the interactions between the fermion field and the spin-0 scalars as   
\begin{equation}\begin{split}\label{higher_dim_ferm_int}
\mathcal{L}_{ferm}= -\mathbb{C}^S_s \, [\bar{\psi}\psi S]\, - \mathbb{C}^S_p \, [\bar{\psi} i \gamma_{5} \psi S]  - \mathbb{C}^{h_1}_s \, [\bar{\psi}\psi h_1]\, - \mathbb{C}^{h_1}_p \, [\bar{\psi} i \gamma_{5} \psi h_1],
\end{split}
\end{equation}
with
\begin{align}\label{eq:higherdimeq}
	   \mathbb{C}^{h_1}_s =  \left( \frac{y_f}{\sqrt{2}} - \frac{C u \alpha }{\Lambda \sqrt{2}} \right) \cos \theta + \frac{C v \alpha }{\Lambda \sqrt{2}}  \sin \theta, \quad 
	   \mathbb{C}^{h_1}_{p} =  \left( \frac{C u }{\Lambda \sqrt{2}} + \frac{y_f \alpha}{\sqrt{2}} \right) \cos \theta - \frac{C v }{\Lambda \sqrt{2}} \sin \theta, \nonumber  \\
	   \mathbb{C}_s^{S} = 	\left(\frac{y_f}{\sqrt{2}} - \frac{C u \alpha }{\Lambda \sqrt{2} } \right) \sin \theta - \frac{C v \alpha }{\Lambda \sqrt{2}} \cos \theta \quad \text{ and } \quad 
	   \mathbb{C}^{S}_p =  \left( \frac{C u }{\Lambda \sqrt{2}} + \frac{y_f \alpha}{\sqrt{2}} \right) \sin \theta  + \frac{C v }{\Lambda \sqrt{2}} \cos \theta.  
	   \end{align}
	   In the above equations, we have obtained the fields $ S  $ and $h_1$ from a mixing of $S_1$ and $h$ with the mixing angle $\theta$. The angle $ \alpha $ has been introduced to define the chiral rotation of the fermions. For the details, please see discussion appendix-\ref{sec:apphigherdimopr}. We can compare the filed $S$ with the new spin-0 scalar and $h_1$ as the SM Higgs.  

Similarly, after expanding the gauge interaction term in eq. \eqref{gauge_higherDim} we obtain  
\begin{equation}
\mathcal{L}_{gauge}=  \mathbb{C}_W^{S}  ~  W^{+\mu} W^{-}_{\mu} S  + \mathbb{C}_Z^{S} ~ Z_{\mu}Z^{\mu} S,
 \end{equation}
with 
 \begin{align}
 \mathbb{C}_W^{S} = \frac{2 m_W^2}{v} \sin \theta + \frac{C' m_W^2}{\Lambda} \cos \theta \quad \text{and}  \quad \mathbb{C}_Z^{S}  = \frac{m_{Z}^2}{v} \sin \theta + \frac{C'm_{Z}^2}{2\Lambda} \cos \theta . 
\end{align}
	 
This is the simplest example of a dimension-5 operator from which we can generate the type of interactions we are interested in. There could be more complex scenarios in which one could generate similar structures, but this is beyond the scope of this paper.

\subsection{Effective Vertices} 
	We know that FCNC processes are loop-suppressed in the SM. Hence, these processes could be highly sensitive to the new interactions beyond the SM. The FCCC processes are the tree-level processes in the SM, and any contribution from an NP scenario will be highly constrained. At the present level of precision, the data on these processes could be useful to constrain the new physics model parameters contributing to these processes at the loop level \cite{Biswas:2021pic}. 
	
	The interaction defined in eq.~\eqref{eq:fermion_coupling} will contribute to the FCNC and FCCC processes at the tree level, there will be only loop-level effects. The new interactions will modify the FCNC and FCCC vertices. In the following, we will discuss these effects.

	\subsubsection{FCNC Processes}\label{FCNC_loop}	
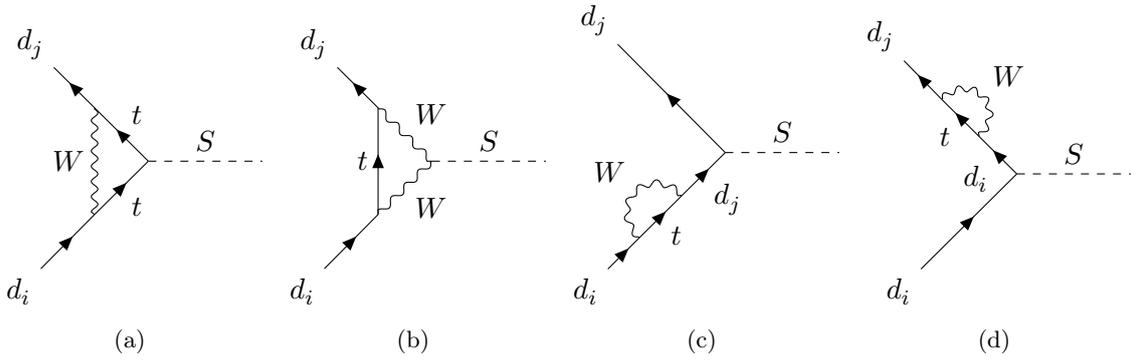
\begin{figure}[t]
		\centering 
		\subfloat[]{\begin{tikzpicture}
			\begin{feynman}
			\vertex (a1){\( d_j\)};
			\vertex [below right=1.2cm of a1](a2);
			\vertex [below right=1cm of a2](a3);
			\vertex [below left=1cm of a3](a4);
			\vertex [below left=1cm of a4](a5){\( d_i\)};
			\vertex [right=1.5cm of a3](a6);
			
			\diagram* {
				(a3) -- [fermion, arrow size=1.2pt,edge label'={\( t\)}] (a2) -- [fermion, arrow size=1.2pt] (a1),
				(a5) --[fermion, arrow size=1.2pt] (a4) -- [fermion, arrow size=1.2pt,edge label'={\( t\)}] (a3),
				(a2) --[boson, edge label'={\(W\)}] (a4),
				(a3) --[scalar,edge label={\(S\)}](a6),
				
			};
			\end{feynman}
			\end{tikzpicture}\label{fig:FCNC_loop1}}\hspace{.1cm} 
		\subfloat[]{\begin{tikzpicture}
			\begin{feynman}
			\vertex (a1){\( d_j\)};
			\vertex [below right=1.2cm of a1](a2);
			\vertex [below right=1cm of a2](a3);
			\vertex [below left=1cm of a3](a4);
			\vertex [below left=1cm of a4](a5){\( d_i\)};
			\vertex [right=1.5cm of a3](a6);
			
			\diagram* {
				(a2) -- [boson,edge label={\( W \)}] (a3),
				(a3) --[boson,edge label={\( W \)}] (a4),
				(a5) --[fermion, arrow size=1.2pt](a4) --[fermion, arrow size=1.2pt, edge label={\(t \)}] (a2) -- [fermion, arrow size=1.2pt](a1) ,
				(a3) --[scalar,edge label={\(S\)}](a6),
				
			};
			\end{feynman}
			\end{tikzpicture}\label{fig:FCNC_loop2}}\hspace{.1cm}
		\subfloat[]{\begin{tikzpicture}
			\begin{feynman}
			\vertex (a1){\( d_i\)};
			\vertex [above right=1cm of a1](a2);
			\vertex [above right=0.8cm of a2](a3);
			\vertex [above right=0.8cm of a3](a4);
			\vertex [above left=2cm of a4](a5){\( d_j\)};  
			\vertex [right=1.5cm of a4](a6);
			
			\diagram* {
				(a1) --[fermion, arrow size=1.2pt](a2) --[fermion, arrow size=1.2pt, edge label'={\(t \)}](a3) --[fermion, arrow size=1.2pt,edge label'={\(d_j\)}](a4) --[fermion, arrow size=1.2pt] (a5),	
				(a2) --[boson,half left,  looseness=2, edge label=\( W\)](a3),
				(a4) --[scalar, edge label=\( S\)](a6),		
			};
			\end{feynman}
			\end{tikzpicture}\label{fig:FCNC_loop3}}\hspace{.1cm}
		\subfloat[]{\begin{tikzpicture}
			\begin{feynman}
			\vertex (a1){\( d_i\)};
			\vertex [above right=2.2cm of a1](a2);
			\vertex [above left=0.7cm of a2](a3);
			\vertex [above left=0.7cm of a3](a4);
			\vertex [above left=0.7cm of a4](a5){\( d_j\)};
			\vertex [right=1.5cm of a2](a6);

			\diagram* {
				(a1) --[fermion, arrow size=1.2pt] (a2) --[fermion, arrow size=1.2pt,edge label= {\(d_i\)}] (a3) --[fermion, arrow size=1.2pt, edge label={\(t\)}] (a4) --[fermion, arrow size=1.2pt] (a5),
				(a4) --[boson, half left, looseness=2, edge label = \(W\)] (a3),
				(a2) --[scalar, edge label=\( S\)] (a6),
			};
			\end{feynman}
			\end{tikzpicture}\label{fig:FCNC_loop4}}
		\caption{ Feynman diagrams depicting the FCNC vertex correction for the $ d_i \to d_j S $ vertex. In these diagrams, all the internal quarks are shown only with the top-quark. Similar diagrams with c-quark and u-quark are also possible.  }
		\label{fig:b_to_s}			
	\end{figure}

	In the SDM considered above, the contribution to the FCNC processes will be through the one-loop diagrams shown in fig. \ref{fig:b_to_s} for the $d_i\to d_j S $ vertex.  
	Each of these diagrams has divergences, and they do not sum up to zero. This is not surprising since our working model is not UV-complete. The divergences from the last two diagrams of fig. \ref{fig:b_to_s} will cancel each other. The contribution from the second diagram, in particular, the divergent contribution, will be proportional to the product of the masses of the external quarks $m_{d_i} m_{d_j}$. Therefore, for the FCNC processes, like $b\to s(d)$, $s\to d$ or $c\to u$, this contribution will be proportional to either $m_s$ or $m_d$ or $m_u$, hence, small compared to what we will obtain from the first diagram. The contribution from the first diagram can be expressed as:
	\begin{equation}\label{eq:lbsS}
	\mathcal{L}_{eff}^{bsS}=\frac{ g^2}{2} V_{td_i}  V^*_{td_j}\bigg[ C_1[\bar{d_j} (m_{d_j} P_L + m_{d_i} P_R)d_i]  + C_2 [\bar{d_j} (m_{d_j} P_L - m_{d_i} P_R)d_i] \bigg] S,
	\end{equation}
	where $ C_1 $and $ C_2 $ are the effective coefficients coming from the loop diagram and contain the divergent pieces. It is not difficult to understand that in a simplified model, the one-loop contributions to flavour-changing transitions could be, in general, UV divergent, see for example, the following refs.~\cite{Batell:2009jf, Freytsis:2009ct, Arcadi:2017kky}. Hence, we could describe our model as an effective theory below some new physics scale $\Lambda$ with the following replacement: $1/\epsilon + \log(\mu^2/m^2) \to \log(\Lambda^2/m^2)$\footnote{Actually, the divergences can be absorbed by the RGE running of the respective coefficients and in the leading logarithmic approximation, the RGE evolution of these Wilson coefficients show that they depend on the scale logarithmically.} and obtained our results as given below
	\begin{eqnarray}\label{eq:RGE}
	C^{\ref{fig:FCNC_loop1}}_{1}(\Lambda) & =& \frac{1}{2}(I_1+I_2) + \frac{1}{16 \pi^2} \frac{m_t^2}{2m_W^2} 3c_s \log \frac{\Lambda^2}{m_t^2}, \nonumber  \\
	C_{2}^{\ref{fig:FCNC_loop1}}(\Lambda) & =& \frac{1}{2}(I_1-I_2) + \frac{1}{16 \pi^2} \frac{m_t^2}{2m_W^2} (-ic_p) \log \frac{\Lambda^2}{m_t^2}.
	\end{eqnarray} 
	and 
	 \begin{eqnarray}\label{eq:FCNC_vertexloop_2}
		C_{1}^{\ref{fig:FCNC_loop2}}(\Lambda) &=& \frac{1}{2} (I_3 + I_{4})+ \frac{1}{16 \pi^2} \frac{m_t^2}{2 M_W^2} \left( -3 c_G \right) \log \frac{\Lambda^2}{m_W^2}\,, \nonumber \\
		C_2^{\ref{fig:FCNC_loop2}}(\Lambda) &=& \frac{1 }{2}(I_3-I_4)\,.
		\end{eqnarray}
		The loop functions : $ I_{1}, I_{2}, I_{3},I_{4} $ are given in the appendix-\ref{apndxA}. 
		Total contribution:
	 \begin{eqnarray}
		C_{1}(\Lambda) &=& C^{\ref{fig:FCNC_loop1}}_{1}(\Lambda) + C_{1}^{\ref{fig:FCNC_loop2}} (\Lambda)\,, \nonumber \\
		C_2(\Lambda) &=&	C_{2}^{\ref{fig:FCNC_loop1}}(\Lambda)+  C_2^{\ref{fig:FCNC_loop2}}(\Lambda)\,.
		\end{eqnarray}
	As mentioned above, the divergences reflect the dependence of our results on the suppression scale $\Lambda$. It is the scale at which one may need to add new degrees of freedom with tree-level FCNCs\footnote{We have presented our analysis for $\Lambda = 1$ and 2 TeV, respectively. However, we have noted that given the precision in the current data, the bounds we have obtained on the coefficients will not change much even if we analyze at a slightly larger scale, for example, at $\Lambda =5$ TeV.}. In a UV-complete theory, the additional new physics at the scale $\Lambda$ is expected to cancel the divergences present and make the theory renormalizable. However, the available data is expected to constrain such an interaction in general. Here, we take an optimistic view and assume that the new high-scale (higher-dimensional operators) contributions in the low-energy observables will have a negligible impact on our analysis. The renormalization group evolution (RGE) over the energy range that we have considered here will not change the coupling structure significantly. In this context, it will be relevant to mention that the cutoff dependence of the low energy logarithmic divergence could reliably reflects the true dependence on the heavy mass scale of a full UV complete theory (please see the discussion in sec. 3 of ref. \cite{Burgess_1993}).
	
	In eq. \eqref{eq:RGE}, the contributions $ I_1, I_2 $ are the finite part of the loop integration, which are given in the appendix-\ref{apndxA}. Numerically, the finite part is small compared to the logarithmic divergent contributions. However, we have kept them for completeness, and at the current level of precision, these contributions do not impact our conclusions. Also, note that in the FCNC processes, we will discuss the effective vertices with right-handed quark current that will be suppressed due to the small mass $m_{d_j}$.

	\subsubsection{FCCC Processes}\label{subsec:FCCC_eff_vertex}

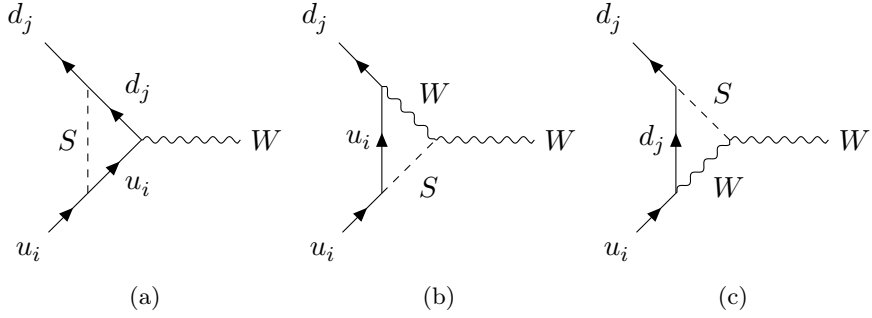
\begin{figure}[t]
		\centering
		\subfloat[]{\begin{tikzpicture}
		\begin{feynman}
		\vertex (a1){\(u_i\)};
		\vertex [above right=1.1cm of a1](a2);
		\vertex [above right=1.cm of a2](a3);
		\vertex [above left=1.cm of a3](a4);
		\vertex [above left=0.8cm of a4](a5){\(d_j\)};
		\vertex [right=1.3cm of a3](a6){\(W \)};			
		
		\diagram* { 
			(a1) --[fermion, arrow size=1.2pt](a2) --[fermion, arrow size=1.2pt, edge label'=\(u_i \)](a3) --[fermion, arrow size=1.2pt, edge label'=\(d_j \)](a4) --[fermion, arrow size=1.2pt](a5),	
			(a2) --[scalar, edge label=\(S \)](a4),
			(a3) --[photon](a6),
		};	
		\end{feynman}
		\end{tikzpicture}\label{fig:FCCC_vertex_1a}}
		\subfloat[]{\begin{tikzpicture}
		\begin{feynman}
		\vertex (a1){\(u_i\)};
		\vertex [above right=1.1cm of a1](a2);
		\vertex [above right=1.cm of a2](a3);
		\vertex [above left=1.cm of a3](a4);
		\vertex [above left=0.8cm of a4](a5){\(d_j\)};
		\vertex [right=1.3cm of a3](a6){\(W \)};			
		
		\diagram* { 
			(a1) --[fermion, arrow size=1.2pt](a2) --[scalar, edge label'=\(S \)](a3) --[boson, edge label'=\(W \)](a4) --[fermion, arrow size=1.2pt](a5),	
			(a2) --[fermion, arrow size=1.2pt, edge label=\(u_i \)](a4),
			(a3) --[boson](a6),
		};	
		\end{feynman}
		\end{tikzpicture}\label{fig:FCCC_vertex_1b}}
		\subfloat[]{\begin{tikzpicture}
		\begin{feynman}
		\vertex (a1){\(u_i\)};
		\vertex [above right=1.1cm of a1](a2);
		\vertex [above right=1.cm of a2](a3);
		\vertex [above left=1.cm of a3](a4);
		\vertex [above left=0.8cm of a4](a5){\(d_j\)};
		\vertex [right=1.3cm of a3](a6){\(W \)};			
		
		\diagram* { 
			(a1) --[fermion, arrow size=1.2pt](a2) --[boson, edge label'=\(W \)](a3) --[scalar, edge label'=\(S \)](a4) --[fermion, arrow size=1.2pt](a5),	
			(a2) --[fermion, arrow size=1.2pt, edge label=\(d_j \)](a4),
			(a3) --[boson](a6),
		};	
		\end{feynman}
		\end{tikzpicture}\label{fig:FCCC_vertex_1c}}
		\caption{Feynman diagrams contributing as a correction to $u_i\to d_j W$ vertex.}
		\label{fig:FCCC_vertex_1}
	\end{figure}

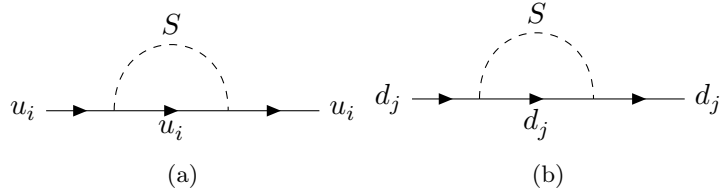
\begin{figure}[t]
		\centering 
		\subfloat[]{\begin{tikzpicture}
		\begin{feynman}
		\vertex (a1){\(u_i\)};
		\vertex [right=1.2cm of a1](a2);
		\vertex [right=1.5cm of a2](a3);
		\vertex [right=1.2cm of a3](a4){\( u_i\)};
		
		\diagram* { 
			(a1) --[fermion, arrow size=1.2pt](a2) --[fermion, arrow size=1.2pt, edge label'=\(u_i \)](a3) --[fermion, arrow size=1.2pt](a4),
			(a2) --[scalar, half left, looseness=2, edge label = \(S\)] (a3),
		};
		\end{feynman}
		\end{tikzpicture}\label{fig:FCCC_self_1a}}
		\subfloat[]{\begin{tikzpicture}
		\begin{feynman}
		\vertex (a1){\(d_j\)};
		\vertex [right=1.2cm of a1](a2);
		\vertex [right=1.5cm of a2](a3);
		\vertex [right=1.2cm of a3](a4){\( d_j\)};
		
		\diagram* { 
			(a1) --[fermion, arrow size=1.2pt](a2) --[fermion, arrow size=1.2pt, edge label'=\(d_j \)](a3) --[fermion, arrow size=1.2pt](a4),
			(a2) --[scalar, half left, looseness=2, edge label = \(S\)] (a3),
		};
		\end{feynman}
		\end{tikzpicture}\label{fig:FCCC_self_1b}}
		\caption{Self-energy corrections to the up and down type quark fields which are relevant for the wave function renormalizations.}
		\label{fig:FCCC_self_1}
	\end{figure}

	In the previous subsection, we have shown the corrections to the FCNC vertices. In the SM, the coupling strength for the $u_i\to d_j W$ charged current interaction is given by $\frac{i g V^*_{ij}}{\sqrt{2}}$ and the interaction is of the type $(V-A)$. In our model, we have the coupling of the spin-0 mediator with the fermion pair and gauge boson. Therefore, we will get corrections to FCCC vertices. We have shown the possible diagrams for the $u_i\to d_j W$ vertex corrections in fig. \ref{fig:FCCC_vertex_1} , where $u_i$ and $d_j$ are the up and down type quarks, respectively. In addition to these diagrams, there will be contributions of the counter terms from the vertex renormalizations. The self-energy correction diagrams relevant to the counter terms are shown in fig. \ref{fig:FCCC_self_1}. The one-loop correction of this charged current vertex due to the interaction given in eq.~\eqref{eq:fermion_coupling} introduces one new $(V+A)$ type interaction in addition to the original $(V-A)$ type interaction. The effective charged current interaction can be written as:
	\begin{equation}
	{\cal L}^{eff}_{u_i\to d_j W} = \frac{-  g V^*_{ij}}{\sqrt{2}}\left[ C_{VL} {\bar d_j}\gamma_{\mu}(1-\gamma_5) u_i  + C_{VR} {\bar d_j} \gamma_{\mu}(1 +\gamma_5)u_i \right] W^{\mu}.
	\label{eq:effvertex}
	\end{equation}  
	
	 Here, the effects of NP coming from loop corrections are introduced in the coefficients $C_L$ and $C_R$, respectively. Hence, we can say that at the tree level (pure SM), $C_{VL} =1$ and $C_{VR}=0$. We have performed the calculation in a unitary gauge using dimensional regularization and found that the one-loop contribution to the charged current vertex is given in figs. \ref{fig:FCCC_vertex_1a}, \ref{fig:FCCC_vertex_1b} and \ref{fig:FCCC_vertex_1a} are in general divergent. The contributions from these three diagrams will not sum up to zero even after adding the counter-term contributions. Following the arguments in the previous subsection, we have estimated the loop corrections to the $u_i \to d_j W$ vertex from the figs. \ref{fig:FCCC_vertex_1a}, \ref{fig:FCCC_vertex_1b} and \ref{fig:FCCC_vertex_1a} which are given as:

\begin{equation}\label{eq:la}
\mathcal{L}_{u_i\to d_j W}^{eff,\ref{fig:FCCC_vertex_1a}} = -\frac{  g V^*_{ij}}{\sqrt{2}} C^{a}_{VR}~\mathbf{[\bar{d_j} \gamma_{\mu} P_R u_i ] } W^{\mu}\,.
\end{equation}
	
\begin{equation}\label{eq:lb}
\mathcal{L}_{u_i\to d_j W}^{eff,\ref{fig:FCCC_vertex_1b}} = -\frac{  g V^*_{ij}}{\sqrt{2}}\left(C^b_{VL}  \mathbf{[\bar{d_j} \gamma_{\mu} P_L u_i ] } + C^b_{VR}  \mathbf{[\bar{d_j} \gamma_{\mu} P_R u_i ] } \right) W^{\mu}\,.
\end{equation}
	
\begin{equation}\label{eq:lc}
\mathcal{L}_{u_i\to d_j W}^{eff,\ref{fig:FCCC_vertex_1c}} =-\frac{  g V^*_{ij}}{\sqrt{2}}\left(C^c_{VL}  \mathbf{[\bar{d_j} \gamma_{\mu} P_L u_i ] } + C^c_{VR}  \mathbf{[\bar{d_j} \gamma_{\mu} P_R u_i ] } \right) W^{\mu}\,.
\end{equation}
		
Here, the leading contributions to the effective coefficients $C^{b,c}_{VL} $ and $C^{a,b,c}_{VR}$ are given as:
\begin{equation}\label{eq:wcaVLVR}
C^{a}_{VR} = \left(-\frac{1}{32 \pi^2} \right)(c_s^2 + c_p^2) m_{d_j} m_{u_i} \bigg( \log\frac{\Lambda^2}{m_{u_i}^2} + \log\frac{\Lambda^2}{m_{d_j}^2} \bigg)\,,
\end{equation}
	\begin{align}\label{eq:wcbVLVR}
	C^b_{VL} &=  \frac{c_s c_G m_{u_i}^2}{16 \pi^2}  \bigg( \frac{1}{2} \log\frac{\Lambda^2}{{M^2_W}} - \log\frac{\Lambda^2}{{M^2_S}} \bigg) + \frac{i c_p c_G m_{u_i}^2}{16 \pi^2}  \bigg( \frac{3}{2} \log\frac{\Lambda^2}{{M^2_W}} - \log\frac{\Lambda^2}{{M^2_S}} \bigg) ,\nonumber \\
	C^b_{VR} &= (c_s c_G  + i c_p c_G  ) \frac{m_{u_i} m_{d_j}}{16 \pi^2}  \bigg( \frac{3}{2} \log\frac{\Lambda^2}{{M^2_W}} - \log\frac{\Lambda^2}{{M^2_S}} \bigg). 
	\end{align}

	\begin{align}\label{eq:wccVLVR}
	C^c_{VL}  &=   \frac{c_s c_G m_{d_j}^2}{16 \pi^2} \left(  \frac{1}{2} \log \frac{\Lambda^2}{M_S^2} - \log\frac{\Lambda^2}{M_W^2} \right) - \frac{i c_p c_G m_{d_j}^2}{16 \pi^2} \left( \frac{3}{2} \log \frac{\Lambda^2}{M_S^2} - \log\frac{\Lambda^2}{M_W^2} \right), \nonumber \\
	C^c_{VR} &=  (c_s c_G - i c_p c_G) \frac{m_{u_i} m_{d_j}}{16 \pi^2} \left( \frac{3}{2} \log \frac{\Lambda^2}{M_S^2} - \log \frac{\Lambda^2}{M_W^2} \right).
	\end{align}
   After extracting the divergent pieces of the self-energy diagrams depicted in figs. \ref{fig:FCCC_self_1a} and \ref{fig:FCCC_self_1b}, we obtain the counter-term of the $u_i \to d_j W$ vertex as follows:    
	\begin{align}\label{eq:lcounter}
	\mathcal{L}_{\rm counter} & =- \frac{g V^*_{ij}}{\sqrt{2}} C^{\rm counter}_{VL} ~\mathbf{[\bar{d_j} \gamma_{\mu} P_L u_i ] W^{\mu} } \,,\nonumber \\ 
	&=  \frac{ g }{\sqrt{2}} \frac{1}{ 32 \pi^2} \bigg(\frac{m^2_{u_i}}{2} (c_s^2 + c_p^2) \log \frac{\Lambda^2}{m^2_{u_i}} + \frac{m^2_{d_j}}{2} (c_s^2 + c_p^2) \log \frac{\Lambda^2}{m^2_{d_j}}\bigg)~\mathbf{[\bar{d_j} \gamma_{\mu} P_L u_i ] } W^{\mu}.
	\end{align}
The total contributions to $C_{VL}$ and $C_{VR}$ in eq.~\eqref{eq:effvertex} will be obtained summing the contributions mentioned in eqs.~\eqref{eq:la}, \eqref{eq:lb}, \eqref{eq:lb} and \eqref{eq:lcounter}, respectively. Therefore, one will obtain:
\begin{equation}
C_{VL} = C^b_{VL} +  C^c_{VL} + C^{\rm counter}_{VL}, \ \ \ \ \  C_{VR} = C^a_{VR} +  C^b_{VR} + C^{c}_{VR}.
\end{equation} 
In the following section, we will discuss the various observables potentially sensitive to these $d_i \to d_j S$ transitions and the new contributions to the FCCC vertices.    
		
\section{Impacts on various observables}
In this section, we will discuss the impact of our model on the different observables in which the new contributions will enter via the corrections previously discussed. We will divide our discussion into two subsections. In one of these subsections, we will discuss the observables in which the new contributions will be via the FCNC vertex $d_i \to d_j S$. In the other subsection, we will discuss various observables via FCCC processes, which will be useful in constraining the new couplings.  
	
	\subsection{Observables sensitive to new $d_i(u_i) \to d_j (u_j) S$ vertex}\label{subsec:FCNC_observables}
	
	In this subsection, we will focus on the observables related to $b\to s$, $b\to d$, $c\to u$, and $d \to s$ FCNC transitions. These transitions include the $ B_s^0 - \bar{B}_s^0, ~ B^0 - \bar{B}^0, ~K^{0}-\bar{K}^0 $ and $ D^{0}-\bar{D}^0 $ mixing. Also, it includes the rare decays of these neutral mesons, like $ b \to s (d) \ell^+ \ell^- $, $d\to s \ell^+\ell^-$, $c\to u \ell^+\ell^-$, $ b \to s (d) \nu^+ \nu^- $, $d\to s \nu^+\nu^-$ and $c \to u \nu^+\nu^-$ decays, respectively.

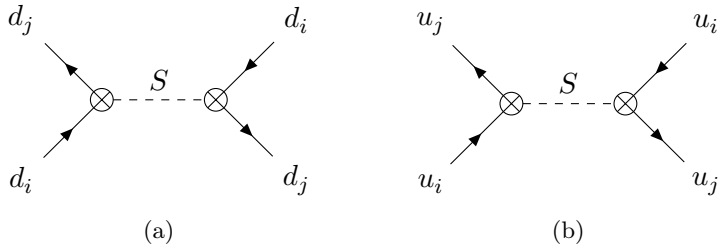
\begin{figure}[h!!!!]
		\centering
	\subfloat[]{\begin{tikzpicture}
	\begin{feynman}\label{fig:downmix}
	\vertex [crossed dot] (b){};
	\vertex [above left=1.5cm of b](a){\( d_j\)};
	\vertex [below left=1.5cm of b](c){\(d_i\)};
	\vertex [crossed dot,right=1.5cm of b](d){};
	\vertex [above right=1.5cm of d](e){\(d_i\)};
	\vertex [below right=1.5cm of d](f){\( d_j\)};
	
	\diagram* {
		(c) -- [fermion, arrow size=1.1pt] (b) -- [fermion, arrow size=1.1pt] (a),
		(b) --[scalar,edge label={\(S\)}] (d),
		(e) --[fermion, arrow size=1.1pt] (d) --[fermion, arrow size=1.1pt] (f),
	};
	\end{feynman}
	\end{tikzpicture}}~~~~~~~~
	\subfloat[]{\begin{tikzpicture}
		\begin{feynman}\label{fig:upmix}
		\vertex [crossed dot] (b){};
		\vertex [above left=1.5cm of b](a){\( u_j\)};
		\vertex [below left=1.5cm of b](c){\(u_i\)};
		\vertex [crossed dot,right=1.5cm of b](d){};
		\vertex [above right=1.5cm of d](e){\(u_i\)};
		\vertex [below right=1.5cm of d](f){\( u_j\)};
		
		\diagram* {
			(c) -- [fermion, arrow size=1.1pt] (b) -- [fermion, arrow size=1.1pt] (a),
			(b) --[scalar,edge label={\(S\)}] (d),
			(e) --[fermion, arrow size=1.1pt] (d) --[fermion, arrow size=1.1pt] (f),
		};
		\end{feynman}
	\end{tikzpicture}}
\caption{Feynman diagrams contributing to neutral meson mixing. The crossed dots represent the effective $d_i(u_i) \to d_j(u_j) S$ vertex discussed in section \ref{FCNC_loop}.} 
\label{fig:mixing_diagrams}
\end{figure}

\paragraph{\underline{Neutral Meson Mixing:} }\label{mixing_input}	In our working model, we have calculated the $d_i(u_i) \to d_j(u_j) S$ vertex where both $d_i (u_i)$ and $d_j (u_j)$ could represent either down- (up-) type quarks of different flavour. The possible Feynman diagrams contributing to various neutral meson mixings are shown in fig. \ref{fig:mixing_diagrams}. Depending on the choices of $d_i$ and $d_j$, the diagram in fig. \ref{fig:downmix} will contribute to $ B_s^0 - \bar{B}_s^0, ~ B^0 - \bar{B}^0, ~K^{0}-\bar{K}^0 $ mixings. Similarly, the contribution to $ D^{0}-\bar{D}^0 $ mixing will arise from fig. \ref{fig:upmix} with a proper choice of $u_i$ and $u_j$, respectively. The mixing amplitude is defined as $\Delta M = 2 |M_{12}|$, where $|M_{12}|$ will be obtained by calculating the dispersive part of the respective diagram given in fig. \ref{fig:downmix}.  
	
As we observe from eq.~\eqref{eq:lbsS}, the contribution of the loop depends on the masses of the external quarks ($ m_i, m_j $), the square of the mass of the particle in the loop ($ m_q^2 $), and most importantly, on the CKM elements ($ V_{qd_i}V^*_{qd_j} $). For $ K^0-\bar{K}^0 $ mixing, the dominant contribution will arise from the top-mediated loop. However, it will be suppressed by both the CKM elements and the masses of the external quarks. Also, the contribution in $ D^{0}-\bar{D}^0 $ mixing will be small since the loop factor will be proportional to $m_b^2$. Also, the loop contribution will be suppressed by the corresponding CKM factor. So, the contributions in our model to $ K^0-\bar{K}^0 $ and $ D^{0}-\bar{D}^0 $ mixing will not be enough to shift their amplitudes much from the respective SM contributions. Hence, we will discuss only the $ B_s^0 - \bar{B}_s^0 $ and $ B^0 - \bar{B}^0 $ mixings.

 For the $ B_q^0 - \bar{B}_q^0 $ ($q = d, s$) mixing the expression for $ \Delta M_q $ is given by :
 \begin{equation}
 \Delta M_q = 2 | M_{12}^q| =  \frac{|\mathcal{M}_q|}{ m_{B_q}}, 
 \end{equation}
 where $ \mathcal{M}_q $ is the amplitude of the mixing process and 
 $ m_{B_q} $ is the mass of the neutral meson. Here, we denote the contribution of new physics to the mixing amplitude as $ \Delta M_{q,NP}$ whereas the same for the standard model is defined as $ \Delta M_{q,SM} $. Therefore, the total contribution can be written as 
	\begin{equation}\label{eq:NP_mixing}
	\begin{split}
	 \Delta M_{q,tot} = \Delta M_{q,SM} + \Delta M_{q,NP} =\Delta M_{q,SM}(1+\Delta_q),
	\end{split}\end{equation}
	where $ \Delta_q \left(= \Delta M_{q,SM}/\Delta M_{q,NP} \right) $. We will estimate $\Delta_q$ and compare it with the respective measurements. For each mixing scenario, we have calculated the percentage of NP allowed in mixing, considering the corresponding errors.

In our NP scenario, for the	$ B_s^0 - \bar{B}_s^0 $ mixing, we obtain the amplitude:  
	\begin{equation}
	\mathcal{M}_s = \left(\frac{2\sqrt{2}G_F m_W^2}{16 \pi^2}\right)^2 |V_{tb}V^*_{ts}|^2 (C_1 m_s [\bar{s}P_Lb ] + C_2 m_b [\bar{s}P_R b]) \frac{1}{M_S^2}(C_1 m_s [\bar{s}P_Lb ] + C_2 m_b [\bar{s}P_R b])  
	\end{equation}
	
	then the contribution of the mass difference between neutral mesons is given by : 
	\begin{align}\label{eq:BsBsbarmix}
	  \Delta M_{s,NP} =& 8G_F^2 m_W^2 |V_{tb}V^*_{ts}|^2 \biggl(  \frac{1}{16 \pi^2 } \biggr)^2  \frac{m_{B_s}^2 }{(m_b + m_s)^2} m_{B_s}^2 f_{B_s}^2 \eta_B \frac{1}{ M_S^2}\times \nonumber \\ 
	  & \left( -\frac{5}{12}B_{2} ( C_1^2 m_s^2 + C_2^2 m_b^2)  + \frac{1}{2}B_{4} m_b m_s (C_1 C_2  + C_2 C_1 )   \right)\,.
	\end{align}
	where following relations have been used \cite{ Saha:2003tq, Hagelin:1992ws, FermilabLattice:2016ipl} :
	\begin{equation} \begin{split}
	&\langle B^0_s|(\bar{b}P_Ls)(\bar{b}P_Ls|\bar{B}^0_s\rangle=\langle B^0_s|(\bar{b}P_Rs)(\bar{b}P_Rs)|\bar{B}^0_s\rangle =-\frac{5}{12}B_{2}(\frac{m_{B_s}}{m_s+m_b})^2 m_{B_s}^2 f_{B_s}^2,   \\&
	\langle B_s^0|(\bar{b}P_Ls)(\bar{b}P_Rs)|\bar{B}^0_s\rangle=\langle B^0_s|(\bar{b}P_Rs)(\bar{b}P_Ls)|\bar{B}^0_s\rangle=\frac{1}{2}B_{4}(\frac{m_{B_s}}{m_s+m_b})^2 m_{B_s}^2 f_{B_s}^2. 
	\end{split} 
	\end{equation}
We can obtain the expressions for $B^0-\bar{B}^0$ similar to eq.~\eqref{eq:BsBsbarmix} with the appropriate replacements.      
	The SM values of mixing observables given by \cite{DiLuzio:2019jyq}:
	
	\begin{subequations}\begin{align}
	&\Delta M_{s}^{SM} = 18.4 ^{+0.7}_{-1.2} ~~ps^{-1}, \nonumber \\&
	\Delta M_{d}^{SM} = 0.533 ^{+0.022}_{-0.036} ~~ps^{-1}.
	\end{align}
	\end{subequations}
	The experimental values are given by \cite{HFLAV:2022}:
	\begin{subequations}
	\begin{align}
	&\Delta M_{s} = 17.765 \pm 0.006 ~~ps^{-1}, \\&
	\Delta M_d = 0.5065 \pm 0.0019 ~~ps^{-1}.
	\end{align}	
    \end{subequations}

\begin{figure}[t]
		\centering
		\begin{tikzpicture}
		\begin{feynman}
		\vertex (a1){\(d_i\)};
		\vertex [crossed dot, above right=2cm of a1](a2){};
		\vertex [above left=2cm of a2](a3){\(d_j\)};
		\vertex [right=2cm of a2](a6);	
		\vertex [above right=1.5cm of a6](a7){\(\ell \)};
		\vertex [below right=1.5cm of a6](a8){\(\ell \)};		
		
		\diagram* { 
			(a1) --[fermion, arrow size=1.1pt](a2) --[fermion, arrow size=1.1pt](a3),
			(a8) --[fermion, arrow size=1.1pt](a6) --[fermion, arrow size=1.1pt](a7),
			(a2) --[scalar, edge label=\( S\)](a6),
		};	
		\end{feynman}
		\end{tikzpicture}
		\caption{Feynman diagram contributing to the quark level processes  $ b \to s \ell \ell, ~b \to d \ell \ell  $ and $ s \to d \ell \ell $ process for this model.}
		\label{fig:btosll}
 	\end{figure}
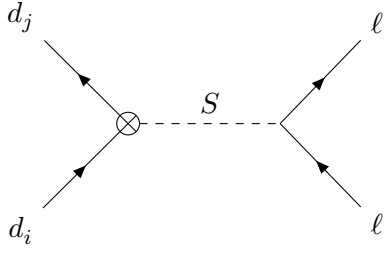

	\paragraph{\underline{Rare Decays:}}\label{para:raredecays} Rare dileptonic decays of neutral mesons are very important probes to new physics since the rate of these decays are suppressed in the SM. As mentioned earlier, we will mainly focus on the FCNC processes: $b\to s \mu^+\mu^-$, $b\to d \mu^+\mu^-$, $s\to d \mu^+\mu^-$. Examples of such decays include $ B_s^0 \to \mu^+ \mu^-, ~B_0 \to \mu^+ \mu^-, ~ K_L(K_S) \to \mu^+ \mu^- $. Our model contributes to these processes via the diagram shown in fig. \ref{fig:btosll}. We have data on the respective branching fractions measured by various collaborations, like Belle-II, LHCb, ATLAS, and CMS. Following are the experimental data of the branching ratios are\cite{ PDG:2022, LHCb:2021trn, LHCb:2020ycd}: 
	\begin{align}\label{eq:rare_exp}
	&\mathcal{B}(B_s \to \mu^+ \mu^-) = (3.09^{+0.46 ~ +0.15}_{-0.43~-0.11}) \times 10^{-9},\nonumber \\&
	\mathcal{B}(B_d \to \mu^+ \mu^-) = (0.12 ^{+0.08}_{-0.07} \pm 0.01) \times 10^{-9},\nonumber \\&
	\mathcal{B}(K_L \to \mu^+ \mu^-) = (6.84 \pm 0.11)\times 10^{-9}, \\&
	\mathcal{B}(K_S \to \mu^+ \mu^-) < 2.1 \times 10^{-10}. \nonumber 
	\end{align}
	Note that, following the reason we mentioned earlier, the NP contributions in $c \to u \mu^+\mu^-$ decays will be small. Hence, we have not considered the data on $ D_{0} \to \mu^+ \mu^-$ decay.  
 	Other important modes relevant to $b\to s \mu^+\mu^-$ FCNC processes include semileptonic decays $B\to K^{(*)}\mu^+\mu^-$ and $B_s \to \phi\mu^+\mu^-$. The wealth of data are available on the differential branching fractions, CP asymmetries and various angular observables measured on these modes by LHCb, Belle, CDF, ATLAS and CMS \cite{CDF:2011tds, LHCb:2013lvw,LHCb:2014cxe,LHCb:2014vgu,LHCb:2015svh,Belle:2016fev,CMS:2017rzx,ATLAS:2018gqc, LHCb:2020gog,LHCb:2021zwz}. Also, precise measurements are available on the ratios of the rates, like $R(K^{(*)}) = \frac{\Gamma(B\to K^{(*)}\mu^+\mu^- )}{\Gamma(B\to K^{(*)}e^+ e^- )}$. The updated value of $ R_{K}, ~R_{K^*} $ provided by LHCb in 2022  which are given below \cite{LHCb:2022qnv}: 
 	
 	\begin{subequations}
 	\begin{align}\label{eq:RKRkst2022}
 	&  R_K^{[0.1-1.1]} = 0.994 ^{+0.090 ~~ + 0.029}_{-0.082 ~~-0.027}\,, \\&
 	R_K^{[1.1-6.0]} = 0.949^{+0.042~~+0.022}_{-0.041 ~~-0.022}\,,\\&
 	R_{K*}^{[0.1-1.1]} = 0.927 ^{+0.093 ~~ + 0.036}_{-0.087 ~~-0.035} \,, \\&
 	R_{K*}^{[1.1-6.0]} = 1.027^{+0.072~~+0.027}_{-0.068 ~~-0.026} \,.
 	\end{align}\end{subequations}
 	In this analysis, we have utilized all these data to constrain the new couplings, the detailed methodology of this global analysis can be seen in our earlier publications \cite{Bhattacharya:2019dot,Biswas:2020uaq}.  
 		
	The low energy effective Hamiltonian describing the $b\to s \ell^+\ell^-$ transitions is \cite{Buras:1995iy,Becirevic:2012fy}
	\begin{equation} \label{eq:Heff}
	{\cal H}_{eff} = - \frac{4\,G_F}{\sqrt{2}} V_{tb}V_{ts}^\ast
	\left[  \sum_{i=1}^{6} C_i (\mu)
	\mathcal O_i(\mu) + \sum_{i=7,8,9,10,P,S} \biggl(C_i (\mu) \mathcal O_i + C'_i (\mu) \mathcal
	O'_i\biggr)\right] \,,
	\end{equation}
	where the twice Cabibbo suppressed contributions ($\propto  V_{ub}V^*_{us} $) have been neglected. The operator basis in which the Wilson coefficients have been computed is~\cite{Bobeth:1999mk, Altmannshofer:2008dz} :
	\begin{align*}\label{eq:basisOps}
	{\mathcal{O}}_{7} &= \frac{e}{g^2} m_b
	(\bar{s} \sigma_{\mu \nu} P_R b) F^{\mu \nu} ,&
	{\mathcal{O}}_{7}^\prime &= \frac{e}{g^2} m_b
	(\bar{s} \sigma_{\mu \nu} P_L b) F^{\mu \nu} , \\
	{\mathcal{O}}_{8} &= \frac{1}{g} m_b
	(\bar{s} \sigma_{\mu \nu} T^a P_R b) G^{\mu \nu \, a} ,&
	{\mathcal{O}}_{8}^\prime &= \frac{1}{g} m_b
	(\bar{s} \sigma_{\mu \nu} T^a P_L b) G^{\mu \nu \, a} , \\
	{\mathcal{O}}_{9} &= \frac{e^2}{g^2} 
	(\bar{s} \gamma_{\mu} P_L b)(\bar{\ell} \gamma^\mu \ell) ,&
	{\mathcal{O}}_{9}^\prime &= \frac{e^2}{g^2} 
	(\bar{s} \gamma_{\mu} P_R b)(\bar{\ell} \gamma^\mu \ell) , \\
	{\mathcal{O}}_{10} &=\frac{e^2}{g^2}
	(\bar{s}  \gamma_{\mu} P_L b)(  \bar{\ell} \gamma^\mu \gamma_5 \ell) ,&
	{\mathcal{O}}_{10}^\prime &=\frac{e^2}{g^2}
	(\bar{s}  \gamma_{\mu} P_R b)(  \bar{\ell} \gamma^\mu \gamma_5 \ell) , \\
	{\mathcal{O}}_{S} &=\frac{e^2}{16\pi^2} m_b
	(\bar{s} P_R b)(  \bar{\ell} \ell) ,&
	{\mathcal{O}}_{S}^\prime &=\frac{e^2}{16\pi^2} m_b
	(\bar{s} P_L b)(  \bar{\ell} \ell) , \\
	{\mathcal{O}}_{P} &=\frac{e^2}{16\pi^2} m_b 
	(\bar{s} P_R b)(  \bar{\ell} \gamma_5 \ell) ,&
	{\mathcal{O}}_{P}^\prime &=\frac{e^2}{16\pi^2} m_b
	(\bar{s} P_L b)(  \bar{\ell} \gamma_5 \ell), \\
	\end{align*}
	where $P_{L,R}=(1\mp \gamma_5)/2$, $\ell = e$ or $\mu$, and the explicit expressions for the QCD penguin operators $\mathcal O_{1-6}$ can be found in the ref. \cite{Bobeth:1999mk}.  Here, the operators $\mathcal{O}_{7,8,9,10}$ are present in the SM, and they may also appear in the extensions of the SM. The rest of the operators will appear only in the NP scenarios. Here, $ C_i $s and the $C'_i$s are the Wilson coefficients corresponding to the operator $ \mathcal{O}_i $ and $ \mathcal{O}'_i$, respectively. 
	
As we have mentioned, the contribution to $b\to s \ell^+\ell^-$ processes in our model will be from the diagrams in fig.~\ref{fig:btosll} with the following replacement $d_i\to b$ and $d_j\to s$. The amplitude of this diagram will depend on the $b\to s S$ effective vertex defined earlier and on the lepton vertex factor defined through the interaction 
\begin{equation}\label{eq:leptonS}
\mathcal{L}^{int}_{\ell\ell S} = m_{\ell}\bar{\ell} (  c_s  + i c_p\gamma_{5}) \ell \,.
\end{equation}
Like before, in the above equation we can take $c_s = \frac{\sqrt{2} g_s}{v}$ and $c_p = \frac{\sqrt{2} g_p}{v}$, and for simplicity we can consider $g_s$ and $g_p$ are the universal couplings, i.e same for quark and lepton interactions with the spin-0 particle. However, in principle, the coupling of quarks and leptons with the spin-0 particle could be different, we will discuss this point later. 

To estimate the NP contributions in $b\to s \ell^+\ell^-$ decays, we need to use eqs.~\eqref{eq:lbsS} and \eqref{eq:leptonS}, respectively. In the low energy limit $q^2 \ll M_S^2$ ($q^2 =$ invariant mass of the leptonic pair), after integrating out the heavy degrees of freedom we obtain from the diagram of fig. \ref{fig:btosll}: 
\begin{align}\label{eq:effHbtosll}
\mathcal{H}_{eff}^{NP} &= \frac{g^2 V_{tb} V^*_{ts} m_{\ell}}{32 \pi^2 M_S^2} \biggl (  C_1 [\bar{s} (m_s P_L + m_b P_R)b ] + C_2 [\bar{s} (m_s P_L - m_b P_R)b]  \biggr ) \biggl (  c_s [\bar{\ell}\ell] + i c_p[\bar{\ell} \gamma_{5}\ell]  \biggr ) \nonumber \\
 & = \frac{ 4 G_F  V_{tb} V^*_{ts} }{\sqrt{2}} \frac{M_W^2 m_{\ell}}{e^2 M_S^2}  \bigg( (C_1 +C_2) \frac{m_s}{m_b} (c_s \mathcal{O}'_S + i c_p \mathcal{O}'_P )  +   (C_1 -C_2) (c_s \mathcal{O}_S  + i c_p \mathcal{O}_P) \bigg).
 \end{align}
A direct comparison of this equation with the most general effective Hamiltonian given in eq.~\eqref{eq:Heff}, we will get 
\begin{align}
C'_S &= \frac{M_W^2 m_{\ell} m_s }{e^2 m_b M_S^2} c_s (C_1(\Lambda) + C_2(\Lambda))\,,\quad & C_S =  \frac{M_W^2 m_{\ell} }{e^2  M_S^2} c_s (C_1(\Lambda) - C_2(\Lambda))\,, \nonumber \\
C'_P &= \frac{M_W^2 m_{\ell} m_s }{e^2 m_b M_S^2} (i c_p) (C_1(\Lambda) + C_2(\Lambda))\,, \quad   &C_P = \frac{M_W^2 m_{\ell} }{e^2  M_S^2} (i c_p) (C_1(\Lambda) - C_2(\Lambda))\,. 
\end{align}
Therefore, in our working model the contributions in $b\to s \ell^+\ell^-$ will be via the four operators $\mathcal{O}_{S,P}$ and $\mathcal{O}'_{S,P}$. Note that all these four operators will contribute to the rates of $B\to K^{(*)}\mu^+\mu^-$ and $B_s \to \phi\mu^+\mu^-$ decays. Later we will discuss the constraints from a global analysis of all the data available in $b\to s\mu^+\mu^-$ processes. 
	
 Following the Hamiltonian given in eq.~\eqref{eq:Heff}, the model independent expression of the branching ratios of $ B_q \rightarrow \mu^+ \mu^- $ decay is given by \cite{Becirevic:2012fy}: 
		\begin{equation}
		\begin{split} \label{eq:BR_formula}
		\mathcal{B}(B_q \rightarrow \mu^+ \mu^-) = &\tau_{B_q} f_{B_q}^2  m_{B_q} \frac{G_F^2 \alpha^2}{64 \pi^3} |V^*_{tq}V_{tb}|^2 \beta_{\mu}(m_{B_q}^2) \left[   \frac{m_{B_q}^2}{m_b^2} |C_s - C'_s|^2 (1-\frac{4m_{\mu}^2}{m_{B_q}^2}) \right.\\&\left.  + \bigg|\frac{m_{B_q}}{m_b}(C_p - C'_p) + 2\frac{m_{\mu}}{m_{B_q}} (C_{10} - C'_{10})\bigg|^2 \right], 
		\end{split}
		\end{equation}
		where, $\beta_\ell(q^2)=\sqrt{  1- { 4 m_\ell^2/q^2}   }$ and the $ B  $ meson decay constant is defined via the matrix element: $ <0| \bar{s}  \gamma_{\mu} P_L b  |B_q(p)> =\frac{i}{2} f_{B_q} p_\mu $. In our model, the contributions will be in $C_S$, $C'_S$, $C_P$ and $C'_P$, not in $C'_{10}$. The SM contribution will be dependent only on $C_{10}$, including the QED corrections the SM predictions are given as \cite{Beneke:2019slt}:
		\begin{subequations}
		\begin{align}
		\mathcal{B}(B_s \rightarrow \mu^+ \mu^-) = (3.66 \pm 0.14)\times 10^{-9}\,,\\  \mathcal{B}(B_d \rightarrow \mu^+ \mu^-) = (1.03 \pm 0.05)\times 10^{-10}\,.
		\end{align}
		\end{subequations}
		Note that these SM predictions are fully consistent with the respective measurements shown in eq.~\eqref{eq:rare_exp}. 
		
	Another important observable is the branching fraction of the rare $ K_{L,S} \to  \mu^+ \mu^-  $ decay. The model-independent effective Hamiltonian will be similar to that given in eq.~\eqref{eq:Heff} with the following replacements $b \leftrightarrow s$ and $s \leftrightarrow d$. With this Hamiltonian, the model-independent expression for the branching ratio of $ K_L \rightarrow \mu^+ \mu^- $ decay is given by \cite{Buras:2013rqa} :  
		\begin{equation} \begin{split}\label{eq:BrKL2mumu}
		\mathcal{B}(K_L \rightarrow \mu^+ \mu^- ) =& \frac{G_F^4 m_W^4}{4 \pi^5} f_K^2  M_K \tau_{K_L} m_{\mu}^2   \sqrt{1- \frac{4 m_{\mu}^2}{M_K^2} }\sin^4\theta_W  \\& \left[ Re(V_{ts}^* V_{td}  \hat{P})^2 + Im(V_{ts}^* V_{td}  \hat{S})^2   \right]\,,
		\end{split}\end{equation}
		with 
		\[ \hat{P} = C_{10} - C'_{10} + \frac{m_K^2}{2 m_{\mu}} \frac{m_s}{m_s + m_d} (C_P - C'_P), \]
		\[ \hat{S} = \sqrt{1- \frac{4 m_{\mu}^2}{M_K^2} }  \frac{m_K^2}{2 m_{\mu}} \frac{m_s}{m_s + m_d} (C_S - C'_S) \,.  \]
		
		Similarly, branching ratio for $ K_S \rightarrow \mu^+ \mu^- $ is given by : 
		\begin{equation} \begin{split}\label{eq:BrKS2mumu}
		\mathcal{B}(K_S \rightarrow \mu^+ \mu^- ) =& \frac{G_F^4 m_W^4}{4 \pi^5} f_K^2  M_K \tau_{K_S} m_{\mu}^2   \sqrt{1- \frac{4 m_{\mu}^2}{M_K^2} }\sin^4\theta_W  \\& \left[\ Im(V_{ts}^* V_{td}  \hat{P})^2 + Re(V_{ts}^* V_{td}  \hat{S})^2   \right].
		\end{split}\end{equation}
	  In our case, we have non-zero contributions in $ C_S$, $C'_S$, $C_P$ and $C'_P$, respectively. 
	 
\paragraph{\underline{Invisible decay:} } \label{section_invisible}
It is important to note that the spin-0 mediator in our model could interact with the dark matter (eq.~\eqref{eq:fermionS}). Therefore, other important channels our model can contribute are the invisible decays of $B$ and $K$ mesons, like $B\to K +  invisible$ and $K \to \pi +  invisible$. In our case, the decays $B \to K\chi\chi$ and $K \to \pi \chi\chi$ will be considered as contributing $B\to K +  invisible$ and $K \to \pi +  invisible$ decays, respectively.

Deacys like $ B \to K \nu \bar{\nu}, ~ K \to \pi \nu \bar{\nu} $, which contains neutrinos as final state particles, are treated as invisible decays in SM since they are not detectable and experimental bounds come from treating them as missing energy. The measured branching fraction of $P \to P'\nu\bar{\nu}$ decay can be expressed as:
	\begin{equation}
	Br[P \to P'\nu\bar{\nu} ]_{Exp} = Br[P \to P' \nu \bar{\nu}]_{SM} + Br[P \to P' + \text{inv}] \,, 
	\end{equation}   
	where $P$ and $P'$ represent the parent and daughter mesons, respectively.  Hence, the measurement of the branching fractions can be utilized to estimate the branching fraction $Br[P \to P' + \text{inv}].$  Available information about branching ratios of invisible decays are \cite{E949:2007xyy,BaBar:2010oqg,Belle:2017oht,KOTO:2020prk}:\\
	\begin{subequations}
	\begin{align}
	&\mathcal{B}(K^+ \rightarrow \pi^+ \nu\bar{\nu})  = (14.7^{+13.0}_{-8.9}) \times 10^{-11}\,,  \\&
	\mathcal{B}(K_L \to \pi^0 \nu\bar{\nu}) < 4.9 \times 10^{-9}\,,  \\&
	\mathcal{B}(B^+ \to K^+ \nu\bar{\nu}) = (2.4 \pm 0.7) \times 10^{-5} \,,    \\&
		\mathcal{B}(B^+ \to \pi^+ \nu\bar{\nu}) = 1.4  \times 10^{-5} \,,    \\&
	\mathcal{B}(B_0 \to K_S \nu\bar{\nu})< 5.6 \times 10^{-5}\,, \\&
	\mathcal{B}(B_0 \to K_0 \nu\bar{\nu})< 2.6 \times 10^{-5}\,, \\&
	\mathcal{B}(B_0 \to \pi \nu\bar{\nu})< 9 \times 10^{-6} \,.
	\end{align}
	\end{subequations}
	In our model, we do not have any diagram contributing to the channel $ P \to P' \nu \bar{\nu}.  $ We will have invisible decay signature from the channel $ P \to p' \bar{\chi} \chi  $ which will contribute via a penguin diagram shown below.	

\begin{figure}[t]
		\centering
		\begin{tikzpicture}
		\begin{feynman}
		\vertex (a1){\(d_i\)};
		\vertex [crossed dot, above right=2cm of a1](a2){};
		\vertex [above left=2cm of a2](a3){\(d_j\)};
		\vertex [right=2cm of a2](a6);	
		\vertex [above right=1.5cm of a6](a7){\(\chi \)};
		\vertex [below right=1.5cm of a6](a8){\(\chi \)};		
		
		\diagram* { 
			(a1) --[fermion, arrow size=1.1pt](a2) --[fermion, arrow size=1.1pt](a3),
			(a8) --[fermion, arrow size=1.1pt](a6) --[fermion, arrow size=1.1pt](a7),
			(a2) --[scalar, edge label=\( S\)](a6),
		};	
		\end{feynman}
		\end{tikzpicture}
		\caption{Feynmann diagram contributing to the invisible decay process $ B \to K \bar{\chi}\chi$. The Crossed dot denotes the effective vertex coming from the loop.  }
		\label{fig:btoschichi}
	\end{figure}
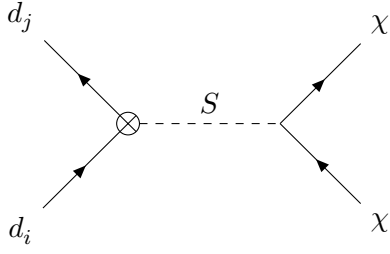

Here we treat the channel $  P \to P' \bar{\chi} \chi $ as invisible decay channel. A representative Feynman diagram of such a process is given in fig. \ref{fig:btoschichi}. For the decay of B meson (both charged and neutral), the underlying quark interaction process is  $ b\to s(d) \bar{\chi} \chi $ and the same for K-meson is $ s\to d \bar{\chi}\chi $. For the decay $ P\to P'\bar{\chi}\chi $ to be kinematically allowed, we will have the bound on the fermionic dark matter mass: $ 2 m_{\chi}\leq (M_P - M_{P'}) $. 
	
The amplitude for the process $ B \to K \bar{\chi}\chi $ will be calculated from the diagram \ref{fig:btoschichi}. In the low energy limit $q^2 \ll M_S^2$ ($M_S$ is heavy), after integrating out the heavy degrees of freedom we obtain   
\begin{align}	
\mathcal{M} &=	-\frac{g^2 V_{tb} V^*_{ts} }{32 \pi^2 M_S^2} \left(  C_1 (m_b+m_s) + C_2 (m_b-m_s)\right) \langle K|\bar{s}b|B \rangle \left(\bar{\chi}(c_{s\chi} + ic_{p\chi}\gamma_{5})\chi\right), \nonumber \\
& =(C_L + C_R)  \langle K|\bar{s}b|B \rangle \left(\bar{\chi}(c_{s\chi} + ic_{p\chi}\gamma_{5})\chi\right).
\end{align}
with 
\begin{equation}
C_L = 	\frac{g^2 V_{tb} V^*_{ts} }{32 \pi^2 M_S^2} (m_b+m_s) C_1(\Lambda) ,\ \ \ \ C_R = \frac{g^2 V_{tb} V^*_{ts} }{32 \pi^2 M_S^2} (m_b-m_s)  C_2(\Lambda) . 
\end{equation}

The matrix element is defined as 
\begin{equation}
	\langle K|\bar{s}b|B \rangle  = \frac{M_{B}^2 - M_{K}^2}{m_b-m_s} f_0(q^2)\,. 
	\end{equation}	
	Hence, the differential decay rate distribution for $ B \to K \bar{\chi}\chi $ decay will be obtained as 
	\begin{equation}\begin{split}
	\frac{d\Gamma_{B\to K \bar{\chi} \chi }}{dq^2} = \frac{1}{256 \pi^3 M_B^3} \lambda^{1/2}(M_B^2,M_K^2,q^2) \sqrt{1-\frac{4m_\chi^2}{q^2}} \left|C_L + C_R\right|^2 \frac{(M_B^2 -M_{K}^2)^2}{(m_b-m_s)^2} \times \\ f^2_{0}(q^2)\left( 2(c_{s\chi}^2 +c_{p\chi}^2)q^2 - 8c_{s\chi}^2 m_{\chi}^2\right).
	\end{split}\end{equation}
	The shape of the form factor will be obtained following the pole expansion 
	\begin{equation}
	f_0(q^2) = \frac{f_0(0)}{1-\beta \frac{q^2}{M^{*2}_B}}\,,
	\end{equation}
	with $ f_{0}(0)=0.33  $ \cite{Becirevic:1999kt,Parrott:2022rgu,Horgan:2013hoa}.
	The branching ratio depends on the couplings $ c_s$ and $c_p $, also, it contains the coupling of DM- mediator interaction $ c_{s\chi}$, $c_{p\chi}$ along with the DM mass, which have to be in the region mentioned above.

\subsection{FCCC Observables}\label{subsec:fccc}
 In the subsection \ref{subsec:FCCC_eff_vertex}, we have shown the contribution in our model to the effective vertices of FCCC. This type of correction will be relevant for the processes in which the SM contribution will be from the tree-level diagrams. Amongst these processes are the different semileptonic and leptonic decays of $B$, $B_s$, $K$ mesons. A wealth of data is available on various observables related to these decays, which we use to constrain the CKM elements. In this analysis, we will use all these observables on the leptonic and semileptonic decays to constrain the new couplings. In our model, there will be a contribution to the anomalous $t\to b W$ vertex. The estimates on these couplings from the data are available in the literature, which we will utilise to constrain the new parameters. In the following we will discuss all such contributions.

	\paragraph{\underline{Anomalous Couplings of $ t\to b W $ vertex:}}
	The general Lagrangian for $ t \to b W_{\mu}^- $ decay is given by \cite{CMS:2020ezf,CMS:2020vac}:
	\begin{equation}\label{t_to_bW_4}
	\mathcal{L}_{tWb} = - \frac{ g}{\sqrt{2}} \bar{b} \gamma_{\mu} \left(V_L P_L + V_R P_R\right) t W_{\mu}^{-} - \frac{ g}{\sqrt{2}} \bar{b} ~\frac{i \sigma_{\mu \nu}q_{\nu} }{m_W} \left(g_LP_L + g_R P_R\right) t W_{\mu}^{-} + h.c.
	\end{equation}
	$ V_{L,R},~ g_{L,R} $ are knows as the anomalous couplings. In Standard Model, we have $ V_L=V^*_{tb} $ and $ V_R,~g_L,~g_R=0 $. The experimental constraints on these couplings are given in table \ref{tab:tbW_exp}.  
	
		\begin{table}[t]
			\begin{center}
				\begin{tabular}{|c |c |c| c|}
					\hline
					\multicolumn{4}{|@{}c|}{95\% CL interval }\\
					\hline
					Coupling & ATLAS \cite{ATLAS:2016fbc} & CMS  \cite{CMS:2016asd,CMS:2020ezf,CMS:2020vac}  & ATLAS+CMS combination\\
					\hline
					Re($V_{\text{R}}$) & $[-0.17, 0.25]$ & $[-0.12, 0.16]$ & $[-0.11, 0.16]$\\
					Re($g_{\text{L}}$) & $[-0.11, 0.08]$ & $[-0.09, 0.06]$ & $[-0.08, 0.05]$\\
					Re($g_{\text{R}}$) & $[-0.03, 0.06]$ & $[-0.06, 0.01]$ & $[-0.04, 0.02]$\\
					\hline
					$ V_L $ \cite{PDG:2022} &\multicolumn{3}{c|}{0.995 $ \pm $ 0.021 }  \\
					\hline
				\end{tabular}
				\caption{Experimental limits on the coefficients of left and right handed vector current in $ t \to b W_{\mu}^{-} $ decay.}
				\label{tab:tbW_exp}
			\end{center}
		\end{table}		
 
	Here, we will get the $ t \to b W_{\mu}^{-} $ decay via the Feynman diagrams of figs. \ref{fig:FCCC_vertex_1} and \ref{fig:FCCC_self_1}, respectively, with the replacement; $u_i \to t$ and $d_j\to b$. Therefore, in our working model, summing up all the contributions from the loops the amplitude looks like this:
	\begin{equation}\label{tbw_eqn}
	\begin{split}
	\mathcal{M}_{tWb}=- \frac{ g V^*_{tb}}{\sqrt{2}} C_{VL} \left[\bar{b}\gamma_{\mu} P_Lt \right]W_{\mu}^- - \frac{ g V^*_{tb}}{\sqrt{2}} C_{VR} [\bar{b}\gamma_{\mu} P_R t]W_{\mu}^- \,.
	\end{split}
	\end{equation}
	So, we will have non-zero contributions to the anomalous couplings $ V_{L} $
 	and $ V_R. $ Comparing eq.~\eqref{tbw_eqn} with eq.~\eqref{t_to_bW_4}, $ V_L  $ and $ V_R $ can be written as:
	\begin{equation}\begin{split}
	& V_R = V^*_{tb}~ C_{VR}\, , \\&
	V_L = V^*_{tb}~(1+ C_{VL})\,. 
	\end{split} \end{equation}
	The factor  `1\rq \,  comes in $ V_L $ from the tree level SM contribution of the $  t \to b W_{\mu}^{-} $ decay. The contributions to $g_L$ and $g_R$ will be negligibly small compared to $V_R$ or $V_L$.

	\paragraph{\underline{Contribution to the Semi-leptonic and leptonic decays:}}
	
	\begin{figure}[t]
		\centering
		\begin{tikzpicture}
		\begin{feynman}
		\vertex (a1){\(d_j\)};
		\vertex [blob,minimum size=0.3cm, above right=2cm of a1,blue](a2){};
		\vertex [above left=2cm of a2](a3){\(u_i\)};
		\vertex [right=2cm of a2](a6);	
		\vertex [above right=1.5cm of a6](a7){\(\ell \)};
		\vertex [below right=1.5cm of a6](a8){\(\nu_{\ell} \)};		
		
		\diagram* { 
			(a1) --[fermion, arrow size=1.1pt](a2) --[fermion, arrow size=1.1pt](a3),
			(a8) --[fermion, arrow size=1.1pt](a6) --[fermion, arrow size=1.1pt](a7),
			(a2) --[boson, edge label=\( W\)](a6),
		};	
		\end{feynman}
		\end{tikzpicture}
		\caption{Feynman diagram contributing to the FCCC semileptonic and leptonic processes. The blue blob represents the effective vertex coming from the loop. }
		\label{fig:FCCC_eff}
	\end{figure}
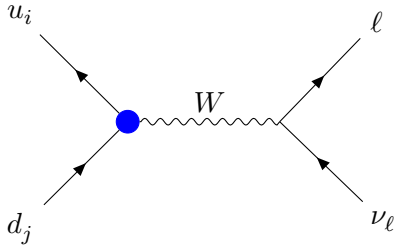

	The corrections to the SM charged current vertices will impact the semileptonic and leptonic decays of $B$, $B_s$, $K$, $D$ and $D_s$ mesons. The available inputs on these decay modes are widely used in the extractions of CKM elements \cite{ParticleDataGroup:2022pth}. We get a contribution to these leptonic and semileptonic decays via the diagram in fig. \ref{fig:FCCC_eff}. Hence, in the low energy limit, after integrating out the heavy degrees of freedom, in our working model the most general Hamiltonian that contains all possible four fermion operators of lowest dimension for $ d_j \to u_i \ell \bar{\nu} $ can be written as   
	\begin{equation}\label{eq:utodceff}
	\mathcal{H}_{eff}=\frac{4 G_F}{\sqrt{2}} V_{u_i d_j}\left[ (1+C_{VL})\mathcal{O}_{VL}  + C_{VR} \mathcal{O}_{VR} \right].
	\end{equation}
The four-fermi operators are given by  
		\begin{align}
		\mathcal{O}_{VL} &=  (\bar{u}_{iL} \gamma^{\mu} d_{jL})(\bar{\ell}_{L} \gamma_{\mu} \nu_{\ell L})\,, \nonumber \\
		\mathcal{O}_{VR} &=  (\bar{u}_{iR} \gamma^{\mu} d_{jR})(\bar{\ell}_{L} \gamma_{\mu} \nu_{\ell L}) \,. 
		\end{align}
For a detail see the discussion in subsection \ref{subsec:FCCC_eff_vertex}. Note that here we have assumed the neutrinos to be left-handed as in the SM. We will get the SM results when $C_{VL} = C_{VR} = 0$. In our case, we don't have NP contribution to the leptonic vertex, since we have taken the couplings to be MLV type (no interactions with the massless neutrinos). We get a non-zero contribution to the Wilson coefficients $ C_{VL} $ and $ C_{VR} $ from the loop calculations. We have noted in subsection \ref{subsec:FCCC_eff_vertex} that the corrections to the charged current vertex $d_j \to u_i W$ are proportional to the product of the masses of the external quarks. In addition, we have loop factors that are proportional to $m_{u_i}^2$ and $m_{d_j}^2$. Therefore, the contributions in kaon decays will be proportional to the masses of $u$, $d$, or $s$ quarks, which will be highly suppressed with respect to the respective tree-level contributions in the SM. For completeness we have kept these constributions in our analysis. 

Note that the semileptonic and leptonic decay rates are used to extract the CKM elements. 
From the effective Hamiltonian given in eq.~\eqref{eq:utodceff}, we obtain the differential rate for $P \to M \ell \nu_\ell$ (P and M are pseudoscalar mesons) decay corresponding to the quark level transition $d_j\to u_i \ell\nu_{\ell}$ as
{\footnotesize
\begin{equation}
\frac{d\Gamma (P \to M \ell \nu_\ell)}{dq^2} = \frac{G_F^2 |V_{u_id_j}|^2}{\pi^3 m_P^3} q^2 \sqrt{\lambda_M (q^2)} \left(1-\frac{m_\ell^2}{q^2}\right) |1+C_{VL} + C_{VR}|^2 \left \{ \left(1+\frac{m_\ell^2}{2q^2}\right) {H_{V,0}^{s}}^2 + \frac{3}{2}\frac{m_\ell^2}{q^2} {H_{V,t}^s}^2 \right \}.
\end{equation}}
Here, the helicity amplitudes are written in terms of the QCD form factors as given below 
\begin{subequations}
	\begin{align}
	H_{V,0}^s(q^2) &  = \sqrt{\lambda_M(q^2) \over q^2} f_+(q^2) \,, \\
	& \nonumber \\
	H_{V,t}^s(q^2) &  = {m_P^2-m_M^2 \over \sqrt{q^2}} f_0(q^2).
	\end{align}
\end{subequations}
Also, the branching fraction for $P\to\ell\nu_{\ell}$ corresponding to the same Hamiltonian is:
\begin{equation}\label{eq:brlep}
\begin{split}
\mathcal{B}(P\to\ell\nu_{\ell}) = & \frac{\tau_P}{8\pi}m_P m_\ell^2 f_P^2 G_F^2 \left(1-\frac{m_\ell^2}{m_P^2}\right)^2 \left|V_{u_i d_j}(1+C_{VL}-C_{VR}) \right|^2.
\end{split}
\end{equation}
We can note from the expressions of the rates that in our working model the new contributions in $P \to M \ell \nu_\ell$ and $P\to\ell\nu_{\ell}$ decays will modify only the vertex factor. Hence, it may impact the overall normalization of the rates but not the shape of the $q^2$ distributions. Therefore, if we extract the CKM elements from these rate distributions, then we will be extracting $|V_{u_id_j}^\prime|= |V_{u_id_j} (1+ C_{VL} \pm C_{VR})|$ instead of $|V_{u_id_j}|$.  Therefore, the CKM elements $|V_{u_id_j}^{\prime}|$ extracted from purely leptonic or $P\to M\ell\nu_{\ell}$ decays, can be directly used to constrain the new parameters along with the Wolfenstein parameters: $A, \lambda, \rho$ and $\eta$ with which we need to parametrize $|V_{u_id_j}|$. In this analysis, to constrain the new couplings, we will use the extracted values of the CKM elements from these semileptonic or leptonic modes. For details on the extractions of the CKM elements, the reader may look at the reviews on this topic available in PDG \cite{ParticleDataGroup:2022pth}.

\subsection{Eletroweak Precision observables}

\begin{figure}[t]
	\centering 
	\subfloat[]{\begin{tikzpicture}
		\begin{feynman}
		\vertex (a1);
		\vertex [right=1.5cm of a1](a2);
		\vertex [right=1.5cm of a2](a3);
		\vertex [right=1.2cm of a3](a4);
		
		\diagram* { 
			(a1) --[boson, edge label=\(W \)](a2) --[boson, edge label'=\(W \)](a3) --[boson, edge label=\(W \)](a4),
			(a2) --[scalar, half left, looseness=2, edge label = \(S\)] (a3),
		};
		\end{feynman}
		\end{tikzpicture}\label{fig:W_self_mass}} \quad
	\subfloat[]{\begin{tikzpicture}
		\begin{feynman}
		\vertex (a1);
		\vertex [right=1.5cm of a1](a2);
		\vertex [right=1.5cm of a2](a3);
		\vertex [right=1.2cm of a3](a4);
		
		\diagram* { 
			(a1) --[boson, edge label=\(Z \)](a2) --[boson, edge label'=\(Z \)](a3) --[boson, edge label=\(Z \)](a4),
			(a2) --[scalar, half left, looseness=2, edge label = \(S\)] (a3),
		};
		\end{feynman}
		\end{tikzpicture}\label{fig:Z_self_mass}}
	\caption{Feynman diagram for self energy correction of $ W $ and $ Z $ boson. }
	\label{fig:selfWZ}
\end{figure}
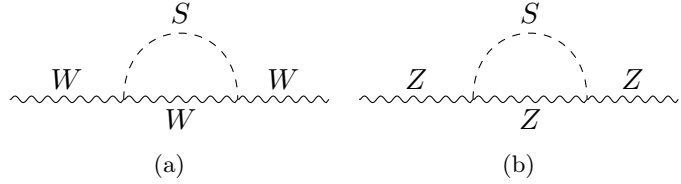

Due to the precise data on the electroweak precision observables (EWPO) at the $W$ and $Z$ poles it is possible to place constraints on the new physics scenarios by studying the loop-level contributions of the new physics to electroweak observables.

\paragraph{\underline{Oblique parameters}:}
Beside various FCNC and FCCC processes that we have mentioned earlier, we will also have contributions in various other observables. For example, our working model will contribute to $W$ and $Z$ boson self energies. The diagrams are shown in figs. \ref{fig:W_self_mass} and \ref{fig:Z_self_mass}, respectively.
The most general expression for the gauge boson ($V = W$ or $Z$) self-energeis can be written as 	
\begin{equation}\label{eq:selfV}
\Sigma_V(q^2) = \bigg(g^{\mu\nu} - \frac{q^{\mu} q^{\nu}}{q^2}\bigg)\Sigma_{V,T}(q^2) + \frac{q^{\mu} q^{\nu}}{q^2}\Sigma_{V,L}(q^2)\,.
\end{equation}
 
Any non-standard contributions to the gauge boson self-energies will contribute to the observables related to the electroweak precision test. Hence, we may get tight constraints on the new physics parameters from the electroweak observables. One of such important observable is the $\rho$ parameter which is defined as: \cite{Ross:1975fq}  	
\begin{equation}
\rho = \frac{G_{NC}}{G_{CC}} = \frac{M_W^2}{c_W^2 M_Z^2} = 1.
\end{equation}
The deviation from unity due to various higher-order corrections in the SM, and in the BSM scenarios can be expressed as:
\begin{equation}
\rho = \frac{1}{1-\Delta \rho},
\end{equation}	
where, 
\begin{equation}
\Delta\rho = (\Delta\rho)^{(1)} + (\Delta\rho)^{(2)} + ....
\end{equation}
The contributions in $\Delta\rho$ can be expressed as the sum of the loop corrections at different orders. The SM contributions to $\Delta\rho$ at one and two levels can be seen from the refs. \cite{Veltman:1977kh,Chanowitz:1978mv,vanderBij:1983bw}. The contributions at the one-loop level can be expressed in terms of the gauge boson self-energies as follows
\begin{equation}\label{eq:rhonp}
(\Delta\rho)^{(1)}_{} = \frac{\Sigma_{Z,T}(0)}{M_Z^2} - \frac{\Sigma_{W,T}(0)}{M_W^2}\,.
\end{equation}
Note that in our working model, we will discuss the corrections due to the NP only at one loop level.

The leading logarithmic contributions to the $W$ and $Z$ boson self energies from the diagrams in figs. \ref{fig:W_self_mass} and \ref{fig:Z_self_mass} are given by  
\begin{align}
\nonumber\Sigma_{W,T}(q^2) &= \frac{c_G^2 M_W^4}{4 \pi^2}\bigg[ \bigg( \frac{3}{8}   - \frac{M_S^2}{8 M_W^2} + \frac{q^2}{24 M_W^2} \bigg)\left( \log \frac{\Lambda^2}{M_S^2} + \log \frac{\Lambda^2}{M_W^2} \right) - \frac{(M_S^2-M_W^2)}{24 M_W^2 q^4} \times \nn \\
&\quad \bigg((M_S^2-M_W^2)^2 - 3 (M_S^2-3 M_W^2) q^2 \bigg) \log\frac{M_S^2}{M_W^2}\bigg],
\end{align}
and 
\begin{align}
\nonumber\Sigma_{Z,T}(q^2) &= \frac{c_G^2 M_Z^4}{4 \pi^2}\bigg[ \bigg( \frac{3}{8}   - \frac{M_S^2}{8 M_Z^2} + \frac{q^2}{24 M_Z^2} \bigg)\left( \log \frac{\Lambda^2}{M_S^2} + \log \frac{\Lambda^2}{M_Z^2} \right)- \frac{(M_S^2-M_Z^2)}{24 M_Z^2 q^4} \times  \nn \\
&\quad \bigg((M_S^2-M_Z^2)^2 - 3 (M_S^2-3 M_Z^2) q^2 \bigg) \log\frac{M_S^2}{M_Z^2}\bigg].
\end{align}

The effect of non-standard contributions to the EWPO is also parametrised in terms of the $[S, T, U]$ parameters \cite{Peskin:1991sw}. Among these parameters, $T$ is related to $\Delta\rho$ calculated at one loop order via the relation
\begin{equation}\label{eq:T}
\Delta\rho^{(1)} = - \alpha_{em} \Delta T. 
\end{equation}
The other parameters are also expressed in terms of the self energies of the gauge Bosons, in our working model which are given by 
\begin{align}\label{eq:SU}
\Delta S &= \frac{4 s_W^2 c_W^2}{\alpha_{em}} \frac{\Sigma_{Z,T}(M_Z^2)- \Sigma_{Z,T}(0)}{M_Z^2}, \nn \\
\Delta U & =\frac{4 s_W^2 }{\alpha_{em}}\bigg( \frac{\Sigma_{W,T}(M_W^2)- \Sigma_{W,T}(0)}{M_W^2} - c_W^2 \frac{\Sigma_{Z,T}(M_Z^2)- \Sigma_{Z,T}(0)}{M_Z^2}\bigg). 
\end{align} 
Another important parameter is $\Delta r$, which is commonly used to indicate the shift in the $W$ mass due to NP. We could extract it by solving the following equation 
\begin{equation}\label{eq:delrMW}
M_W^2 \left(1- \frac{M_W^2}{M_Z^2} \right) = \frac{\pi \alpha_{em}}{\sqrt{2} G_F} \frac{1}{1- \Delta r}.
\end{equation}

\begin{table}[t]
	\footnotesize
	\centering
	\renewcommand{\arraystretch}{1.4}
	\begin{tabular}{|c|c|c|c|}
		\hline
		Measured Value $M_W$ & Reference & $ \Delta r $ & $ \delta(\Delta r ) $ eq.~\eqref{eq:deltar} \\
		\hline
		$ 80.357 \pm 0.006 $ GeV &  SM \cite{PDG:2022} &  $ -0.03068 \pm 0.00040 $ & $ - $\\
		$ 80.4335 \pm 0.0094 $ GeV & CDF \cite{CDF:2022hxs} & $ -0.03526 \pm 0.00059 $ & $  (-4.58316 \pm 0.66899) \times 10^{-3}$ \\
		$ 80.354 \pm 0.030 $ GeV & LHCb\cite{LHCb:2021bjt} & $ -0.03050\pm 0.00190 $ & $ (0.17915 \pm 1.92090) \times 10^{-3}  $ \\
		$ 80.360 \pm 0.016 $ GeV & ATLAS\cite{ATLAS:2023fsi} & $ -0.03086 \pm 0.00097$ & $ (-0.17919 \pm 1.02076) \times 10^{-3} $ \\
		$ 80.367 \pm  0.023 $ GeV & D0 \cite{D0:2012kms} &  $ -0.03128 \pm  0.00156 $ & $ (-0.59747 \pm 1.42068) \times 10^{-3} $  \\
		\hline
	\end{tabular}
	\caption{SM and experimental values of $ M_{W} $ from different experiments like CDF, LHCb, ATLAS, D0 along with the values of $ \Delta r $ and $ \delta(\Delta r) $, which is calculated using eq.~\eqref{eq:deltar}.}\label{tab:mwdelr}
\end{table}

In the oblique approximation, the NP contributions in $\Delta r$  can be expressed in terms of the $[S,T,U]$ parameters via the equation
\begin{equation}\label{eq:delr}
\Delta r = \frac{\alpha_{em}}{s_W^2} \left( - \frac{1}{2} \Delta S + c_W^2 \Delta T + \frac{c_W^2-s_W^2}{4 s_W^2} \Delta U  \right).
\end{equation}
From eq.~\eqref{eq:delrMW} we can estimate the measured values of $\Delta r$ using the measured values of $M_Z$ and $M_W$. The NP parameters can then be extracted using eqs.~\eqref{eq:T}, \eqref{eq:SU} and \eqref{eq:delr}. In table \ref{tab:mwdelr}, we have estimated the values of $\Delta r$ in the SM and using the measured values of $M_W$ by the different experiments. Finally, in each of these cases we have estimated 
\begin{equation}\label{eq:deltar}
\delta(\Delta r) = (\Delta r)_{Exp} - (\Delta r)_{SM},
\end{equation} 
where $(\Delta r)_{SM}$ and $(\Delta r)_{Exp}$ are the SM and the measured values of $\Delta r$. Therefore, $\delta(\Delta r)$ will be sensitive to beyond the SM contributions in $\Delta S$, $\Delta U$ and $\Delta T$ given in eqs.~\eqref{eq:SU} and \eqref{eq:T}, respectively. Note that the $\delta(\Delta r)$ obtained using the measurement of $M_W$ by CDF \cite{CDF:2022hxs} is not consistent with zero while the rest of the estimates are consistent with zero. In the analysis, We have used the weighted mean of the other measurements.

\paragraph{\underline{Z-pole Observables}:}

\begin{figure}[t]
		\footnotesize
		\centering 
		\subfloat[]{\begin{tikzpicture}
			\begin{feynman}
			\vertex (a1){\(Z\)};
			\vertex [right=1.3cm of a1](a2);
			\vertex [below right=1.3cm of a2](a3);
			\vertex [above right=1.3cm of a2](a4);
			\vertex [right=1.3cm of a3](a5){\(f\)};
			\vertex [right=1.3cm of a4](a6){\(f\)};
			
			\diagram* { 
				(a1) --[boson](a2),
				(a2) --[scalar, edge label'={\(S\)}](a3),
				(a2) --[boson, edge label={\(Z\)}](a4),
				(a5) --[fermion, arrow size=1.1pt](a3) --[fermion, arrow size=1.1pt, edge label={\(f\)}](a4) --[fermion, arrow size=1.1pt](a6),			
			};	 	
			\end{feynman}
			\end{tikzpicture}\label{fig:Zpole_1}}
		\subfloat[]{\begin{tikzpicture}
			\begin{feynman}
			\vertex (a1);
			\vertex [right=1.3cm of a1](a2);
			\vertex [right=1cm of a2](a3);
			\vertex [right=1cm of a3](a4);
			\vertex [above right=1.3cm of a4](a5){\(f\)};
			\vertex [below right=1.3cm of a4](a6){\(f\)};
			
			\diagram* {
				(a1) --[boson, edge label={\(Z\)}](a2) --[boson, edge label'={\(Z\)}](a3) --[boson, edge label={\(Z\)}](a4),
				(a2) --[scalar, half left, looseness=2, edge label = \(S\)](a3),
				(a6) --[fermion, arrow size=1.1pt](a4) --[fermion, arrow size=1.1pt](a5);
				
			}; 
			\end{feynman}
			\end{tikzpicture}\label{fig:Zpole_2}}
		\subfloat[]{\begin{tikzpicture}
			\begin{feynman}
			\vertex (a1);
			\vertex [right=1.3cm of a1](a2);
			\vertex [above right=0.5cm of a2](a3);
			\vertex [above right=0.5cm of a3](a4);
			\vertex [above right=0.6cm of a4](a5){\(f\)};
			\vertex [below right=1.5cm of a2](a6){\(f\)};
			
			\diagram* {
				(a1) --[boson, edge label={\(Z\)}](a2),
				(a6) --[fermion, arrow size=1.1pt](a2) --[fermion, arrow size=1.1pt,edge label'={\(f\)}](a3) --[fermion, arrow size=1.1pt, edge label'={\(f\)}](a4) --[fermion, arrow size=1.1pt](a5),
				(a3) --[scalar, half left, looseness=2, edge label={\(S\)}](a4),
			};	 	
			
			\end{feynman}
			\end{tikzpicture}\label{fig:Zpole_3}}
		\caption{All the diagrams arising from this model which are constributing to $ Z- $ pole observables of the process $ Z \to f \bar{f}. $}\label{fig:Z-pole_Feynman}
	\end{figure}
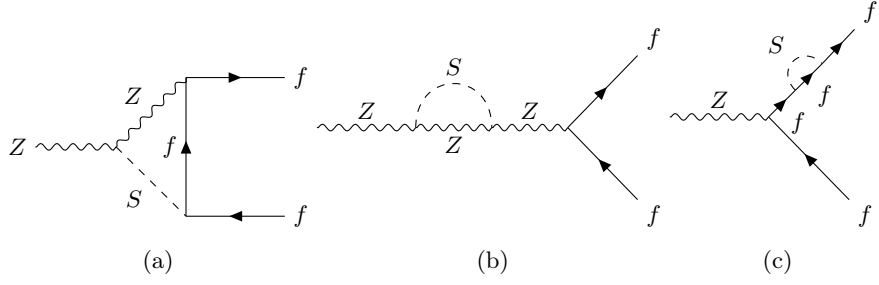

Our working model also contributes to the process $ Z \to  \bar{f} f $ ($f$ = SM fermions). The corresponding diagrams are shown in fig. \ref{fig:Z-pole_Feynman}. As a result, we could obtain the constraints on the model parameters from the measured values of the $Z$-pole observables.
The one loop order correction to $Z\to \bar{f} f $ vertex in our model is given in fig. \ref{fig:Zpole_1}. The counter-term diagrams are shown in figs. \ref{fig:Zpole_2} and \ref{fig:Zpole_3}, respectively. Following are the contributions from these three diagrams: 
\begin{gather}\label{eq:zff1}
\mathcal{L}_{Zff}^{\ref{fig:Zpole_1}} = \frac{g}{\cos \theta_W} \left(\frac{c_G m_{f} m_z^2}{16 \pi^2} \right)  \frac{a_f m_{f}}{2 m_z^2} \biggl[     \left( ic_p \log \frac{\Lambda^2}{m_{f}^2} - c_s \log \frac{\Lambda^2}{M_S^2} \right)[\bar{u}(p_3)] \gamma_{\mu} P_L v(p_2)] \nonumber  \\ 
+  \left( c_s \log \frac{\Lambda^2}{M_S^2} + i c_p \log \frac{\Lambda^2}{m_{f}^2}\right)[\bar{u}(p_3)] \gamma_{\mu} P_R v(p_2)]  \biggr] Z^{\mu},
\end{gather}

	\begin{gather}\label{eq:zff2}
	\mathcal{L}_{Zff}^{\ref{fig:Zpole_2}} = \frac{g}{\cos \theta_W } \frac{c_G^2 m_z^2}{192 \pi^2} \left( \log \frac{\Lambda^2}{M_S^2} + \log \frac{\Lambda^2}{m_z^2}\right) 
	\biggl(  \frac{v_f + a_f}{2} \left[\bar{u}(p_3) \gamma_{\mu}P_L v(p_2)\right] \nonumber \\
	  + \frac{v_f - a_f}{2} \left[\bar{u}(p_3) \gamma_{\mu}P_R v(p_2)\right] \biggr) Z^{\mu},
	\end{gather}
	and 
	\begin{gather}\label{eq:zff3}
	\mathcal{L}_{Zff}^{\ref{fig:Zpole_3}} = -\frac{g}{\cos \theta_W}  \frac{(c_s^2 + c_p^2)m_f^2}{128 \pi^2} \left( \log \frac{\Lambda^2}{M_S^2} + \log\frac{\Lambda^2}{m_f^2}\right)   \biggl(  \frac{v_f + a_f}{2} \left[\bar{u}(p_3) \gamma_{\mu}P_L v(p_2)\right] \nonumber \\ + \frac{v_f - a_f}{2} \left[\bar{u}(p_3) \gamma_{\mu}P_R v(p_2)\right] \biggr)Z^{\mu}.
	\end{gather}
	In the above equations the couplings in the SM are defined as
	\begin{equation}
	v_f= I_W^{(3)} - 2 Q \sin^2 \theta_W \quad\text{and} \quad a_f= I_W^{(3)}.
	\end{equation} 
	
The decay width of $Z\to f\bar{f}$ process can be written as \cite{Soni:2010xh,Bernabeu:1990ws,CarcamoHernandez:2005ka,Martinez:2014lta} : 
\begin{align}\label{eq:decaywidth_Zpole}
\Gamma_{tot}(Z \to f \bar{f}) = \frac{N_c^b}{48} \frac{\alpha}{s_W^2 c_W^2} m_Z \sqrt{1-\mu_{f}^2}& \biggl( |g_{af}|^2 (1-\mu_{f}^2) + |g_{vf}|^2(1+\frac{\mu_{f}^2}{2})\biggr)(1+\delta_{f}^0)(1+\delta_{b}) \nonumber \\&  (1+\delta_{QCD})(1+\delta_{QED}) (1+\delta^{f}_{\mu}) \,,
\end{align}
with $ \mu_{f}^2 = \frac{4 m_{f}^2}{m_{Z}^2}$. Other terms like $ \delta_{f}^0, ~ \delta_{b}, ~\delta_{QCD}, ~\delta_{QED}, ~\delta_{f}^{\mu} $ are the various corrections to the leading order rate, for a detail see the refs.  \cite{Soni:2010xh,Bernabeu:1990ws}. The variable $g_{af}$ and $g_{vf}$ are the effective axial and vector couplings of fermion pair with Z-boson, respectively. In our case, the NP contributions will modify the vertex factor i.e  \begin{gather}
\nonumber g_{af} \to a_f + \Delta a_{f}^{NP} \,, \\  g_{vf} \to v_f + \Delta v_f^{NP} \,. 
\end{gather}
We can extract $\Delta a_{f}^{NP}$ and $\Delta v_f^{NP}$ from the eqs.~\eqref{eq:zff1}, \eqref{eq:zff2} and \eqref{eq:zff3}, respectively. 

	LEP, SLAC obtained values for various branching ratios in order to have a better control of the systematic uncertainties. There are different observables related to this process, commonly known as the $Z$-pole observables, which have been measured with a reasonably good accuracy \cite{PDG:2022}. Among those observables, we have taken the following ratios of the decay rate \cite{Bernabeu:1996zh,Papavassiliou:2000pq}
	\begin{equation}\label{eq:RZobs}
	R_{\ell} = \frac{\Gamma_{had}}{\Gamma_{\ell}} \quad, ~~R_{c} = \frac{\Gamma_{c}}{\Gamma_{had}} \quad, ~~R_{b} = \frac{\Gamma_{b}}{\Gamma_{had}}. 
	\end{equation}  
	We have  
	 $ \Gamma_{had} = \Gamma_{u}+\Gamma_{d} + \Gamma_{s} + \Gamma_{c}+\Gamma_{b}$.
In our working model, the effective vertex functions for $ Z \to f \bar{f} $ decays depend on the respective fermion masses. Hence, the most important contribution will come from $ Z \to b \bar{b} $ or $ Z \to \tau \bar{\tau} $ decays. In the presence of contributions from the BSM, the more general expressions for $ R_{b} $ and $R_{\ell}$ can be written as \cite{Bernabeu:1996zh,Papavassiliou:2000pq}
\begin{align}
R_{b} = \frac{\Gamma_{b}}{\Gamma_{had}} = \frac{\Gamma_b^{SM} + \delta \Gamma_{b}^{NP}}{\Gamma_{had}^{SM} + \delta \Gamma_{b}^{NP} }= R_{b}^{SM} \frac{1+\delta_{b}^{NP}}{1+R_{b}^{SM} \delta_{b}^{NP}}\,,
\end{align}
Where, $ \delta_{b}^{NP}   \approx \frac{\delta \Gamma_{b}^{NP}}{\Gamma_{b}^{SM}}$. Similarly,
\begin{align}
R_{\ell} = \frac{\Gamma_{had}}{\Gamma_{\ell}} = \frac{\Gamma_{had}^{SM} + \delta \Gamma_{b}^{NP}}{\Gamma_{\ell}^{SM} + \delta \Gamma_{\ell}^{NP}}=R_{\ell}^{SM}\left( 1+ \frac{\delta \Gamma_{b}^{NP}}{\Gamma_{had}^{SM}}- \frac{\delta \Gamma_{\ell}^{NP}}{\Gamma_{\ell}^{SM}}  \right)\,.
\end{align}	
	
The measured value of these ratios are given in table \ref{tab:CKM-updated-obs}. 
	The SM values are given below \cite{Freitas:2014hra,PDG:2022} :
\begin{eqnarray}
 & R^{SM}_{e} = 20.736 \pm 0.010, \quad   R^{SM}_{\mu} = 20.736 \pm 0.010, \quad R^{SM}_{\tau }=20.781 \pm 0.010, \nonumber \\&
    R^{SM}_{b} = 0.21582 \pm 0.00002, \quad  R^{SM}_{c}= 0.17221 \pm 0.00003.
\end{eqnarray}

Measurements of production and decay asymmetries of $e^+e^- \to Z \to f\bar{f}$ with the polarised electron beam are helpful to extract the effective electroweak $Zf\bar{f}$ coupling. At Born level, the differential cross section for the process $e^+e^- \to Z \to f\bar{f}$ can be expressed as a function of the polar angle $\theta$ of the fermion relative to the electron beam direction \cite{SLD:2004kjl},
\begin{equation}\label{eq:diffdist}
\frac{d\sigma_f}{d\cos\theta} \propto (1- A_e P_e )(1+\cos^2\theta) + 2 A_f (A_e - P_e)\cos\theta.
\end{equation}
Here, $P_e$ is the longitudinal polarization of the electron beam. Therefore, from a fit to the above differential cross section separately for predominantly left- and right-handed beam, we can simultaneously determine the initial and final state asymmetries $A_e$ and $A_f$, respectively. Note that $P_e < 0$ represents the left-handed beam electrons and $P_e > 0$ mostly right-handed beam electrons, and $\theta$ is the polar angle of the fermion relative to the electron beam direction.  Also, from eq.~\eqref{eq:diffdist} we could extract the left-right asymmetry \cite{SLD:1997qsa,ParticleDataGroup:2022pth}
\begin{equation}
A_{LR} = \frac{\sigma_{e_L}- \sigma_{e_R} }{\sigma_{e_L}+\sigma_{e_R}} = A_e \,,
\end{equation}
and the polarised forward-backward asymmetry 
\begin{equation}
A_{FB}^f = \frac{\sigma_{\cos\theta >0}- \sigma_{\cos\theta<0} }{\sigma_{\cos\theta >0}+ \sigma_{\cos\theta<0}} =\frac{3}{4} |P_e| A_f \,,
\end{equation}
Hence, the measurements of these asymmetries will allow one to extract $A_e$ and $A_f$ independently. In the above equation, the asymmetries $A_f$ can be written as 
 \begin{equation}
 A_{f} = 2 \frac{g_{af}~ g_{vf}}{(1- \mu_f^2 ) g_{af}^2 + (2 + \mu_f^2 )g_{vf}^2/2 } \,,
 	\end{equation}
which are directly sensitive to the effective vertices. In this analysis, we have considered the measured values of $A_{\tau}$, $A_{\mu}$, $A_{e}$, $A_{b}$ and $A_{c}$. For the light fermions, the above expression for the asymmetry will be 
\begin{equation}
 A_{f} = 2 \frac{g_{af} ~g_{vf} }{ g_{af}^2 + g_{vf}^2 }\,,
\end{equation}
 One could also extract these asymmetries, from the measurements of forward-backward asymmetries $A_{FB}^f$ formed with an unpolarized electron beam ($P_e = 0$). Such a forward-backwards asymmetry will be equal to $\frac{4}{3} A_e A_f$. In our analysis, the inputs on these forward-backward asymmetries will not add, numerically, any additional information, hence, we have not included them.

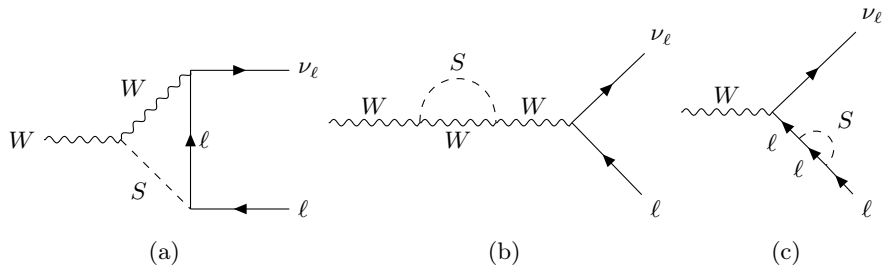
\begin{figure}[t]
		\footnotesize
		\centering 
		\subfloat[]{\begin{tikzpicture}
		\begin{feynman}
		\vertex (a1){\(W\)};
		\vertex [right=1.3cm of a1](a2);
		\vertex [below right=1.3cm of a2](a3);
		\vertex [above right=1.3cm of a2](a4);
		\vertex [right=1.3cm of a3](a5){\(\ell\)};
		\vertex [right=1.3cm of a4](a6){\(\nu_{\ell}\)};
		
		\diagram* { 
			(a1) --[boson](a2),
			(a2) --[scalar, edge label'={\(S\)}](a3),
			(a2) --[boson, edge label={\(W\)}](a4),
			(a5) --[fermion, arrow size=1.1pt](a3) --[fermion, arrow size=1.1pt, edge label'={\(\ell\)}](a4) --[fermion, arrow size=1.1pt](a6),			
		};	 	
		\end{feynman}
		\end{tikzpicture}\label{fig:W_booson_univ_1}}
		\subfloat[]{\begin{tikzpicture}
			\begin{feynman}
			\vertex (a1);
			\vertex [right=1.2cm of a1](a2);
			\vertex [right=1cm of a2](a3);
			\vertex [right=1cm of a3](a4);
			\vertex [above right=1.3cm of a4](a5){\(\nu_{\ell}\)};
			\vertex [below right=1.3cm of a4](a6){\(\ell\)};
			
			\diagram* {
				(a1) --[boson, edge label={\(W\)}](a2) --[boson, edge label'={\(W\)}](a3) --[boson, edge label={\(W\)}](a4),
				(a2) --[scalar, half left, looseness=2, edge label = \(S\)](a3),
				(a6) --[fermion, arrow size=1.1pt](a4) --[fermion, arrow size=1.1pt](a5);
				
			}; 
			\end{feynman}
			\end{tikzpicture}\label{fig:W_booson_univ_2}}
		\subfloat[]{\begin{tikzpicture}
			\begin{feynman}
			\vertex (a1);
			\vertex [right=1.2cm of a1](a2);
			\vertex [above right=1.5cm of a2](a6){\(\nu_{\ell}\)};
			\vertex [below right=0.5cm of a2](a3);
			\vertex [below right=0.5cm of a3](a4);
			\vertex [below right=0.5cm of a4](a5){\(\ell\)};
			
			\diagram* {
				(a1) --[boson, edge label={\(W\)}](a2),
				(a5) --[fermion, arrow size=1.1pt](a4) --[fermion, arrow size=1.1pt,edge label={\(\ell\)}](a3) --[fermion, arrow size=1.1pt, edge label={\(\ell\)}](a2) --[fermion, arrow size=1.1pt](a6),
				(a3) --[scalar, half left, looseness=2, edge label={\(S\)}](a4),
			};	 	
			
			\end{feynman}
			\end{tikzpicture}\label{fig:W_booson_univ_3}}
		\caption{Feynman diagrams contributing to the process $ W_{\mu} \to \ell \nu_{\ell} $. }\label{fig:W_booson_univ}
	\end{figure}

\paragraph{\underline{Branching fractions of $W \to \ell\nu$ decays:}}
This model will also contribute to the decay of $ W $ boson to the leptons. The corresponding Feynman diagrams at one-loop are shown in fig. \ref{fig:W_booson_univ}. The contributions from the diagrams can be given as: 

\begin{align}
\footnotesize
\mathcal{L}^{\ref{fig:W_booson_univ_1}}_{W\to \ell \bar{\nu}} & =   \frac{g}{\sqrt{2}} \frac{1}{16 \pi^2} \biggl[ \left( (   c_{p} c_{G} -i c_s c_{G} ) \frac{m_{\ell}^2}{2} \log\frac{\Lambda^2}{M_S^2}  \right) + \bigg(  ( 5 i c_s  + 17 c_p) M_W^4 \nonumber \\& 
+ M_S^2 \left( 4 (  2 c_p - i c_s)M_W^2 - (c_p - i c_s)M_S^2 \right) \frac{c_{G} m_{\ell}^2 }{2 M_W^4} \log\frac{M_S^2}{M_W^2} \bigg)  \\&
+ \left( M_S^2 - 2 M_W^2 \right)\left( (c_p - i c_s) M_S^2 - 6 (3 c_p+i c_s)M_W^2 \right)   \frac{c_{G} m_{\ell}^2 }{8 M_W^4} \log\frac{M_S^2}{m_{\ell}^2} \biggr] \left[\bar{\ell} \gamma_{\mu}P_L \nu_{\ell}\right]W^{\mu}, \nonumber
\end{align}

\begin{align}
\footnotesize
\mathcal{L}^{\ref{fig:W_booson_univ_2}}_{W\to \ell \bar{\nu}} &  =  \frac{g}{\sqrt{2}} \frac{c_G^2 M_W^2}{192 \pi^2} \left( \log \frac{\Lambda^2}{M_S^2} + \log \frac{\Lambda^2}{M_W^2}\right)  \left[\bar{\ell} \gamma_{\mu}P_L \nu_{\ell}\right] W^{\mu}\,,
\end{align}

\begin{align*}
\mathcal{L}^{\ref{fig:W_booson_univ_3}}_{W\to \ell \bar{\nu}} &  =  \frac{g}{\sqrt{2}}   \frac{(c_s^2 + c_p^2)m_f^2}{128 \pi^2} \left( \log \frac{\Lambda^2}{M_S^2} + \log\frac{\Lambda^2}{m_f^2}\right)  \left[\bar{\ell} \gamma_{\mu}P_L \nu_{\ell}\right] W^{\mu}\,. 
\end{align*}

The observable corresponds to this process which is considered: 
\begin{equation}
R(\ell_{1}/\ell_{2}) = \frac{\mathcal{B}(W \to \ell_{1} \nu_{\ell_{1}})}{\mathcal{B}(W \to \ell_{2} \nu_{\ell_{2}})}\,.  
\end{equation}

In the SM, this ratio $ R(\tau/\mu) $ is unity if we neglect the small phase-space effects due to the masses of the final state charged lepton \cite{PDG:2022, dEnterria:2020cpv}. The experimental values measured by various experiments are given by : 
\begin{subequations}
\begin{align}
R(\tau/\mu) &= 0.992 \pm 0.013 ~~\text{\cite{ATLAS:2020xea}} \,, \\
R(\mu/e) &= 1.009 \pm 0.009  ~~\text{\cite{CMS:2022mhs}  }\,,\\
R(\tau/e) &= 0.994 \pm 0.021 ~~\text{\cite{CMS:2022mhs}},.
\end{align}
\end{subequations}

\section{Analysis and Results}\label{sec:analysis_result}

We have learned from the previous sections that our simplified model impacts various observables associated with the FCNC and FCCC heavy flavour decays and meson mixings. Also, it has an impact on the $W$- and $Z$-pole observables. In this section, we will try to understand the constraints on the model parameters from the available data on these processes. We have so much data to analyse, which might make the analysis complicated. Therefore, in the beginning, we looked for the inputs that provide tighter constraints on the parameters of our model so that in the later part of our analysis, we could discard the less relevant data sets. We have separately studied the impact of the data on the semileptonic FCNC processes $B \to K^{(*)}\mu^+\mu^-$,  $B_s\to \phi\mu^+\mu^-$. Also, we have independently studied the constraints from the data on the meson mixing amplitudes $\Delta M_d$, $\Delta M_s$, and the rare decays $B (B_s) \to \mu^+\mu^-$, $K_{L/S} \to \mu^+\mu^-$, respectively. In addition, we have checked the impact of the measurements in FCCC processes, $W$, and $Z$-pole observables. We have noted that the constraints obtained from the data on semileptonic $B \to K^{(*)}\mu^+\mu^-$ and $B_s\to \phi\mu^+\mu^-$ decays are relatively less stringent. In the following paragraphs, we will discuss the results of different analyses.

\begin{table}[t]
	\begin{center}
		\rowcolors{1}{blue!5}{blue!3!red!6!green!4}
		\renewcommand{\arraystretch}{1.9}
		\begin{tabular}{ccccc}
			\toprule
			$\Lambda [\text{TeV}]$  &  $M_S[\text{GeV}]$  &  
			$c_s \rm [ \text{GeV}^{-1} ] $  &  $c_p \rm [ \text{GeV}^{-1} ] $  &  $c_G \rm [ \text{GeV}^{-1}] $  \\
			\hline
			\hline
			\cellcolor{blue!5}  &  $250$  &  $\text{0.43(181)}$  &  $\text{-0.51(44)}$ 
			&  $\text{-0.26(122)}$  \\
			\cellcolor{blue!5}  &  $500$  &  $\text{0.87(363)}$  &  $\text{-1.02(886)}$ 
			&  $\text{-0.52(245)}$  \\
			\multirow{-3}{*}{\cellcolor{blue!5}1}  &  $800$  &  $\text{-1.39(581)}$  &  $\text{1.63(142)}$  & 
			$\text{0.84(392)}$  \\
			\hline
			\hline 
			\cellcolor{blue!5}  &  $250$  &  $\text{0.37(152)}$  &  $\text{-0.43(38)}$ 
			&  $\text{-0.24(113)}$  \\
			\cellcolor{blue!5}  &  $500$  &  $\text{0.73(307)}$  &  $\text{-0.86(75)}$ 
			&  $\text{-0.49(227)}$  \\
			\cellcolor{blue!5}  &  $800$  &  $\text{1.17(491)}$  &  $\text{-1.38(120)}$  & 
			$\text{-0.78(363)}$  \\
			\multirow{-3}{*}{\cellcolor{blue!5}2}  &  $1000$  &  $\text{-1.47(615)}$  &  $\text{1.72(150)}$  
			&  $\text{0.97(454)}$  \\
			\bottomrule
		\end{tabular}
		\caption{Fit results of the couplings $ c_s, c_p \text{ and } c_G $ from a fit to the available data on $B\to K^{(*)}\mu^+\mu^-$ and $B_s\to \phi\mu^+\mu^-$ decays and on $R(K^{(*)})$.}
		\label{tab:b2s_global}
	\end{center}
\end{table}

\paragraph{\underline{Fit to semileptonic $ b \rightarrow s \ell^+ \ell^- $ decays:}} We will discuss the constraints obtained from the available data on the differential rates and the angular observables in $B\to K^{(*)}\mu^+\mu^-$ and $B_s \to \phi\mu^+\mu^-$ decays. The NP contributions to these decays from our simplified model are discussed in the paragraph rare decays in subsection \ref{para:raredecays}. In addition, we have considered the inputs on $R(K)$ and $R(K^*)$, which are given in eq.~\eqref{eq:RKRkst2022}. In this part of the analysis, we have not included the inputs on the branching fractions of the rare leptonic decays (eq.~\eqref{eq:rare_exp}).   

As we have discussed earlier, we will get new physics contribution to these decays via the operators  $\mathcal{O}_s$, $\mathcal{O}_s^{'}$,$ \mathcal{O}_p$, $\mathcal{O}_p^{'}$. We performed a fit using a \emph{Mathematica\textsuperscript \textregistered} package~\cite{optex}  to all the available data on differential rates, isospin asymmetry, angular observables in $ B \to K\mu^+\mu^-$, $B \to K^*\mu^+\mu^-$ and $B_s\to \phi\mu^+\mu^-$ decays. Here, the NP contribution is taken only in the muon sector since the NP contributions in $b\to s e^+e^-$ decays are very small. As we have mentioned earlier, the relevant inputs on the differential rates and the angular observables are taken from the refs. \cite{CDF:2011tds, LHCb:2013lvw,LHCb:2014cxe,LHCb:2014vgu,LHCb:2015svh,Belle:2016fev,CMS:2017rzx,ATLAS:2018gqc, LHCb:2020gog,LHCb:2021zwz}. The methodology of the fit is 
same as discussed in \cite{Biswas:2021pic}.  

The fit results are shown in table \ref{tab:b2s_global} for different values of the scalar mass $M_S$. Also, to note the dependence of the results with the cut-off scale $\Lambda$, we have done the fit for $ \Lambda = 2, 1$ TeV, respectively. Although a couple of data used in the fit are inconsistent with the respective SM predictions, the constraints on the NP parameters are very relaxed or practically unconstrained. This shows that the angular observables or the rates in these decays are insensitive to the WCs $C_{S}^{(\prime)}$ and $C_{P}^{(\prime)}$.    

\begin{figure*}[t]
	\begin{center}
		\subfloat[]{\includegraphics[scale=0.1]{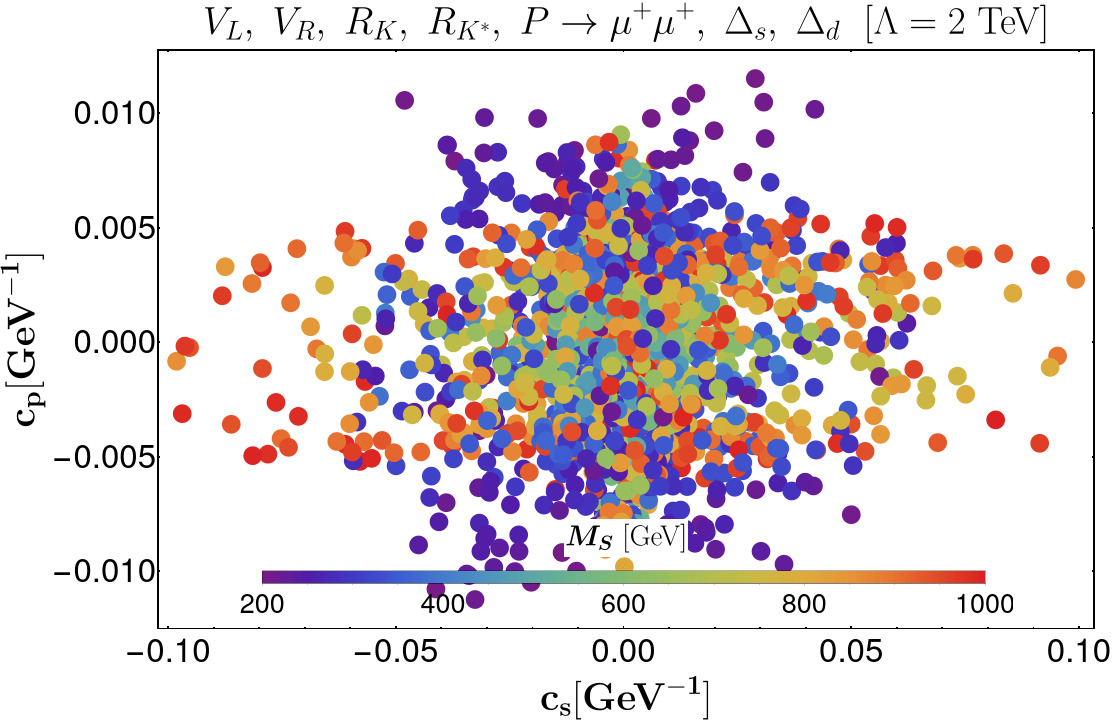}}~~~~
		\subfloat[]{\includegraphics[scale=0.1]{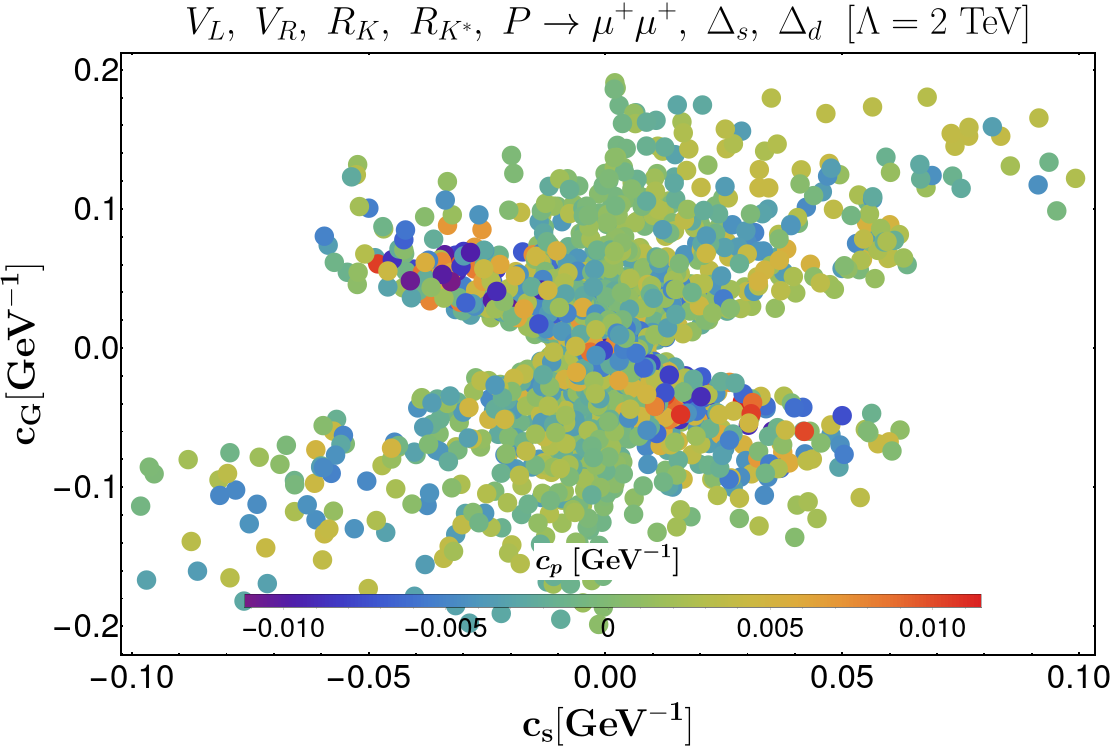}\label{fig:rare_RK_tbW_2TeV_2}}\hspace{0.0001cm}
	\end{center}
	\caption{Allowed parameter space in $ c_{s}-c_{p} $ (left) and $ c_s-c_{G} $(right) planes obtained from a scan of the observables mentioned on the top of the figures. Also, the scalar mass $ M_S $ (left) and pseudoscalar coupling $ c_{p} $ (right) have been varied (shown in colour bars).} \label{fig:scan1_rare_mixing_RK_VLVR}   
\end{figure*}

In another analysis, we have done a parameter scan. In that analysis, we have included the data on the purely leptonic rare decays, like $\mathcal{B}(B_s \to \mu^+\mu^-)$, $\mathcal{B}(B_d\to \mu^+\mu^-)$, $\mathcal{B}( K_{L/S} \to \mu^+\mu^-)$ and the $B_{s/d}-\bar{B}_{s/d}$ mixing amplitudes. The corresponding measured values or the upper limits we have shown in the respective paragraphs in subsection \ref{subsec:FCNC_observables}. Also, we have included the available inputs on $V_L$ and $V_R$ obtained from the study of $t\to b W$ decays, which we have shown in table \ref{tab:tbW_exp}, respectively. Note that on $V_L$, we have the measured value with a statistical error, while for $V_R$, we only have allowed ranges obtained from different other measurements. In addition, we have included the measured values of $R(K^{(*)})$. 
For this scan, we have varied the parameters over the ranges as given below
\begin{equation}\label{eq:scan1}
|c_s| \le 1 \, \text{GeV}^{-1},\ \ |c_p| \le 1 \, \text{GeV}^{-1},\ \ |c_G| \le 1 \, \text{GeV}^{-1}, \ \  200\ \text{GeV} \le M_S \le 1000\ \text{GeV}.
\end{equation}
For all these inputs, we have generated numbers in equal intervals and for $c_s$ , $c_p$ and $ c_{G} $ this interval is $0.0001 \rm GeV^{-1}$ while that for $M_S = 0.5$ GeV. The results of the scan are shown in fig. \ref{fig:scan1_rare_mixing_RK_VLVR}. We have shown the correlations between scalar, pseudoscalar, and gauge couplings. We present the results for $\Lambda = 2$ TeV. We have obtained similar parameter spaces for $\Lambda = 1$ TeV, which we have not shown separately.      

The allowed parameters are consistent with "zero", which is as per expectations since the data used in this analysis are consistent with the respective SM predictions. The  allowed ranges are $|c_s| \le 0.1$ $\text{GeV}^{-1}$, $|c_p| \le 0.01$ $\text{GeV}^{-1}$ and $|c_G| < 0.2$ $\text{GeV}^{-1}$, respectively. We can see from the plots that there are more concentrations of allowed points in the region $|c_s|\le 0.05$. A few scattered points are in the region $|c_s| > 0.05$ for relatively higher mass values $M_S (\gtrsim 700)$ GeV. Also, we note that for very small values of $c_s$, relatively large values of $c_p$ and $c_G$ are allowed. We can naively understand these behaviours from the given expressions of WCs in eqs. \eqref{eq:wcaVLVR}, \eqref{eq:wcbVLVR} and \eqref{eq:wccVLVR}, respectively. The dominating cotributions in the expressions of $C_{VL}$ and $C_{VR}$ are coming from the terms $\propto (c_s^2 + c_p^2)$ and $\propto c_s c_G$, respectively. Also, solutions are allowed when both $|c_s|$ and $|c_G|$ have relatively large values. These points are related to the solutions $|c_s| > 0.05$ when $M_S \gtrsim 700$ GeV. This behaviour is due to the terms $\propto c_s c_G\, \log [\Lambda/M_S]$ in the expressions of WCs in \eqref{eq:wcbVLVR} and \eqref{eq:wccVLVR}, respectively. For the values of $M_S$ close to 1000 GeV, these logarithms will become relatively small compared to those for the smaller values of $M_S$. Therefore, to satisfy the given inputs, we need both the $|c_s|$ and $|c_G|$ to be large. Note that the bounds on $|c_s|$ is also related to the bounds on $c_s^2 + c_p^2$. Therefore, independently, it can not be very large. It is important to mention that among the rare leptonic decays, the $\mathcal{B}(B_{s/d}\to \mu^+\mu^-)$ constraints the data more strongly as compared to $\mathcal{B}(K_{L/S} \to \mu^+\mu^-)$. This data set will be important for the later part of the analysis. We will include these inputs in a bigger data set.

\begin{table}[t]
	\footnotesize
	\centering
	\renewcommand{\arraystretch}{1.25}
	\begin{tabular}{|c|c|c|c|}
		\hline
		Observable & Value & Observable  & Value \\
		\hline
		$|V_{ud}|$ (nucl) & $0.97373 \pm 0.00009 \pm 0.00053 $ \cite{Xayavong:2021pkp} &	$|\varepsilon_K|$ & $(2.228 \pm 0.011) \times 10^{-3}$ \cite{PDG:2022}  \\
		$|V_{us}|f_+^{K \to \pi}(0)$ & $0.2165 \pm 0.0004$ \cite{ParticleDataGroup:2022pth} & 	sin~$2\beta$ & $0.699 \pm 0.017$ \cite{HFLAV:2022} \\
		$|V_{cd}|_{\nu N}$ & $0.230 \pm 0.011$ \cite{PDG:2022} & 	$\phi_s$ & $-0.057 \pm 0.021$ \cite{CKMFitter:2021}    \\
		$|V_{cs}|_{W \to c\bar{s}}$ & $0.94^{+0.32}_{-0.26} \pm 0.13$ \cite{ParticleDataGroup:2022pth} & $\alpha$ & $(85.2^{+4.8}_{-4.3})^\circ$ \cite{HFLAV:2022}  \\
		$|V_{ub}|_{excl}$ & $(3.91 \pm 0.13)\times 10^{-3}$ \cite{Biswas:2022yvh} & $\gamma$ & $(66.2^{+3.4}_{3.6} )^\circ $ \cite{HFLAV:2022}  \\
		$|V_{cb}|_{B\to D}$ & $(40.84 \pm 1.15) \times 10^{-3}$ \cite{Jaiswal:2017rve} &  	$ V_L $ & $ 0.995 \pm 0.021 $ \cite{PDG:2022}  \\
		$\frac{\mathcal{B}(\Lambda_p \to p \mu^- \bar{\nu}_\mu)_{q^2 > 15}}{\mathcal{B}(\Lambda_p \to \Lambda_c \mu^- \bar{\nu}_\mu)_{q^2 > 7}}$ & $(0.947 \pm 0.081)\times 10^{-2}$ \cite{LHCb:2015eia} & $ \Delta_s $ & $ -0.0345 \pm 0.0498 $ \cite{DiLuzio:2019jyq, LHCb:2021moh}    \\
		$\mathcal{B}(B^- \to \tau^- \bar{\nu}_\tau)$ & $(1.09 \pm 0.24) \times 10^{-4}$ \cite{HFLAV:2019} & $ \Delta_d $ & $ -0.0497 \pm 0.0518 $  \cite{DiLuzio:2019jyq,HFLAV:2022}   \\
		$\mathcal{B}(D_s^- \to \mu^- \bar{\nu}_\mu)$ & $(5.51 \pm 0.16) \times 10^{-3}$ \cite{HFLAV:2019} & 	$ R_b $ & $ 0.21629 \pm  0.00066 $ \cite{PDG:2022}     \\
		$\mathcal{B}(D_s^- \to \tau^- \bar{\nu}_\tau)$ & $(5.52 \pm 0.24) \times 10^{-2}$ \cite{HFLAV:2019}  &   	$ R_c $ & $ 0.1721 \pm  0.0030 $ \cite{PDG:2022}    \\
		$\mathcal{B}(D^- \to \mu^- \bar{\nu}_\mu)$ & $(3.77 \pm 0.18) \times 10^{-4}$  \cite{HFLAV:2019}  & 	$ R_e $ & $ 20.804 \pm 0.050 $ \cite{PDG:2022}     \\
		$\mathcal{B}(D^- \to \tau^- \bar{\nu}_\tau)$ & $(1.20 \pm 0.27) \times 10^{-3}$ \cite{HFLAV:2019} & 	$ R_{\mu} $ &$ 20.784 \pm 0.034 $ \cite{PDG:2022}  \\
		$\mathcal{B}(K^- \to e^- \bar{\nu}_e)$ & $(1.582 \pm 0.007) \times 10^{-5}$ \cite{PDG:2022}  & $ R_{\tau} $ & $ 20.764 \pm 0.045 $  \cite{PDG:2022} \\
		$\mathcal{B}(K^- \to \mu^- \bar{\nu}_\mu)$ & $0.6356 \pm 0.0011$ \cite{PDG:2022} & $ A_e  $ & $ 0.1515 \pm 0.0019 $ \cite{PDG:2022}    \\
		$\mathcal{B}(\tau^- \to K^- \bar{\nu}_\tau)$ & $(0.6986 \pm 0.0085) \times 10^{-2}$ \cite{HFLAV:2019} & $ A_{\mu} $ & $ 0.142 \pm 0.015 $  \cite{PDG:2022}   \\
		$\frac{\mathcal{B}(K^- \to \mu^- \bar{\nu}_\mu)}{\mathcal{B}(\pi^- \to \mu^- \bar{\nu}_\mu)}$ & $1.3367 \pm 0.0029$ \cite{PDG:2022} & 	$ A_{\tau} $ & $ 0.143 \pm 0.004 $ \cite{PDG:2022}   \\
		$\frac{\mathcal{B}(\tau^- \to K^- \bar{\nu}_\tau)}{\mathcal{B}(\tau^- \to \pi^- \bar{\nu}_\tau)}$ & $(6.467 \pm 0.84) \times 10^{-2}$ \cite{HFLAV:2019}  &  	$ A_{s} $ & $ 0.90 \pm 0.09 $ \cite{PDG:2022}  \\
		$\mathcal{B}(B_s \to \mu^+ \mu^-)$ & $(3.09 ^{+0.46 ~~ + 0.15}_{-0.43 ~~-0.11}) \times 10^{-9}$ \cite{LHCb:2021vsc} &  	$ A_c $ & $ 0.670 \pm 0.027 $ \cite{PDG:2022}  \\
		$\mathcal{B}(B_0 \to \mu^+ \mu^-)$ & $(0.12^{+0.08}_{-0.07}\pm 0.01)\times 10^{-9} $ \cite{LHCb:2021vsc} &  	$ A_b $ & $ 0.923 \pm 0.020 $ \cite{PDG:2022} \\			
		$|V_{cd}|f_+^{D \to \pi}(0)$ & $0.1426 \pm 0.0018$ \cite{HFLAV:2022} & & \\
		$|V_{cs}|f_+^{D \to K}(0)$ & $0.7180 \pm 0.0033$  \cite{HFLAV:2022} & & \\			
		\hline
	\end{tabular}
	\caption{List of observables from FCNC, FCCC, and the Z-pole observables used in the global fit. }
	\label{tab:CKM-updated-obs}
\end{table}

\subsection{Fit to all the relevant FCCC, FCNC, W- and Z-pole observables}\label{all_flav_fit}

In this subsection, we will present the result of the fits to all the relevant data related to FCNC and FCNC processes discussed in subsection \ref{subsec:FCNC_observables} and \ref{subsec:fccc}, respectively. Also, we have included the W-pole and the $Z$-pole observables. The details of various inputs used in the fit are presented in tables \ref{tab:CKM-updated-obs} and \ref{tab:mwdelr}, respectively. The other relevant theory inputs can be seen in table \ref{tab:theoryinputs} in the appendix. In addition, we have included the data on $R(K^{(*)})$ given in eq.~\eqref{eq:RKRkst2022} and we have not included the data on the differential rates and the angular observables of $B\to K^{(*)} \mu^+\mu^-$ and $B_s \to\phi\mu^+\mu^-$ decays which have negligible impact on our conclusions. 

We have analysed the data for the cases $ c_s \ne c_p$, $ c_s = c_p$, and $c_s=0$, respectively, and presented the fit results for a few fixed values of $M_S$. Alongside the fits, we have scanned the parameter spaces using all the available data mentioned above. The results of the scan will be helpful to understand the correlations among the NP parameters. Also, we will be able to understand the dependence of the fit results on the scalar mass $M_S$. 

\begin{table}[t]
	\centering
	\rowcolors{1}{blue!5}{blue!3!red!6!green!4}
	\renewcommand{\arraystretch}{2.0}
	\begin{tabular}{c c c c c c }
		\toprule
		\shortstack{$ \Lambda $\\ $\rm
			[TeV] $}  & \shortstack{$M_S$ \\ $ \rm [GeV] $ }   &  \shortstack{$c_s \times 10^2$\\ $ \rm [GeV^{-1}] $}  &  \shortstack{$c_p  $\\ $ \rm [GeV^{-1}] $}  &  \shortstack{$c_G \times 10^3 $\\ $ \rm [GeV^{-1}] $} & \shortstack{$ \Delta M_W \times 10^2$\\ $ \rm [GeV] $ }   \\
		\hline
		\hline
		\cellcolor{blue!5}  &  $250$  &  $-0.090\pm 1.747$  &  $0.0\pm 0.018$  &  $2 .770\pm 0.259$  &  $4 .657\pm 0.869$  \\
		\cellcolor{blue!5}  &  $500$  &  $0 .162\pm 2.014$  &  $0.0\pm 0.020$  &  $3 .432\pm 0.321$  &  $4 .654\pm 0.869$  \\
		\multirow{-3}{*}{\cellcolor{blue!5} $ 1$}  &  $800$  &  $0 .566\pm 3.616$  &  $0.0\pm 0.033$  &  $4 .487\pm 0.419$  &  $4 .645\pm 0.869$  \\
		\hline
		\hline
		\cellcolor{blue!5} &  $250$  &  $-0.130\pm 1.493$  &  $0.0\pm 0.015$  &  $2 .305\pm 0.215$  &  $4 .658\pm 0.869$  \\
		\cellcolor{blue!5}  &  $500$  &  $-0.018\pm 1.560$  &  $0.0\pm 0.015$  &  $2 .646\pm 0.247$  &  $4 .656\pm 0.869$  \\
		\cellcolor{blue!5}  &  $800$  &  $0 .107\pm 1.693$  &  $0.0\pm 0.016$  &  $3 .048\pm 0.285$  &  $4 .653\pm 0.869$  \\
		\multirow{-4}{*}{\cellcolor{blue!5} $ 2$} &  $1000$  &  $0 .197\pm 1.829$  &  $0.0\pm 0.017$  &  $3 .330\pm 0.311$  &  $4 .651\pm 0.869$  \\
		\bottomrule
	\end{tabular}
	\caption{Fit results of the parameters $ c_s, ~c_p  $ and $ c_{G} $ for different combinations of $ \Lambda $ and $ M_S, $ all the observables of table \ref{tab:CKM-updated-obs} along with $ \delta(\Delta r) $ (CDF, LHCb, ATLAS, D0) are taken into account, with a p-value of $ \sim 3 \%  $ for 45 d.o.f. The last column shows the prediction of the value of $ \Delta M_W $ from this fit result for each case. }\label{tab:combinefitCDF}
\end{table}

 \begin{table}[t]
 	\centering
 	\rowcolors{1}{blue!5}{blue!3!red!6!green!4}
 	\renewcommand{\arraystretch}{2.0}
 	\begin{tabular}{c c c c c c }
 		\toprule
 		\shortstack{$ \Lambda $ \\ $ \text{[TeV]}$ }  & \shortstack{$M_S$ \\ $ \rm [GeV] $ }   &  \shortstack{$c_s \times 10^2$\\ $ \rm [GeV^{-1}] $}  &  \shortstack{$c_p  $\\ $ \rm [GeV^{-1}] $}  &  \shortstack{$c_G \times 10^3 $\\ $ \rm [GeV^{-1}] $} & \shortstack{$ \Delta M_W \times 10^2$\\ $ \rm [GeV] $ }   \\
 		\hline
 		\hline
 		\cellcolor{blue!5} &  $250$  &  $0 .027\pm 1.734$  &  $0.0\pm 0.018$  &  $-0.820\pm 1.413$  &  $0 .409\pm 1.408$  \\
 		\cellcolor{blue!5}  &  $500$  &  $-0.048\pm 1.953$  &  $0.0\pm 0.020$  &  $-1.013\pm 1.755$  &  $0 .406\pm 1.405$  \\
 		\multirow{-3}{*}{\cellcolor{blue!5} $ 1$}  &  $800$  &  $-0.172\pm 2.239$  &  $0.0\pm 0.022$  &  $-1.333\pm 2.275$  &  $0 .410\pm 1.399$  \\
 		\hline
 		\hline
 		\cellcolor{blue!5}  &  $250$  &  $0 .038\pm 1.457$  &  $0.0\pm 0.015$  &  $-0.684\pm 1.174$  &  $0 .410\pm 1.409$  \\
 		\cellcolor{blue!5} &  $500$  &  $0 .005\pm 1.558$  &  $0.0\pm 0.016$  &  $-0.782\pm 1.353$  &  $0 .407\pm 1.408$  \\
 		\cellcolor{blue!5}  &  $800$  &  $0 .032\pm 1.662$  &  $0.0\pm 0.017$  &  $0 .899\pm 1.546$  &  $0 .405\pm 1.392$  \\
 		\multirow{-4}{*}{\cellcolor{blue!5} $ 2$}  &  $1000$  &  $0 .058\pm 1.726$  &  $0.0\pm 0.017$  &  $0 .982\pm 1.690$  &  $0 .404\pm 1.391$  \\
 		\bottomrule
 	\end{tabular}
 	\caption{Fit results of the parameters $ c_s, ~c_p  $ and $ c_{G} $  for different combinations of $ \Lambda $ and $ M_S, $ when all the observables of table \ref{tab:CKM-updated-obs} are considered along with  weighted mean data of $ \delta(\Delta r) $  observable from LHCb, ATLAS and D0 experiment \cite{LHCb:2021bjt,ATLAS:2017rzl,ATLAS:2023fsi,D0:2012kms}.  The p-value for the fit is $ \sim 30.49 \% $ with 42 d.o.f. The last column shows the prediction of the $ \Delta M_W $ value from this fit result for each case.  }\label{tab:combinefitwithout_CDF}
 \end{table}
 \begin{table}[htbp]
 	\centering
 	\rowcolors{1}{blue!5}{blue!3!red!6!green!4}
 	\renewcommand{\arraystretch}{1.8}
 	\begin{tabular}{*{4}{c}}
 		\toprule
 		\cellcolor{red!6} & \multicolumn{3}{c}{Predictions: SM and NP (for $\Lambda = 2$ TeV)} \\
 		\multirow{-2}{*}{\cellcolor{red!5} Obs.}	&  SM & $M_S = 500$ GeV & $M_S = 1000$ GeV \\		
 		\hline\hline
 		$R_b$  &  $\text{0.215820(20)}$  &  $\text{0.215814(20)}$  &  $\text{0.215812(20)}$  \\
 		$R_c$  &  $\text{0.172210(30)}$  &  $\text{0.172199(30)}$  &  $\text{0.172195(30)}$  \\
 		$R_e$  &  $\text{20.7360(100)}$  &  $\text{20.7306(100)}$  &  $\text{20.7286(101)}$  \\
 		$R_{\mu }$  &  $\text{20.7360(100)}$  &  $\text{20.7310(100)}$  &  $\text{20.7290(100)}$  \\
 		$R_{\tau }$  &  $\text{20.7810(100)}$  &  $\text{20.7743(100)}$  &  $\text{20.7724(100)}$  \\
 		$A_e$  &  $\text{0.146800(300)}$  &  $\text{0.146830(300)}$  &  $\text{0.146840(300)}$  \\
 		$A_{\mu }$  &  $\text{0.146800(300)}$  &  $\text{0.146830(300)}$  &  $\text{0.146840(300)}$  \\
 		$A_{\tau }$  &  $\text{0.146800(300)}$  &  $\text{0.146830(300)}$  &  $\text{0.146840(300)}$  \\
 		$A_s$  &  $\text{0.934700(0)}$  &  $\text{0.934741(8)}$  &  $\text{0.934755(10)}$  \\
 		$A_c$  &  $\text{0.667700(100)}$  &  $\text{0.667802(102)}$  &  $\text{0.667836(103)}$  \\
 		$A_b$  &  $\text{0.935600(0)}$  &  $\text{0.935668(13)}$  &  $\text{0.935690(17)}$  \\
 		\bottomrule
 	\end{tabular}
 	\caption{Prediction of the Z-pole ratio and asymmetric observables for the case $ c_s \neq c_p \neq 0. $ Here, the prediction is shown for $ \Lambda = 2 \rm TeV. $}
 	\label{tab:predZpole}
 \end{table}

We have presented our results based on whether or not we have included in the fit the estimate of $\delta (\Delta r)$ by the CDF (table \ref{tab:mwdelr}). In table \ref{tab:combinefitCDF}, we have presented the result of a fit, which includes all the data we have mentioned above. The p-value of the fit is $\approx$ 3\%, which is a statistically allowed fit. The results are obtained for $\Lambda =1$ and 2 TeV, respectively, for a few values of $M_S$. Note that for $\Lambda =2$ TeV, for $M_S \approx 1000$ GeV for both the $c_p$ and $c_s$, the maximum allowed values are  $< 0.02$ GeV$^{-1}$. On the other hand, for $\Lambda =1$ TeV, the allowed values could be as large as $\approx 0.03$ GeV$^{-1}$ for higher values of $M_S$. For lower values of $M_S$, the maximum allowed values will be reduced. On the other hand, the gauge coefficient $c_G$ has non-zero allowed values, though very small. The $W$- and $Z$-pole observables are highly sensitive to $c_G$. The non-zero values of $c_G$ are allowed due to the data on $M_W$ by CDF, which deviates from the SM. Note that with the increasing value of $M_S$, the bounds on $c_G$ are a little relaxed, as compared to those bounds obtained for lower values of $M_S$. For the values of $M_S$ we have considered in this analysis, the allowed value of $c_G$ is of order $\mathcal{O}(10^{(-3)})$. In table \ref{tab:combinefitwithout_CDF}, we have presented the results of a fit in which we have not taken the input on $\delta(\Delta r)$ from CDF. We note an increase in the quality of the fit, which is indicated by an increase in the $p$ value of the fits. The constraints on $c_p$ and $c_p$ do not change much, and we have similar observations. However, the allowed values of $c_G$ now become zero consistent, and the maximum allowed values are of order $\mathcal{O}(10^{-3})$. The explicit numbers can be seen from the table. 
 
Using the results of the fits, we have predicted the respective values of $\Delta M_W$, which is a shift of the value of $M_W$ from the corresponding SM predictions. For the fit results, which include the data on $M_W$ from CDF, we note a slight non-zero shift $\Delta M_W \approx 0.05$ GeV. However, from the results of the fits without the CDF input, we have not observed a non-zero shift in $\Delta M_W$, it is fully consistent with the SM. In addition, we have predicted all the Z-pole observables, shown in table \ref{tab:predZpole} using the results of the fit given in table \ref{tab:combinefitwithout_CDF}. We note that the predicted values of all the Z-pole observables are fully consistent with the respective data and the SM predictions. This indicates that in our simplified model, it is possible to explain the observed deviation in $M_W$ by CDF and the measured precise values of the Z-pole observables.    
 
\begin{figure}[t]
 	\begin{center}
 		\subfloat[]{\includegraphics[scale=0.1]{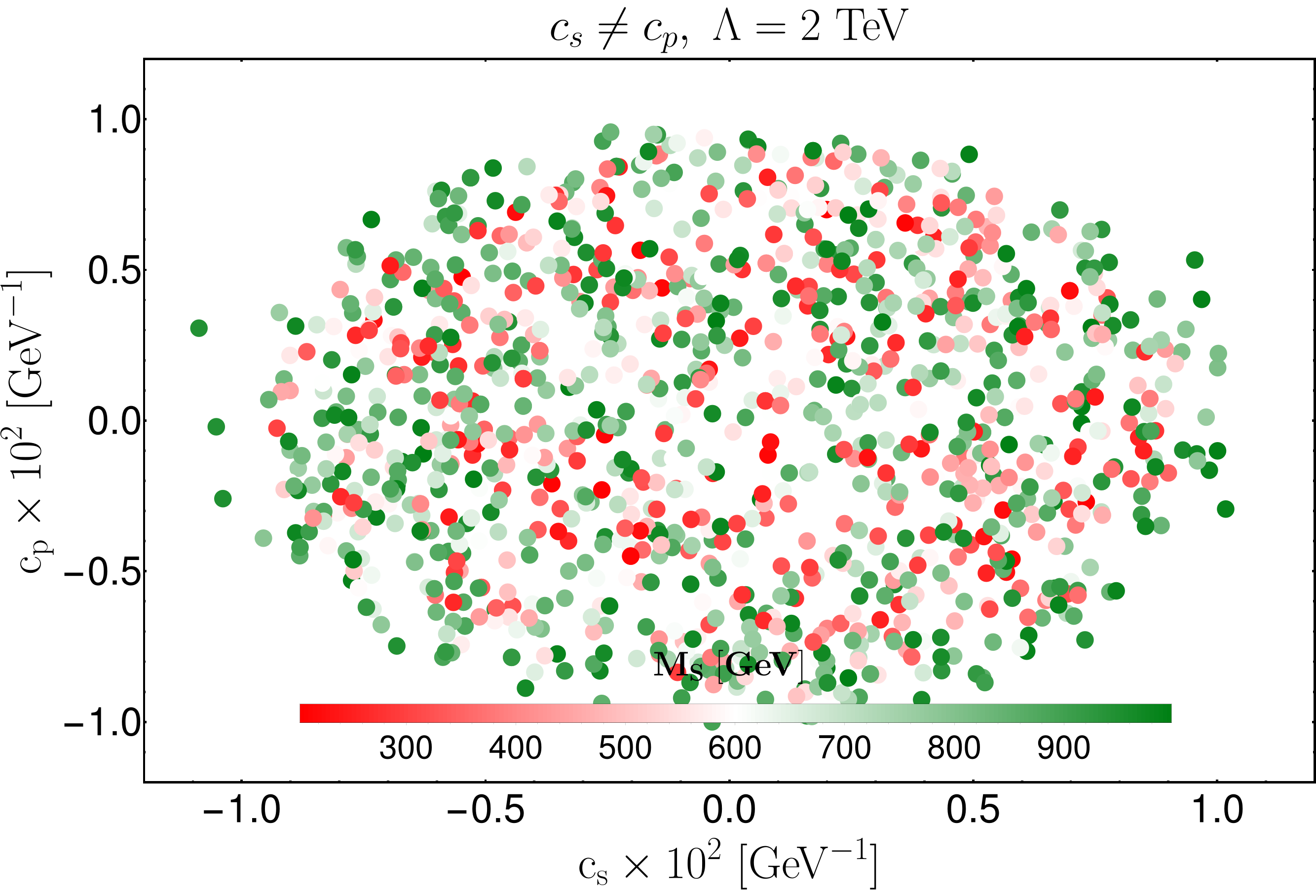}\label{fig:cs_neq_cp_cs_cp_2TeV_1}} \hspace{0.0001cm}
 		\subfloat[]{\includegraphics[scale=0.1]{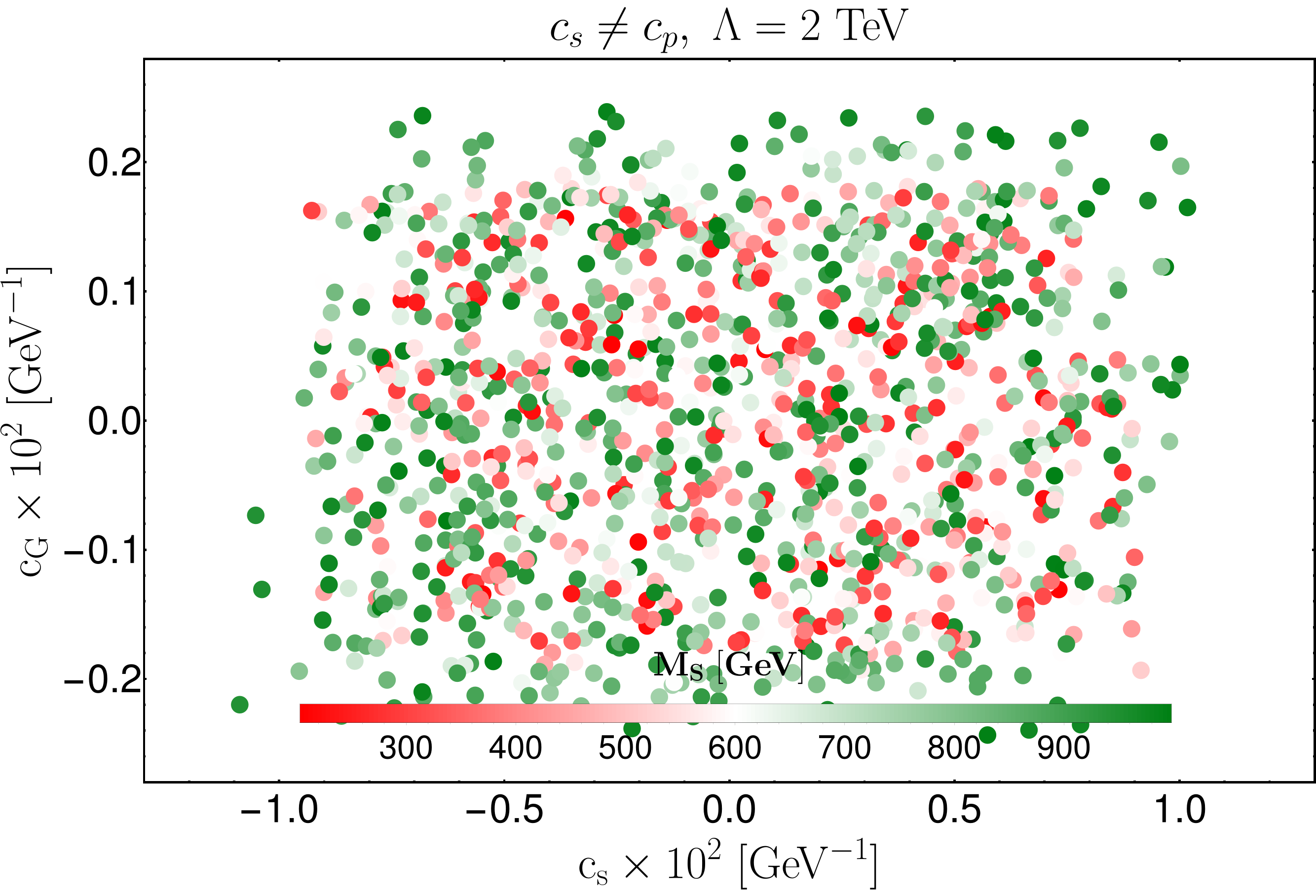}\label{fig:cs_neq_cp_cs_cp_2TeV_2}} \hspace{0.0001cm}
 	\end{center}
 	\caption{The allowed parameter space in the $c_s$-$c_p$ and $c_p$-$c_{G}$ planes for $\Lambda = 2~\rm TeV$ is determined by considering all flavour and electroweak observables. } \label{fig:combscan}   
\end{figure} 
 
As we have mentioned earlier, to get an idea about the relevant correlations between the couplings, we have done the scan using the data given in tables \ref{tab:CKM-updated-obs}, \ref{tab:mwdelr} and eq.~\eqref{eq:RKRkst2022}, respectively. For the scan, we have varied the parameters over the ranges defined in eq. \eqref{eq:scan1}. The results of the scan are given in fig. \ref{fig:combscan}. In these scans, we have dropped the input on the $\delta(\Delta r)$ from CDF. Within the allowed regions of $c_p$, $c_s$, and $c_G$, we do not see any strong correlation between these three couplings. Also, the allowed ranges are consistent with the fit results given in table \ref{tab:combinecseqcpwithout_CDF}. We notice a slight increase in the allowed regions of the couplings for higher values of $M_S$. In addition, the correlation between $c_s$ and $c_p$ indicates that for $c_s=0$, the allowed range of $c_p$ will not change. Hence, the analysis with $c_s=0$ will give us the same allowed/fitted values of $c_p$ as we have obtained in tables \ref{tab:combinecseqcp_CDF} and \ref{tab:combinecseqcpwithout_CDF}, respectively. We have not presented those results separately.

 \begin{figure}[t]
 	\begin{center}
 		\subfloat[]{\includegraphics[scale=0.1]{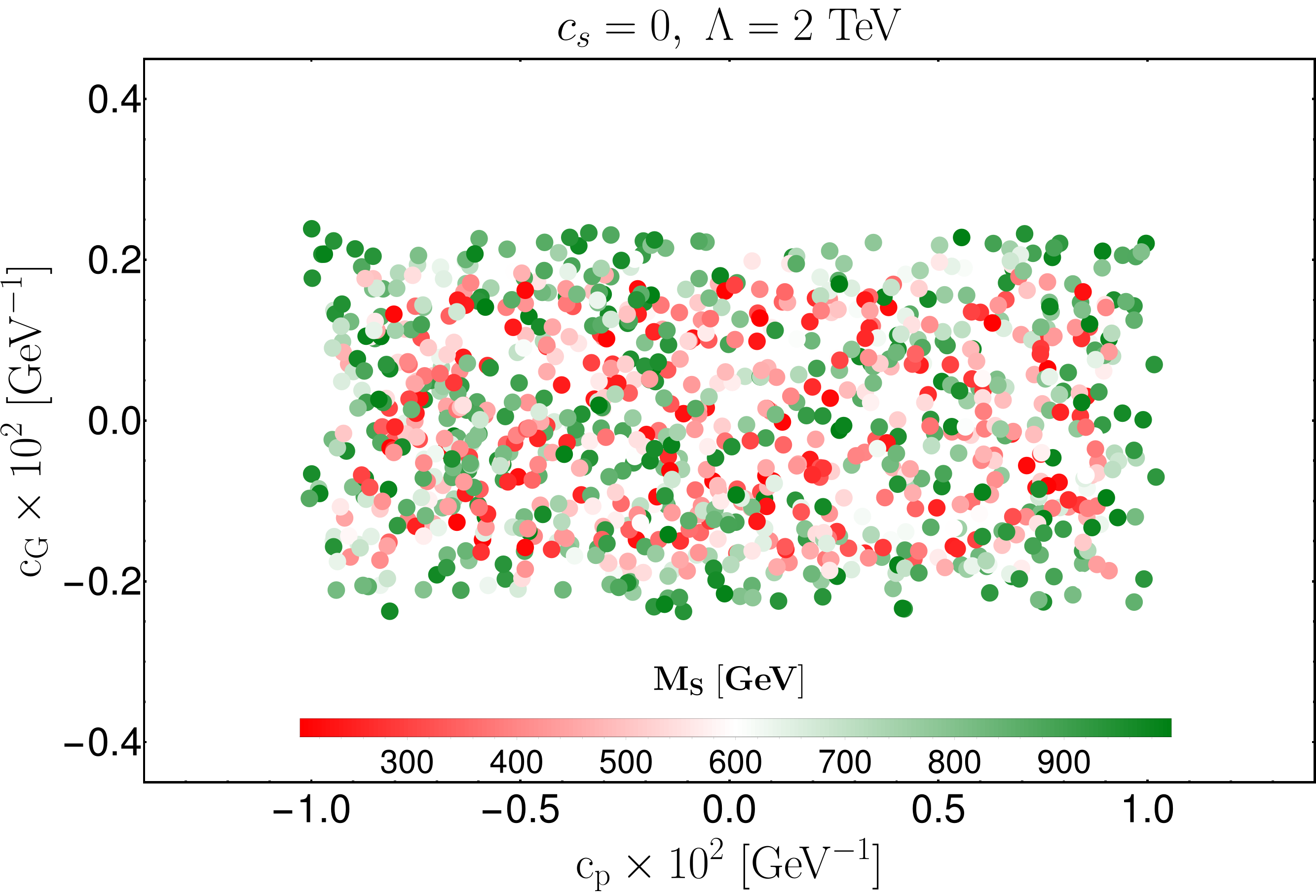}\label{fig:cs_zero_param_space_2TeV}}~~~~
 		\subfloat[]{\includegraphics[scale=0.1]{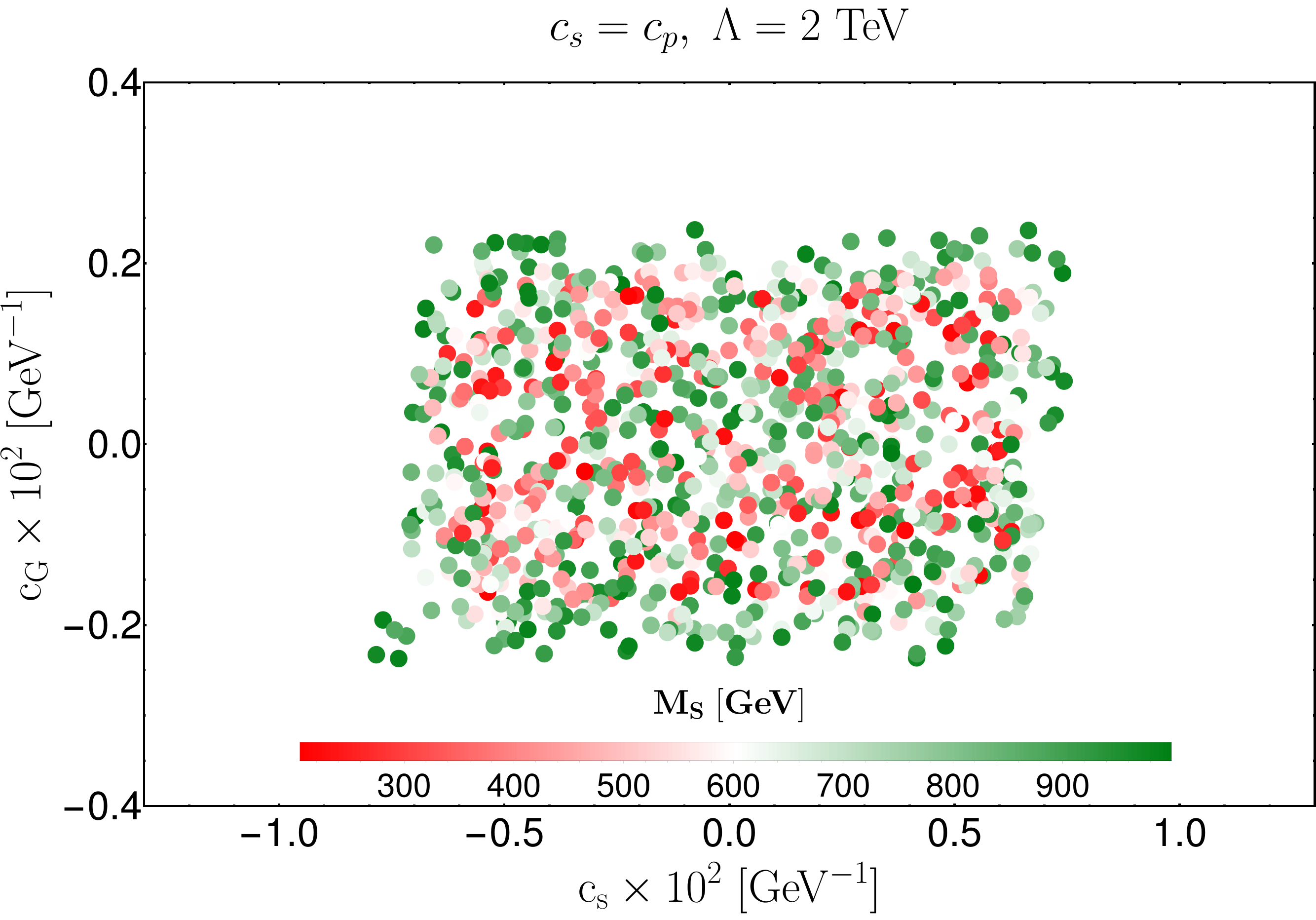}\label{fig:cs_eq_cp_param_space_2TeV}} \hspace{0.0001cm}
 	\end{center}
 	\caption{Allowed parameter space in $ c_p-c_G $ plane, when we have set $ c_s =0 $ (left) and $c_s =c_p$ (right). To generate these plots, we have varied $M_S$ over a range of 200 to 1000 GeV. } \label{fig:combine_scan_notrandom}   
 \end{figure} 
 
 We have separately presented the results of the analyses of the scenario $\bf c_s = c_p$ in tables \ref{tab:combinecseqcp_CDF} and \ref{tab:combinecseqcpwithout_CDF}, respectively, with and without the inputs on $\delta(\Delta r)$ from CDF. The other inputs are similar to the one discussed above. The constraints on $c_p (=c_s)$ and $c_G$ remain same as before. Also, in such a case, we will be able to explain $\Delta M_W$ and all the Z-pole observables simultaneously. In fig. \ref{fig:combine_scan_notrandom}, we have shown the correlations between $c_p$ and $c_G$ in the scenarios $c_s=0$ (left plot) and $c_p=c_s$ (right plot). Like before, we do not see any noticeable correlations between these coefficients, and the allowed regions are similar to those obtained from the fits.

\section{Phenomenology of Dark Matter and Flavour }

	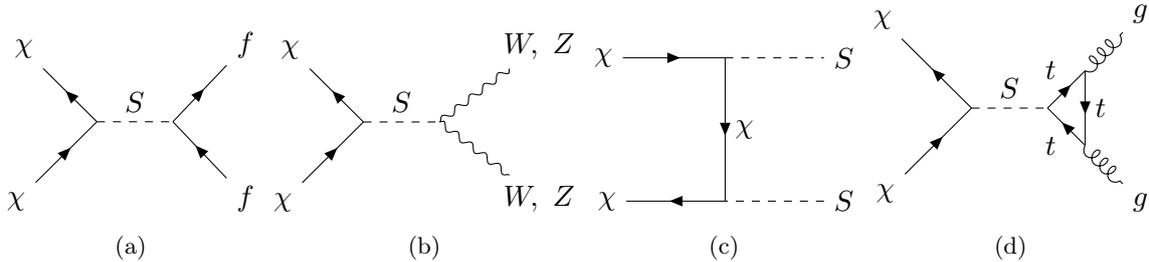
\begin{figure}[t]
		\centering
		\subfloat[]{\begin{tikzpicture}
			\begin{feynman}
			\vertex (a1){\(\chi \)};
			\vertex [above right=1.5cm of a1](a2);
			\vertex [above left=1cm of a2](a3){\(\chi \)};
			\vertex [right=1cm of a2](a6);	
			\vertex [above right=1cm of a6](a7){\(f \)};
			\vertex [below right=1cm of a6](a8){\(f \)};		
			
			\diagram* { 
				(a1) --[fermion, arrow size=1.1pt](a2) --[fermion, arrow size=1.1pt](a3),
				(a8) --[fermion, arrow size=1.1pt](a6) --[fermion, arrow size=1.1pt](a7),
				(a2) --[scalar, edge label=\( S\)](a6),
			};	
			\end{feynman}
			\end{tikzpicture}\label{fig:DM_ann1}}
		\subfloat[]{\begin{tikzpicture}
			\begin{feynman}
			\vertex (a1){\(\chi \)};
			\vertex [above right=1.5cm of a1](a2);
			\vertex [above left=1cm of a2](a3){\(\chi \)};
			\vertex [right=1cm of a2](a6);	
			\vertex [above right=1cm of a6](a7){\(W,~Z \)};
			\vertex [below right=1cm of a6](a8){\(W,~Z\)};		
			
			\diagram* { 
				(a1) --[fermion, arrow size=1.1pt](a2) --[fermion, arrow size=1.1pt](a3),
				(a8) --[boson](a6) --[boson](a7),
				(a2) --[scalar, edge label=\( S\)](a6),
			};	
			\end{feynman}
			\end{tikzpicture}\label{fig:DM_ann2}}
		\subfloat[]{\begin{tikzpicture}
			\begin{feynman}
			\vertex (a1){\( \chi\)};
			\vertex [right=1.6cm of a1](a2);
			\vertex [right=1.3cm of a2](a3){\( S\)};
			\vertex [below=1.9cm of a2](a4);
			\vertex [left=1.3cm of a4](a5){\(\chi \)};
			\vertex [right=1.3cm of a4](a6){\(S\)};
			
			\diagram* { 
				(a1) --[fermion, arrow size=1.1pt](a2),
				(a2) --[scalar](a3),
				(a6) --[scalar](a4) --[fermion, arrow size=1.1pt](a5),
				(a2) --[fermion, arrow size=1.1pt, edge label={\(\chi \)}](a4),
				
			};
			\end{feynman}
			\end{tikzpicture}\label{fig:DM_ann3}}
		\subfloat[]{\begin{tikzpicture}
			\begin{feynman}
			\vertex (a1){\(\chi \)};
			\vertex [above right=1.6cm of a1](a2);
			\vertex [above left=1.3cm of a2](a3){\(\chi \)};
			\vertex [right=1.cm of a2](a6);	
			\vertex [above right=0.7cm of a6](a7);
			\vertex [below right=0.7cm of a6](a8);	
			\vertex [above right=0.7cm of a7](a9){\(g \)};
			\vertex [below right=0.7cm of a8](a10){\(g \)};
			
			\diagram* { 
				(a1) --[fermion, arrow size=1.1pt](a2) --[fermion, arrow size=1.1pt](a3),
				(a8) --[fermion, arrow size=1.1pt, edge label=\( t\)](a6) --[fermion, arrow size=1.1pt, edge label=\( t\)](a7),
				(a7) --[fermion, arrow size=1.1pt, edge label=\( t\)](a8) --[gluon](a10),
				(a7) --[gluon](a9),
				(a2) --[scalar, edge label=\( S\)](a6),
			};	
			\end{feynman}
			\end{tikzpicture}\label{fig:DM_ann4}}
		\caption{Feynman diagrams depicting the annihilation channels of the DM are significant for determining relic density.}
		\label{fig:Feyn_DM}
	\end{figure}
	
	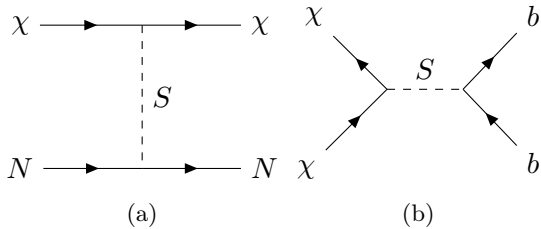
\begin{figure}
		\centering
		\subfloat[]{\begin{tikzpicture}
			\begin{feynman}
			\vertex (a1){\( \chi\)};
			\vertex [right=1.6cm of a1](a2);
			\vertex [right=1.3cm of a2](a3){\( \chi\)};
			\vertex [below=1.9cm of a2](a4);
			\vertex [left=1.3cm of a4](a5){\(N \)};
			\vertex [right=1.3cm of a4](a6){\(N\)};
			
			\diagram* { 
				(a1) --[fermion, arrow size=1.1pt](a2)--[fermion, arrow size=1.1pt](a3),
				(a5) --[fermion, arrow size=1.1pt](a4) --[fermion, arrow size=1.1pt](a6),
				(a2) --[scalar, edge label={\(S\)}](a4),
				
			};
			\end{feynman}
			\end{tikzpicture}\label{fig:DM_DD}}
		\subfloat[]{\begin{tikzpicture}
			\begin{feynman}
			\vertex (a1){\(\chi \)};
			\vertex [above right=1.5cm of a1](a2);
			\vertex [above left=1cm of a2](a3){\(\chi \)};
			\vertex [right=1cm of a2](a6);	
			\vertex [above right=1cm of a6](a7){\(b \)};
			\vertex [below right=1cm of a6](a8){\(b \)};		
			
			\diagram* { 
				(a1) --[fermion, arrow size=1.1pt](a2) --[fermion, arrow size=1.1pt](a3),
				(a8) --[fermion, arrow size=1.1pt](a6) --[fermion, arrow size=1.1pt](a7),
				(a2) --[scalar, edge label=\( S\)](a6),
			};	
			\end{feynman}
			\end{tikzpicture}\label{fig:DM_ID}}	
		\caption{Feynman diagrams contribute to the direct detection process (left) and to the indirect detection for annihilation to $ b\bar{b} $ (right).}
	\end{figure}

	In our simplified model, we consider a fermionic DM with a spin-0 mediator given in eqs.~\eqref{eq:model} and \eqref{eq:fermionS}, respectively. As we know, the WIMP-type dark matter gives the current relic density by the freeze-out method. The relevant diagrams for Dark matter pair annihilation to the SM particles are given in fig. \ref{fig:Feyn_DM}, respectively. The mediator $ S $ interacts with SM fermions and gauge bosons. Hence, the dominant channel for DM annihilation will be $ \bar{\chi}\chi \rightarrow  \bar{f}f$ and $ \bar{\chi}\chi \rightarrow  V V$ (V for $ W_{\mu}, Z_{\mu} $). DM can also annihilate to gluon via a penguin loop shown in fig. \ref{fig:DM_ann4}. For $ m_{\chi} \leq M_{S} $, DM will mostly annihilate to these channels. For the case, $ m_{\chi} \geq M_{S}  $, a significant contribution will come from the t-channel annihilation diagram of DM, i.e., $ \chi \bar{\chi} \to S S .$  At resonance, contribution from s-channel annihilation diagrams will be more. We have specified earlier that in order to maintain MFV, we use the mass-dependent coupling of the spin-0 mediator with SM fermions (similar to the SM-Yukawa). So, the coupling with the top quark is maximum. Using that fact, we will get another channel contributing to our case $ \chi \bar{\chi} \to g g. $ The effective coupling is given by \cite{Buckley:2014fba}: 
	\begin{equation}\label{gluon_coup}
	\mathcal{L}_{gluon} = \frac{\alpha_s}{8 \pi} \biggl (  c_s \tau [1+ (1-\tau)f(\tau) ]G^{\mu \nu} G_{\mu \nu} +  2 c_p \tau f(\tau) G^{\mu \nu} \tilde{G}_{\mu \nu} \biggr) S \,,
	\end{equation}
	Where, $ \tau = \frac{4 m_t^2}{M_S^2} $ and 
	\[ f(\tau) = \begin{cases} 
	\arcsin^2 \frac{1}{\sqrt{\tau}} & \text{if } ~\tau \geq 1 \\
	-\frac{1}{4} (\log \frac{1+\sqrt{1-\tau}}{1-\sqrt{1-\tau}} - i \pi )^2\,  & \text{if } ~\tau < 1
	\end{cases} \]
	This type of contribution will play an important role when $M_\chi < m_t$. In our case, most contributions will come from $ t \bar{t} \text{ and } ~VV$ processes; we still consider it.

As expected, in our model, we will also get a non-zero contribution to the scattering cross-section of the DM with the nucleon, which is relevant for the direct detection process, where we study the recoil energy of the detector nuclei. The upper bounds of such processes are obtained in various measurements, among which the most stringent bound comes from XenonnT \cite{XENON:2023cxc}, LUX-ZEPLIN \cite{LZ:2022lsv} and Panda4X-T\cite{PandaX:2023ejt}. In this work, we have used the bound from XenonnT and LUX-ZEPLIN, which are more stringent than PandaX-4T. Also, the study of gamma-ray annihilation spectrum in indirect detection gives bound on DM annihilation cross-section rate to SM particle pairs ($ b \bar{b}, ~W^+ W^-, ~\tau^+ \tau^- $ etc) the collaborations like Fermi-LAT \cite{Fermi-LAT:2015att,Fermi-LAT:2016afa}, High Energy Stereoscopic System (H.E.S.S) \cite{HESS:2016mib} and Cherenkov Telescope Array (CTA) \cite{Silverwood:2014yza} provide bounds on that.      

\begin{figure}[t]
			\begin{center}		
				\subfloat[]{\includegraphics[scale=0.11]{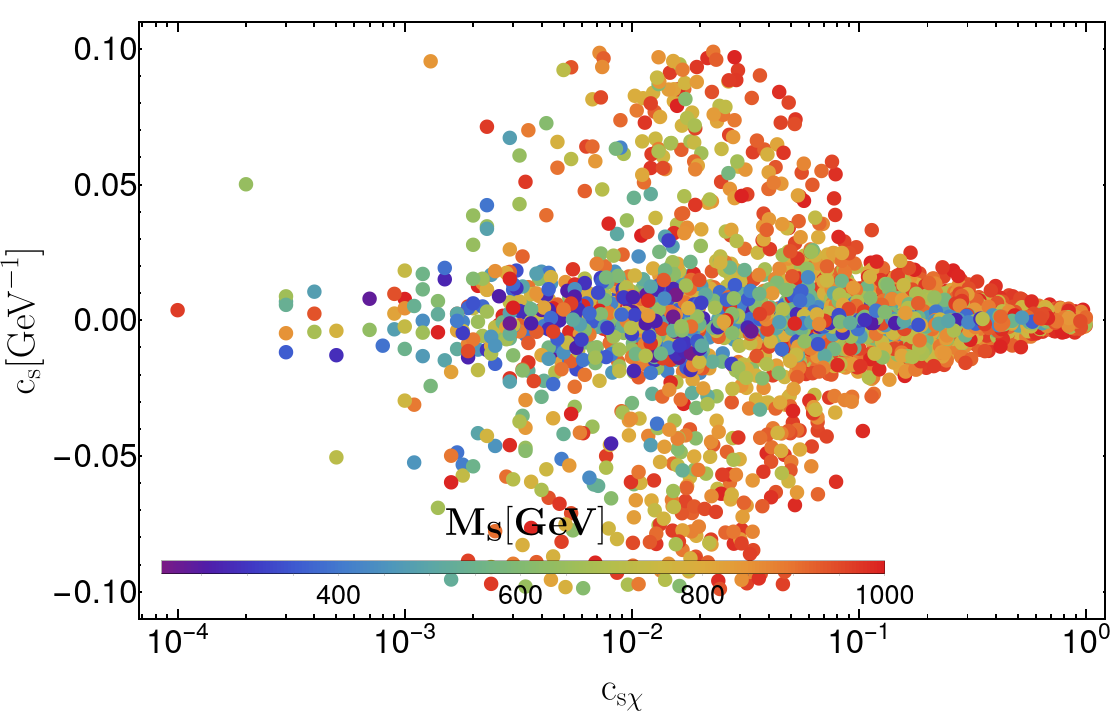}\label{fig:csX_cs_DM}}~~
				\subfloat[]{\includegraphics[scale=0.11]{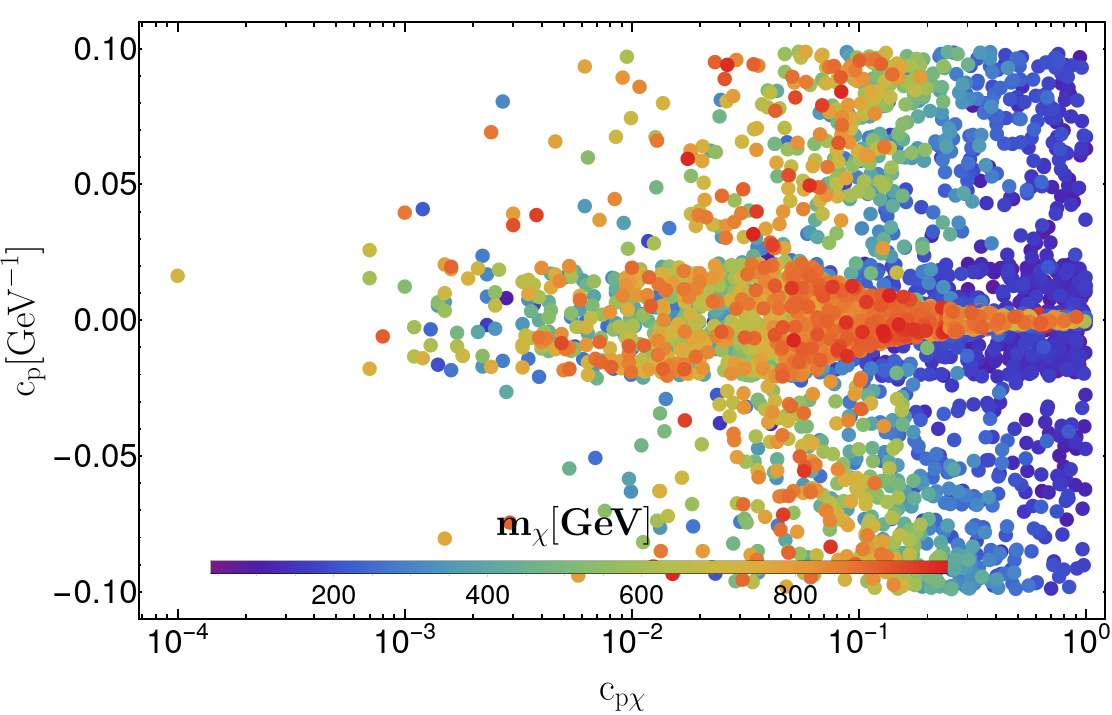}\label{fig:cpX_cp_DM}}\\
				\subfloat[]{\includegraphics[scale=0.11]{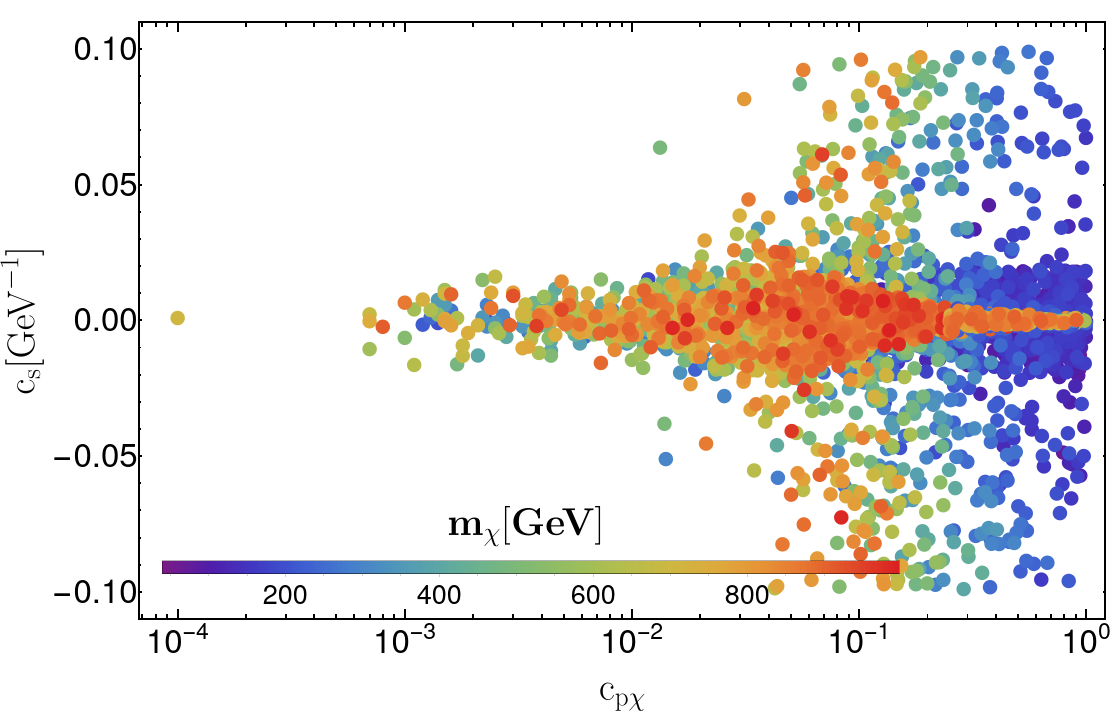}\label{fig:cpX_cs_DM}}~~
				\subfloat[]{\includegraphics[scale=0.11]{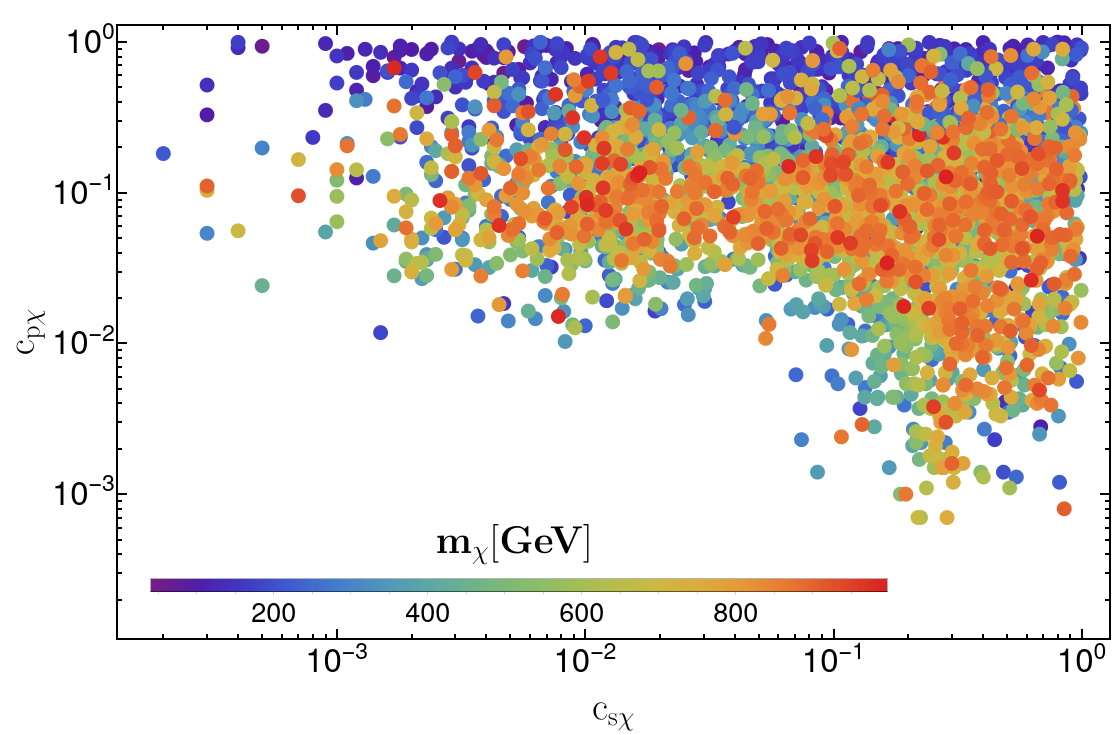}\label{fig:csX_cpX_DM}}\\
				\subfloat[]{\includegraphics[scale=0.11]{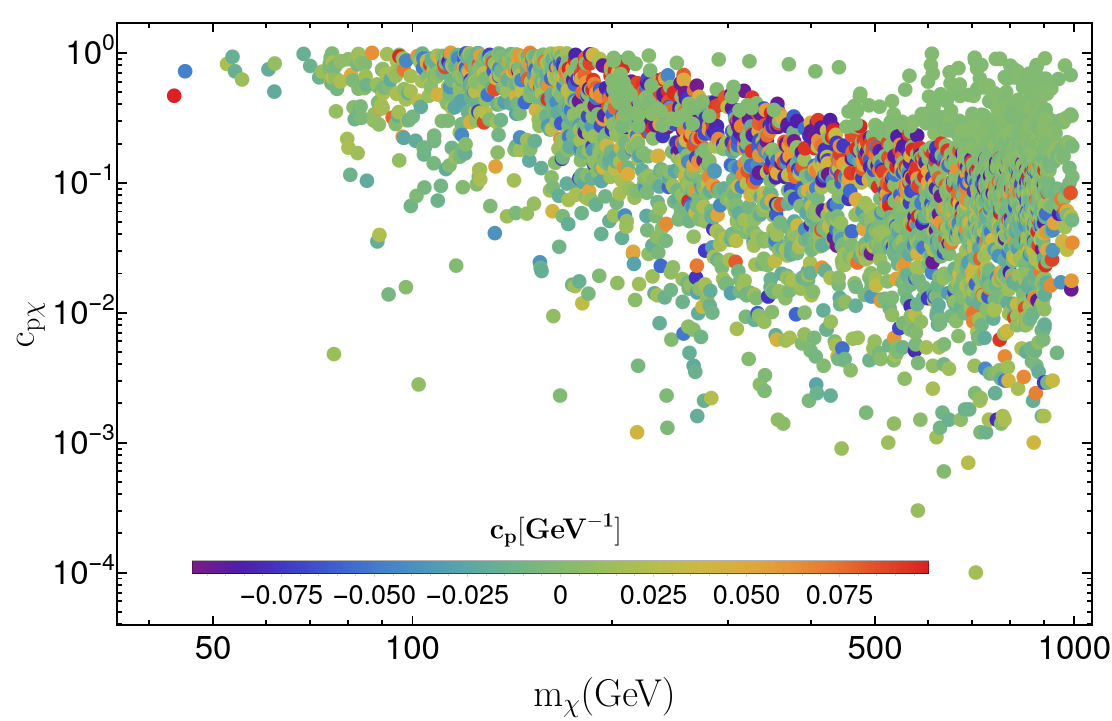}\label{fig:mchi_cpX_DM}}
				\subfloat[]{\includegraphics[scale=0.11]{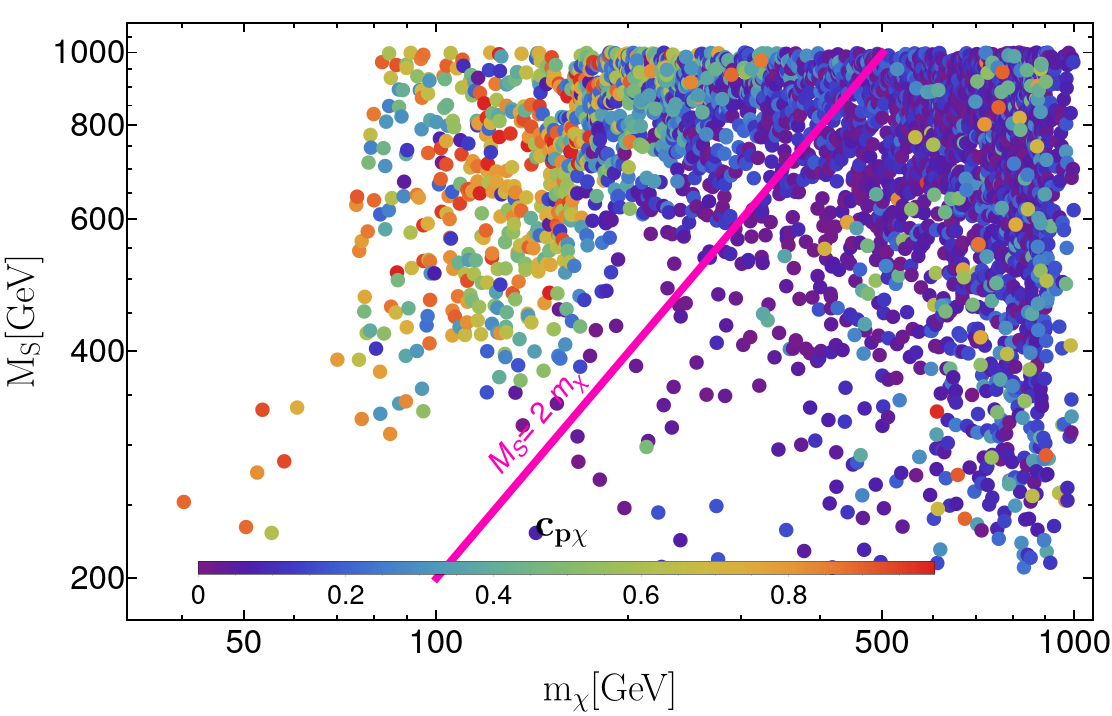}\label{fig:ms_mchi_DM}}~~		
			\end{center}
			\caption{Correlation among the couplings $ c_s, ~c_p, ~c_{s\chi}, ~c_{p\chi}$ and $ ~c_G $ All the points in the plots are allowed by observed relic density data and direct detection cross-section bound provided by LUZ-ZEPLIN 2022. } \label{fig:corr_DM}   
		\end{figure}

\subsection{Results of the analysis}\label{subsec:bounds_univcoup} 

We have analysed the data on relic density and the direct detection cross-section separately and in combination with the flavour data and obtained the constraints on the new physics parameters. At first, we have treated the scalar mass $M_S$, the DM mass $m_{\chi}$ and all the relevant couplings, like $c_{s\chi}$, $c_{p\chi}$, $c_s$, $c_p$ and $c_G$ as free parameters and obtained the bounds from the data on relic density and direct detection cross-section. For the scan, we have created a huge number of benchmark points by randomly varying all these free parameters over the ranges mentioned below. The benchmark points are distributed uniformly over the entire range of parameters scanned.
	
Since we have prior knowledge of the allowed ranges of $c_s$, $c_p$ and $c_G$ from the analysis of the subsection \ref{all_flav_fit}, guided by those bounds, we have generated numbers randomly within the ranges (in units of GeV$^{-1}$) 
	\begin{equation}
	|c_s| \le 0.1,\ \  |c_p| \le 0.1\ \ \text{and}\ \  |c_G| \le 0.1,
	\end{equation}
	with a uniform interval of $0.0001$. Note that in subsection \ref{all_flav_fit} we have obtained (in units of $ \rm GeV^{-1} $)
\begin{equation}
|c_s| \le 0.01,\ \  |c_p| \le 0.01\ \ \text{and}\ \  |c_G| \le 0.002.
\end{equation}
In addition, we have allowed the couplings $c_{s\chi}$ and $c_{p\chi}$ to vary over the ranges\footnote{Initially, we did the scan by varying both the couplings $c_{s\chi}$ and $c_{p\chi}$ of DM to the scalar $S$ over the range $[-1.0, 1.0]$. However, we found no solutions for their $\le 10^{-4}$ values. Also, the negative solutions that are allowed will not provide any additional information. The allowed parameter spaces will be identical to those obtained for their corresponding positive values. This is because the relic density or the direct detection cross section depends on the square of $c_{s\chi}$ or $c_{p\chi}$.} 
 \begin{equation}
 0.0001 \le c_{s\chi} \le 1.0,\ \  0.0001 \le c_{p\chi} \le 1.0,
 \end{equation}
and created points with a uniform interval of $0.0001$. Finally, the masses are varied over the ranges
\begin{equation}
200\ \text{GeV} \le M_S \le 1000\ \text{GeV},\ \  100\ \text{GeV} \le m_{\chi} \le 1000 \ \text{GeV},
\end{equation}
and generated points with a uniform interval of $0.5$ GeV.  

The results of the scan are presented as correlations between different variables in fig. \ref{fig:corr_DM}. To understand these correlations, we must note that the scalar current contribution to the $s$-channel annihilation cross-section is the velocity-suppressed p-wave contribution (for details, please see appendix-\ref{apndxC}). However, the pseudoscalar current contribution to the s-channel cross-section will be the s-wave contribution, which is not velocity-suppressed. In our analysis, the s-wave annihilation cross section will be $\propto c^2_{p\chi}(c_s^2 + c_p^2)$ and velocity suppressed p-wave contribution is $\propto c^2_{s\chi}(c_s^2 + c_p^2)$. Also, these cross-sections are proportional to the square of the dark matter mass and inversely proportional to $(M^2_S-4 m^2_{\chi})^2$. On the other hand, in our simplified model, the dominating contribution to spin-independent direct detection cross section is proportional to $c^2_{s\chi} c^2_{s}$, the rest of the contributions are velocity suppressed. Hence, we will get a tight bound on this product from the direct detection bound. Below, in the items, we will make a few remarks on the correlations between parameters from fig. \ref{fig:corr_DM}.

\begin{itemize}
	\item In fig. \ref{fig:csX_cs_DM}, we have shown the correlation between $c_{s\chi}$, $c_s$ and $M_S$ (in colour band). As mentioned in the above paragraph, the strong bound on the product $c^2_{s\chi} c^2_{s}$ will come from the spin-independent direct detection cross-section. We note that for values $c_{s\chi} > 0.1$, the allowed values of $c_s$ will be highly constrained and small, also in this region, $M_S \gtrsim 600$ GeV. This is because the direct detection cross-section decreases with the increase of mediator mass. Hence, for relatively higher values of $M_S$, the direct detection bound will allow relatively larger values of $c_{s\chi}$. In the region $c_{s\chi} < 0.1$, the allowed values of $c_s$ are relatively relaxed, and the allowed values of $c_s$ will be of order $10^{-3}$ or less only when $M_S < 500$ GeV.  
	
	\item In fig. \ref{fig:cpX_cp_DM} we have shown the correlation between $c_{p\chi}$, $c_p$ and $m_{\chi}$. Note that for $m_{\chi}> 500$ GeV, the allowed values of $c_{p\chi}$ will be $> 0.3$ only when $c_p$ is of order $10^{-3}$ or less. However, for $c_{p\chi} < 0.1$, the allowed values of $c_p$ have a wide range, and they could be as large as order one. For $m_{\chi}< 500$ GeV the large values $(\mathcal{O}(1))$ of both $c_{p\chi}$ is allowed by the data on relic density. In such a situation, relatively higher values of $c_p$ are also allowed. A similar observation holds for the correlation between $c_s$ and $c_{p\chi}$, which we have shown in fig. \ref{fig:cpX_cs_DM}.    
	
	\item In fig. \ref{fig:csX_cpX_DM}, we have shown the correlations between $c_{s\chi}$, $c_{p\chi}$ and $m_{\chi}$ (colour band). Note that for $c_{s\chi} \lesssim 0.1 $, the allowed values of $c_{p\chi} > 0.02$. Also, in such a situation, $c_{p\chi}$ could be of order one when $m_{\chi} \lesssim 350$ GeV. The allowed solutions for $m_{\chi} > 400$ GeV prefers values of $c_{p\chi} \approx 0.1$. In the regions $c_{p\chi} \gtrsim 0.1$ and $c_{s\chi} \gtrsim 0.1$, depending on the values of $c_p$ and $c_s$ we have allowed solutions for $100 < m_\chi < 1000$ (in GeV), which can also be seen in fig. \ref{fig:mchi_cpX_DM}. In addition, there exists a solution in the region $ 0.01 \lesssim  c_{p\chi} \lesssim 0.001$ for $c_{s\chi} > 0.1$ and $m_{\chi} \gtrsim 400$ GeV. We can see from figs. \ref{fig:cpX_cp_DM}, \ref{fig:cpX_cs_DM} and \ref{fig:mchi_cpX_DM} this allowed region belongs to the values of $c_s$ of order $10^{-3}$ and for $c_p \lesssim 0.03 \rm ~GeV^{-1}$. For such small values of $c_{p\chi}$, the contribution to the relic from the velocity-suppressed scalar current could be important. 
	
	\item In fig. \ref{fig:ms_mchi_DM} we have shown the correlations between $M_S$, $m_\chi$ and $c_{p\chi}$ (colour band). Note that for lower values of the DM masses ($m_{\chi} \lesssim 300$ GeV), there are concentrations of allowed solutions near $M_S \gtrsim 500$ GeV and $c_{p\chi} > 0.4$. Actually, this region belongs to $M_S > 2 m_\chi$. On the other hand, for $c_{p\chi} \lesssim 0.4$, the greater concentration of the allowed solutions will be for $M_S> 400$ GeV and $m_{\chi} > 300$ GeV. Note that in this region, the factor $(M_S^2- 4 m_{\chi}^2)^2$ is small, and the annihilation cross-section will increase, hence to satisfy the data on a relic, the product $c_{p\chi}^2 (c_p^2 + c_s^2)$ should decrease. Also, solutions exist in the region $m_\chi > M_S$ but with a low concentration of allowed points.       
\end{itemize}

It is to be noted that for the range of the mediator mass $ M_{s} \geq 100 ~GeV, $ the dominating annihilation channels will be $ \chi \bar{\chi} \to f \bar{f}, V V $, where $ V $ stands for vector boson. More precisely, for $ m_{\chi} \geq m_{t} $, $ \chi \bar{\chi} \to t \bar{t}$, will contribute most to the relic density. For $ m_{\chi} \leq  m_t, $ channels like $ \chi \bar{\chi} \to b \bar{b}, ~W^+ W^-, ~Z Z $ will have maximum contribution. Also, for $ m_{\chi} \geq M_S, $ t-channel annihilation will play a significant role. The corresponding solutions can be seen in fig. \ref{fig:ms_mchi_DM}. 

	\begin{figure}[t]
		\begin{center}	
			\subfloat[]{\includegraphics[scale=0.11]{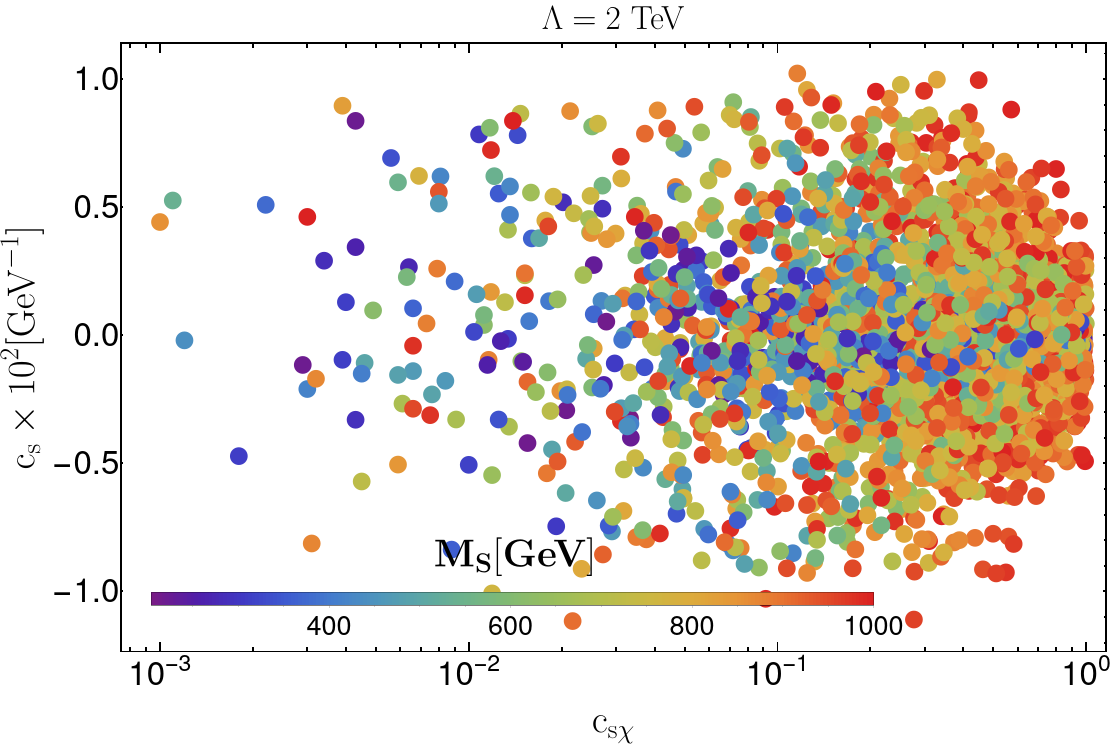}\label{fig:csX_cs_DM_flavor}}~~
			\subfloat[]{\includegraphics[scale=0.11]{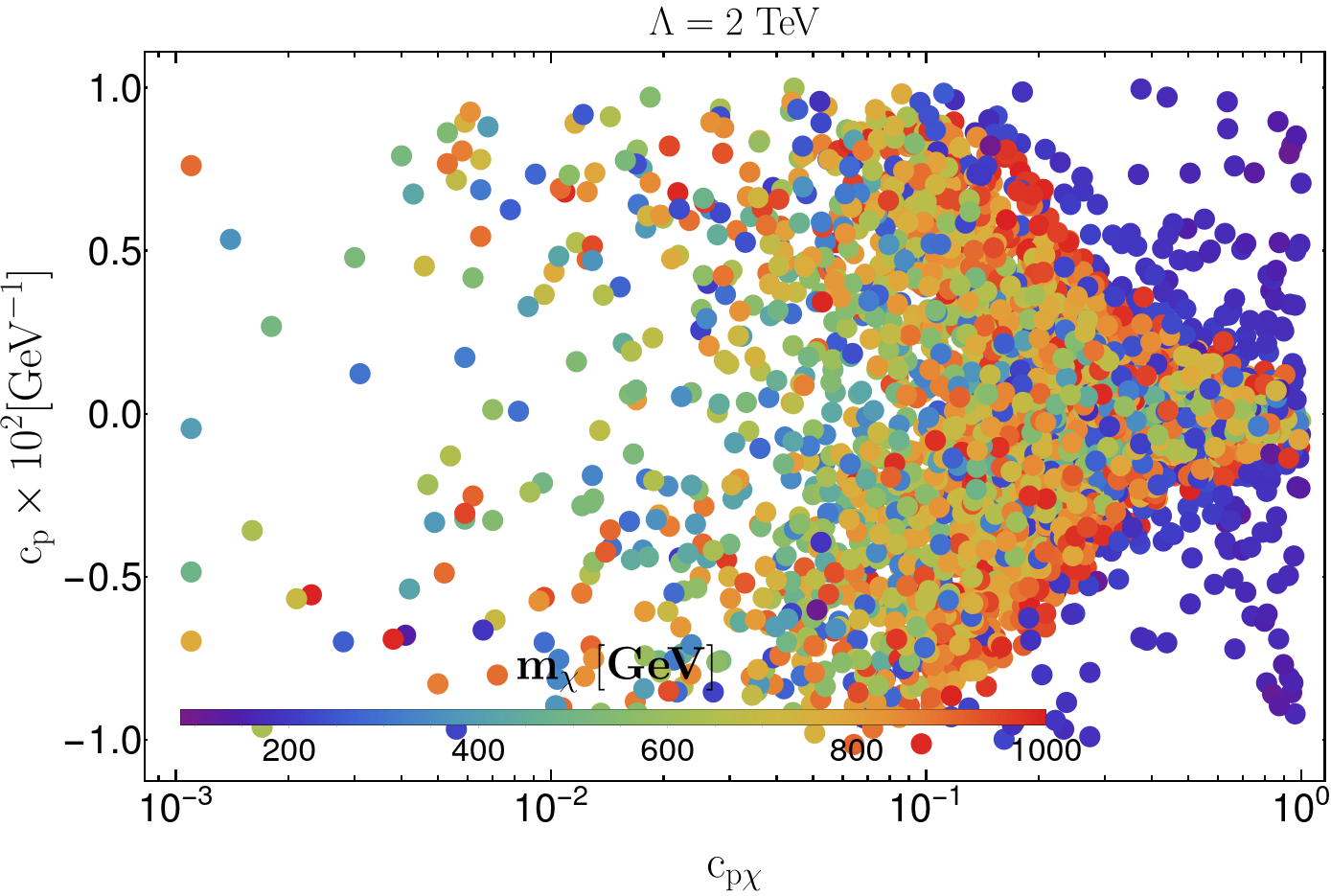}\label{fig:cpX_cp_DM_flavor}}\\
			\subfloat[]{\includegraphics[scale=0.11]{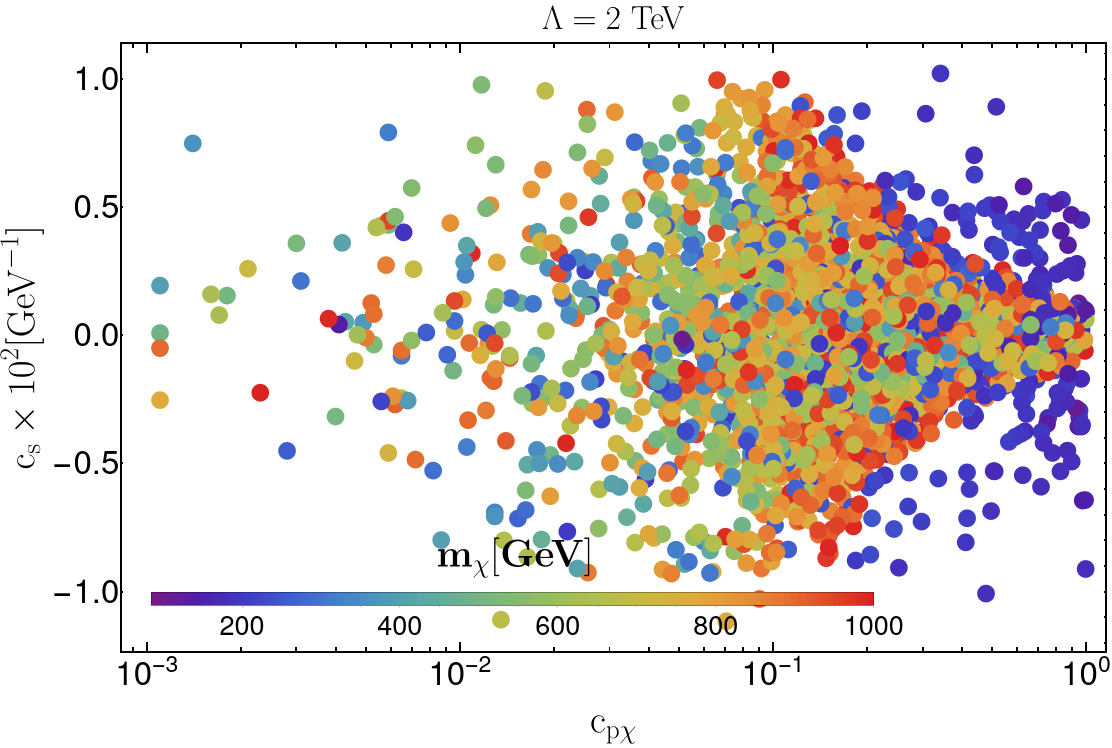}\label{fig:cpX_cs_DM_flavor}}~~	
			\subfloat[]{\includegraphics[scale=0.11]{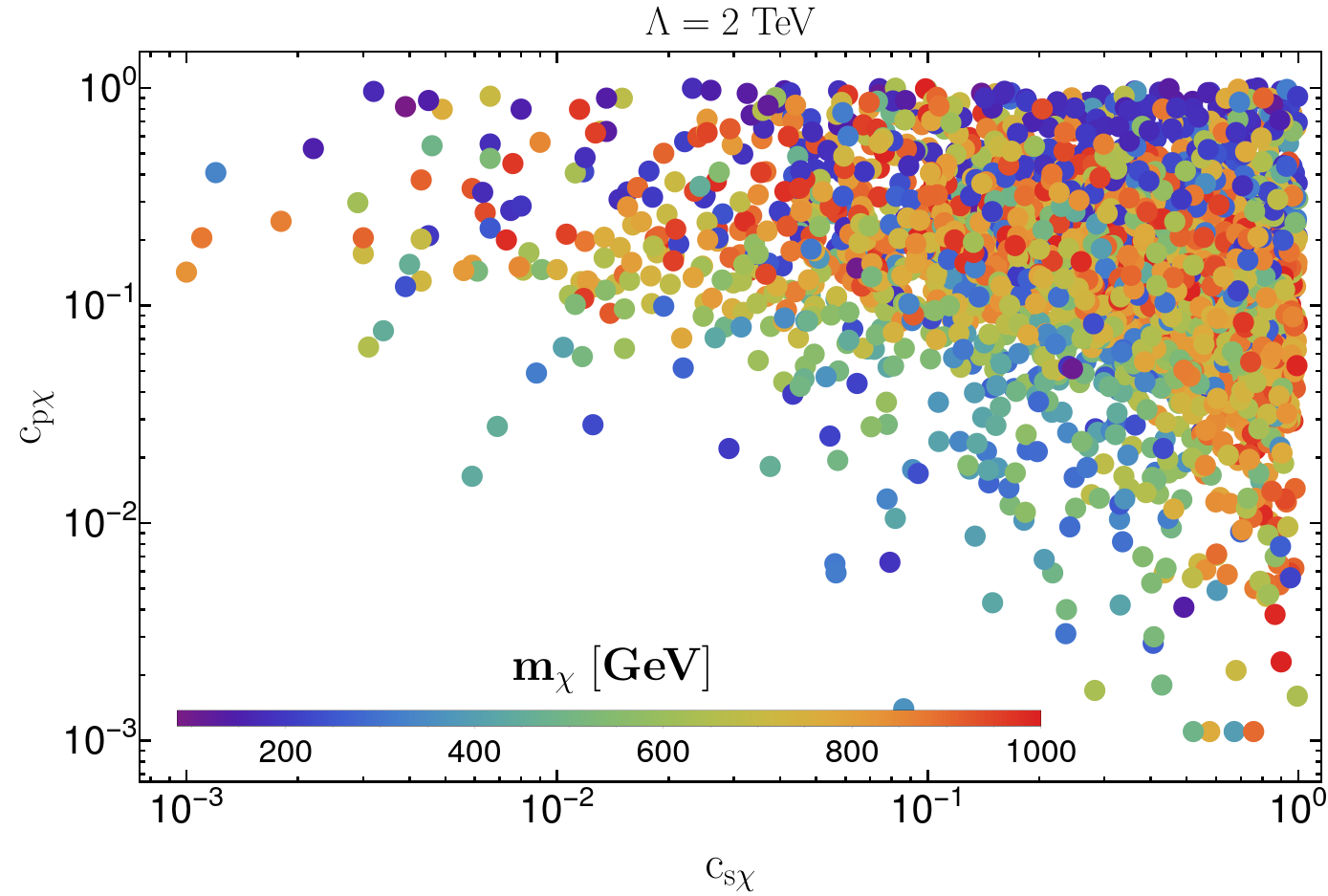}\label{fig:csX_cpX_DM_flavor}}\\	
			\subfloat[]{\includegraphics[scale=0.11]{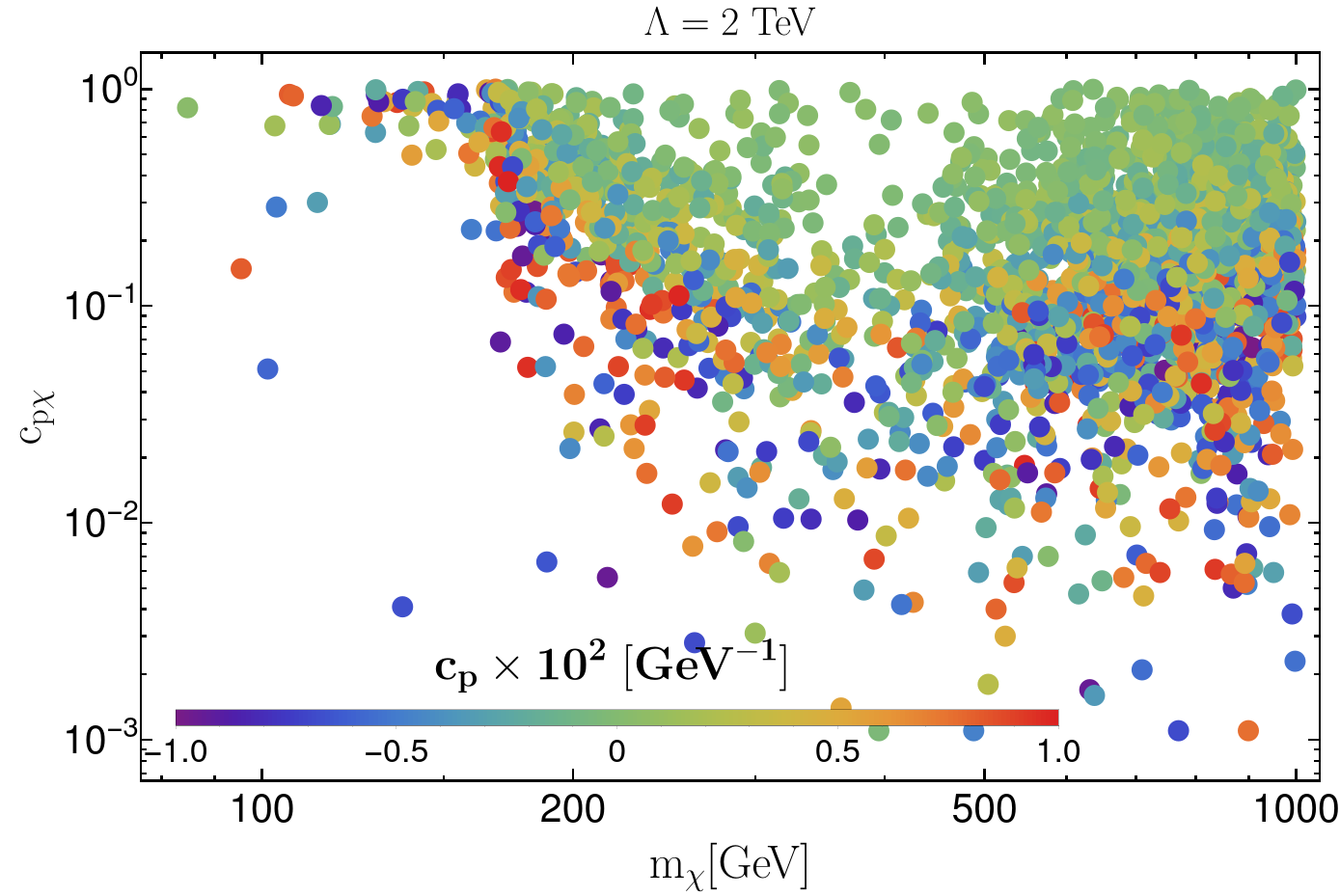}\label{fig:mchi_cpX_cp_DM_flavor}}~~	
			\subfloat[]{\includegraphics[scale=0.11]{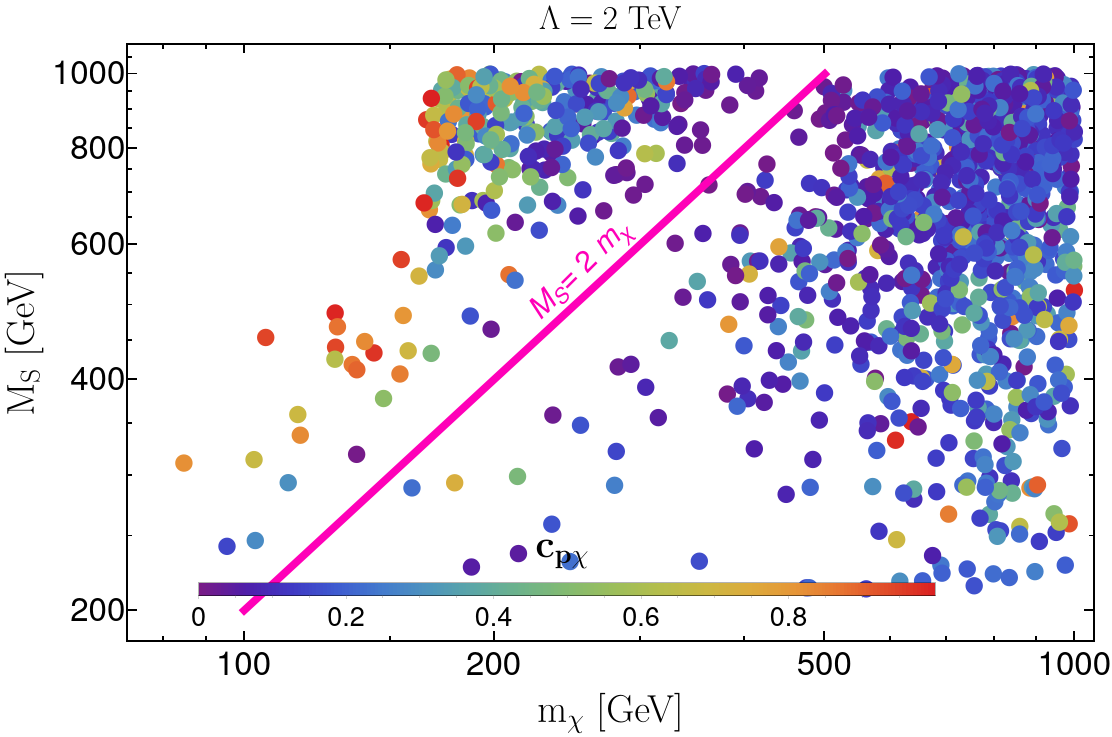}\label{fig:ms_mchi_DM_flavor}}~~			
		\end{center}
		\caption{Correlation among the couplings $ c_s, ~c_p, ~c_{s\chi}, ~c_{p\chi}$ and $ ~c_G $ and DM and mediator mass, from a simultaneous analysis where all the observables relevant to us, taken into account including flavour changing charged and neutral current, electroweak precision observables as well as dark sector constraints.} \label{fig:corr_flavor_DM}   
	\end{figure}

\begin{figure}[t]
	\begin{center}		
		\subfloat[]{\includegraphics[scale=0.11]{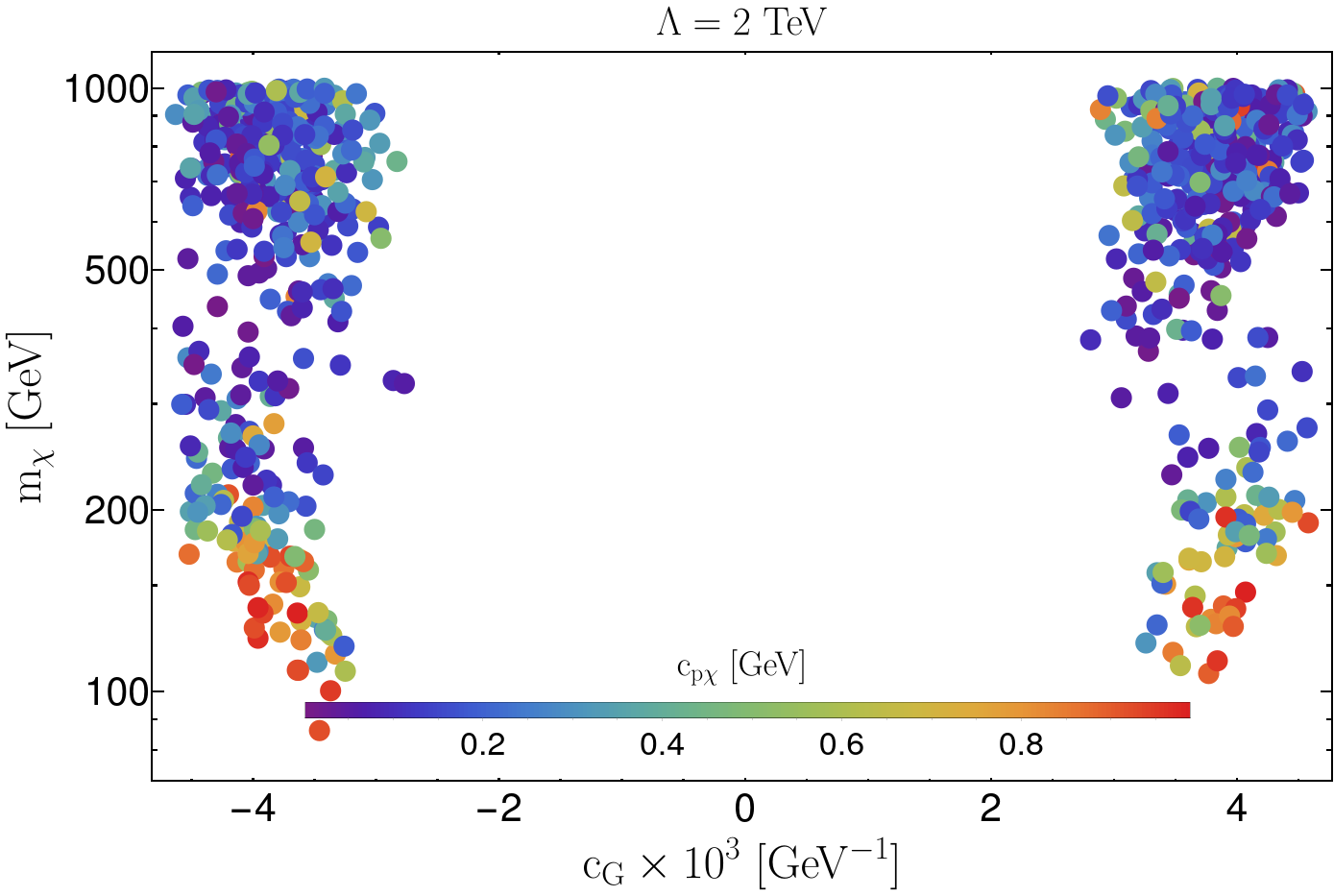}\label{fig:cpX_cG_DM_flavor_CDF}}~~~
		\subfloat[]{\includegraphics[scale=0.11]{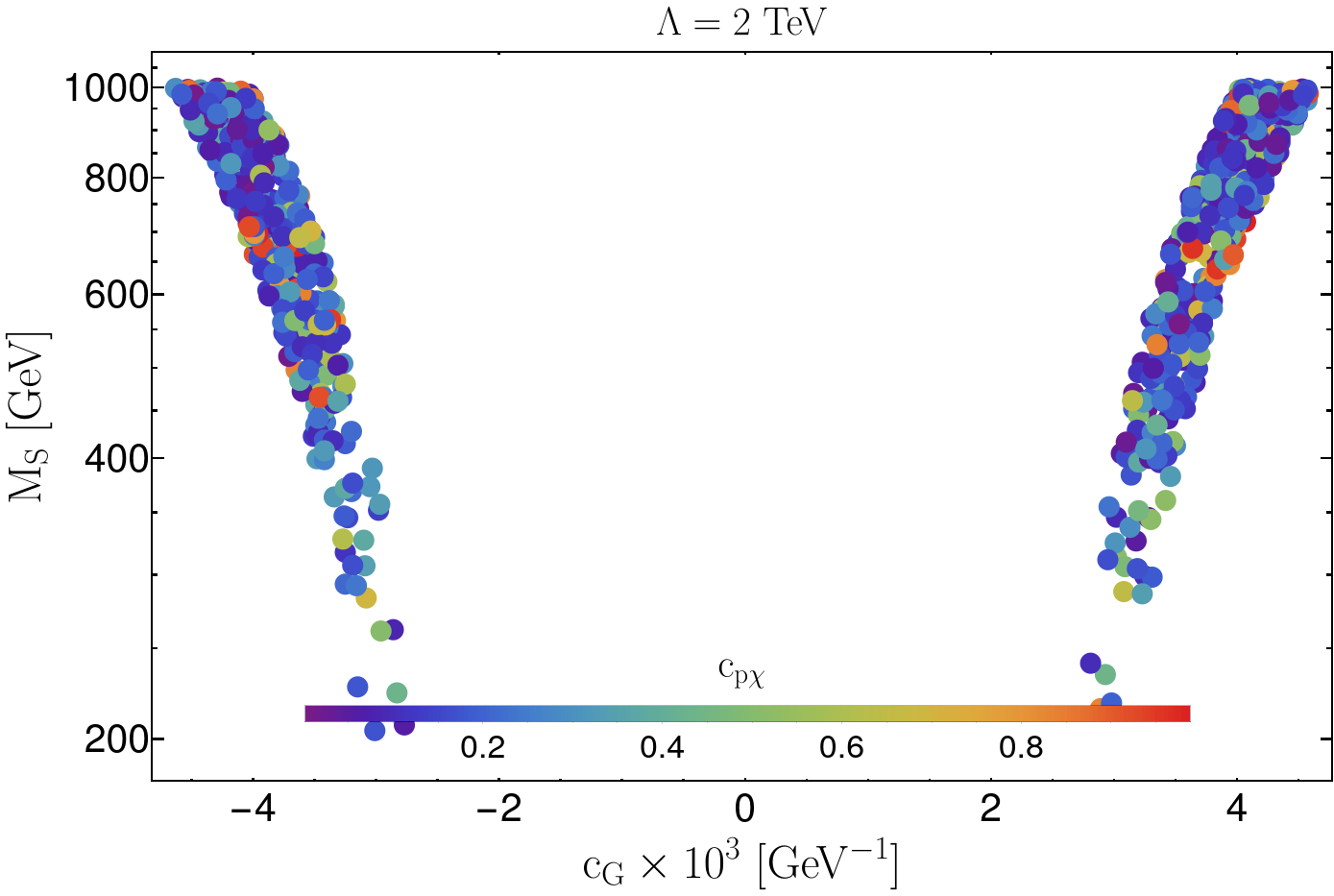}\label{fig:MS_mchi_DM_flavor_CDF}}\\
	\end{center}
	\caption{Correlation among the gauge coupling $ c_{G} $ and DM and mediator masses, when we have taken the input of $ \delta(\Delta r) $ only by CDF into account, along with all the other observables mentioned above.} \label{fig:corr_flavor_DM_CDF}   
\end{figure}

The correlated parameter spaces satisfying the relic density and direct detection bounds are then used to scan the flavour constraints discussed earlier. In this scan, we will not get any direct bounds on the $c_{s\chi}$, $c_{p\chi}$ and $m_{\chi}$. However, these three variables are correlated with each other and with $c_s$, $c_p$ and $M_S$, respectively. Hence, constraints on $c_s$, $c_p$ and $M_S$ from the flavour data will indirectly put constraints on $c_{s\chi}$, $c_{p\chi}$ and $m_{\chi}$. From the scan, we noticed that approximately 25\% of the correlated allowed points of the DM analysis survived the flavour constraints. Hence, many of the allowed solutions for the DM mass and its coupling to the mediator will be discarded. In fig. \ref{fig:corr_flavor_DM}, we have shown the allowed parameter spaces and their correlations, which satisfy the data on the observables related to FCNC, FCCC processes, the W- and Z-pole observables, the relic density, and the direct detection cross-section. Note that to generate these parameter spaces, we do not include the CDF data on W-mass, which shows a large deviation from the rest. However, we have done a separate scan including only this data on $W$ mass, which we have shown in fig. \ref{fig:corr_flavor_DM_CDF}, respectively.

 As we have seen earlier, the analysis, including data other than the relic and direct detection cross-section, restricts the parameter spaces for $c_s$, $c_p$, and $c_G$. We have studied the correlations between these parameters, and the results are similar to those presented in plots of fig. \ref{fig:combscan}. In addition, as expected, the correlations between these parameters do not depend much on the variation of the DM mass. In the analysis without the CDF data on the W-mass, the bound is consistent with the bound obtained in the analysis without the relic and the direct detection cross-section. Also, in the allowed region, $c_{G}$ does not have any noticeable correlations with the other parameters. Therefore, we will not show them separately. The allowed value of the gauge coupling  $|c_G|$ is $\lesssim 0.002 \rm ~GeV^{-1}$.

In the items below we will summarised a few important points of the correlation plots in fig. \ref{fig:corr_flavor_DM}. 
\begin{itemize}
	  \item Like before, in fig. \ref{fig:csX_cs_DM_flavor}, we have shown the correlations between $c_{s\chi}$, $c_{s}$ and $M_S$ (in colour band). We have more allowed solutions for $M_S \gtrsim 500$ GeV for $c_{s\chi} > 0.01$. Also, in this region of $M_S$ for $0.01 < c_{s\chi} \lesssim 0.2$, the maximum allowed values of $|c_s|$ could be near to $0.01 \rm ~GeV^{-1}$. However, for $c_{s\chi} \gtrsim 0.3$ the allowed region of $|c_s|$ is shrinking towards the values $< 0.005 \rm ~GeV^{-1}$, and for a solution $c_{s\chi} \approx 1.0$, the allowed values will be $|c_s| \lesssim 0.001$. These constraints on $c_s$ are further stricter in the region $M_S < 400$ GeV. In this limit, the value $|c_s|$ is $< 0.005$ even when $0.01 < c_{s\chi} \lesssim 0.1$ which will reduce further for higher values of $c_{s\chi} > 0.1$.  
	
	\item In figs. \ref{fig:cpX_cp_DM_flavor} and \ref{fig:cpX_cs_DM_flavor}, we have shown the correlations of $c_{p\chi}$ with $c_p$ and $c_s$ (the variation in $m_{\chi}$ is shown in colour bands), respectively. Here, also, we see the high density of the allowed solutions for $c_{p\chi} \gtrsim 0.05$ for the region of DM mass we have considered. For $c_{p\chi} \lesssim 0.1$, the allowed values of both the $|c_s|$ and $|c_p|$ will be $\lesssim 0.01 \rm ~GeV^{-1}$. However, one should note that there is a correlation between $c_s$ and $c_p$, which we have shown in fig. \ref{fig:combscan}. Note that for values of $|c_s|$ around $0.01 \rm ~GeV^{-1}$, $|c_p|$ will have solutions $\lesssim 0.005 \rm ~GeV^{-1} $ and vice versa. For $M_S \gtrsim 500$ GeV in the region $c_{p\chi} > 0.5$ the allowed regions of $c_s$ and $c_p$ are gradually shrinking with the increasing values of $c_{p\chi}$, and approaches the values $|c_{s(p)}| \lesssim 0.001 \rm ~GeV^{-1} $ when $c_{p\chi}$ approaches towards $1.0$.  For $m_{\chi} < 400$ GeV, $c_p$ or $c_s$ could achive a value around $0.005 \rm ~GeV^{-1} $ even though $c_{p\chi} \approx 1.0$.        
	
	\item We can make a similar observation as in item-2 after a close inspection of the correlation between $c_{p\chi}$ and $m_\chi$ in fig. \ref{fig:mchi_cpX_cp_DM_flavor}. We note that for $m_{\chi} > 250$ GeV, $c_{p\chi} \gtrsim 0.1$ are allowed when $c_p < 0.005 \rm ~GeV^{-1}$ and it could be less than $0.001 \rm ~GeV^{-1}$ when $c_{p\chi}$ obtain a value close to $1.0$. In this region of the DM mass $c_p \lesssim 0.01$ when the allowed values of $c_{p\chi}$ is around $0.1$. The density of solutions for $m_\chi < 175$ reduced considerably. On the other hand, for $175 < m_{\chi} < 250$ GeV we can see the values $|c_p| \approx 0.01 \rm ~GeV^{-1} $ will be allowed even when $c_{p\chi} \approx 1.0$.  
	
	\item In fig. \ref{fig:csX_cpX_DM_flavor}, we have shown the correlation between $c_{s\chi}$, $c_{p\chi}$ and $m_{\chi}$. The nature of the correlation is similar to the one obtained earlier in fig. \ref{fig:csX_cpX_DM}. The difference is that many points are not allowed by the flavour and other data used in this analysis. We see that for both $c_{s\chi}$ and $c_{p\chi}$, the allowed solutions are $> 0.01$.
	
	\item We have also shown the correlation between $M_S$ and $m_{\chi}$ in fig. \ref{fig:ms_mchi_DM_flavor}, which is very similar to what we obtained earlier. However, the combined data now discarded many allowed points of fig. \ref{fig:ms_mchi_DM}. We have solutions for $M_S > 2 m_\chi$ and $M_S < 2 m_\chi$. However, the density of solutions close to the resonance region $M_S \approx 2 m_{\chi}$ has reduced now. Note that for $m_\chi < 500$ GeV, to explain the relic density close to the resonance region, we need relatively smaller values of $c_{p\chi}$ or $c_s$ and/or $c_p$ which are not allowed by the data other than DM which we can see from the figs. \ref{fig:cpX_cp_DM_flavor} and \ref{fig:cpX_cs_DM_flavor}, respectively. Also, we observe an increase in the density of solutions for $m_{\chi} > 500$ GeV and $0.01 <  c_{p\chi} \lesssim 0.8$, which is as per the other observed correlations. Another interesting point is that earlier, we had solutions for $m_\chi < 150$ GeV and $M_S > 500$ GeV, which are missing in fig. \ref{fig:ms_mchi_DM_flavor}. It is due to the reduced allowed points for relatively higher values of $c_{p\chi}$ ($\gtrsim 0.1$) in this DM mass region. We have also noticed this in figs. \ref{fig:cpX_cp_DM_flavor} and \ref{fig:cpX_cs_DM_flavor}. 
\end{itemize} 	

To generate the correlations in fig. \ref{fig:corr_flavor_DM}, we have not included the data on $W$ mass measurement from the CDF \cite{CDF:2022hxs}, which is largely deviated from the other measurements from ATLAS and LHC. Hence, a scan of all the data of $W$ mass measurement will not give us any allowed points. Hence, we have done a separate scan where all the other inputs are used, but for the $W$ mass, we have included only the data from CDF. The results of the scan are shown in fig. \ref{fig:corr_flavor_DM_CDF} where we have shown only the correlations and the allowed points between $c_G$ with the DM mass and the mediator mass. The allowed parameter spaces for the other parameters have not changed with respect to what we have obtained above. Hence, we have shown them separately. As was seen in table \ref{tab:combinecseqcp_CDF}, we obtain tight bounds on $c_G$. We have solutions throughout the given ranges of $m_\chi$ and $M_S$; we notice only a very slight shift in the allowed value of $c_G$ between the higher and lower values of $M_S$ which is mainly coming from the $W$ mass data.  

We have done similar studies for the cases $c_s =0$ and $c_s =c_p$, and we have not noticed any significant changes in the correlations between the relevant parameters. The correlations applicable to these scenarios are more or less identical to the ones presented in the fig. \ref{fig:corr_flavor_DM}. Therefore, we will not present them separately.

\paragraph{\underline{Indirect Detection Bound on the Parameter Space}:}
The limit on DM self annihilation to fermion and boson pairs has been reported by Fermi-collaboration with six years of observation of 15 dwarf spheroidal galaxies \cite{Fermi-LAT:2015att}.  They also have projected sensitivity for 45 dSphs of 16 years of observation \cite{Fermi-LAT:2016afa}. There are also bounds from Fermi-LAT, H.E.S.S telescope, and also the projected bounds from Cherenkov Telescope Array (CTA) to the streaming gamma rays from Galactic Centre which can put bound to different annihilation channels on DMs like $ \chi \bar{\chi} \to b \bar{b}, ~\tau^+ \tau^- , ~W^+ W^-, ~ZZ, ~gg. $ In this work, we have used the projected bound from Fermi-LAT for the annihilation to the fermion pair and the observational bound of Fermi-LAT to the boson pair. 

\begin{figure}[htb!]
	\begin{center}
		\subfloat[]{\includegraphics[scale=0.1]{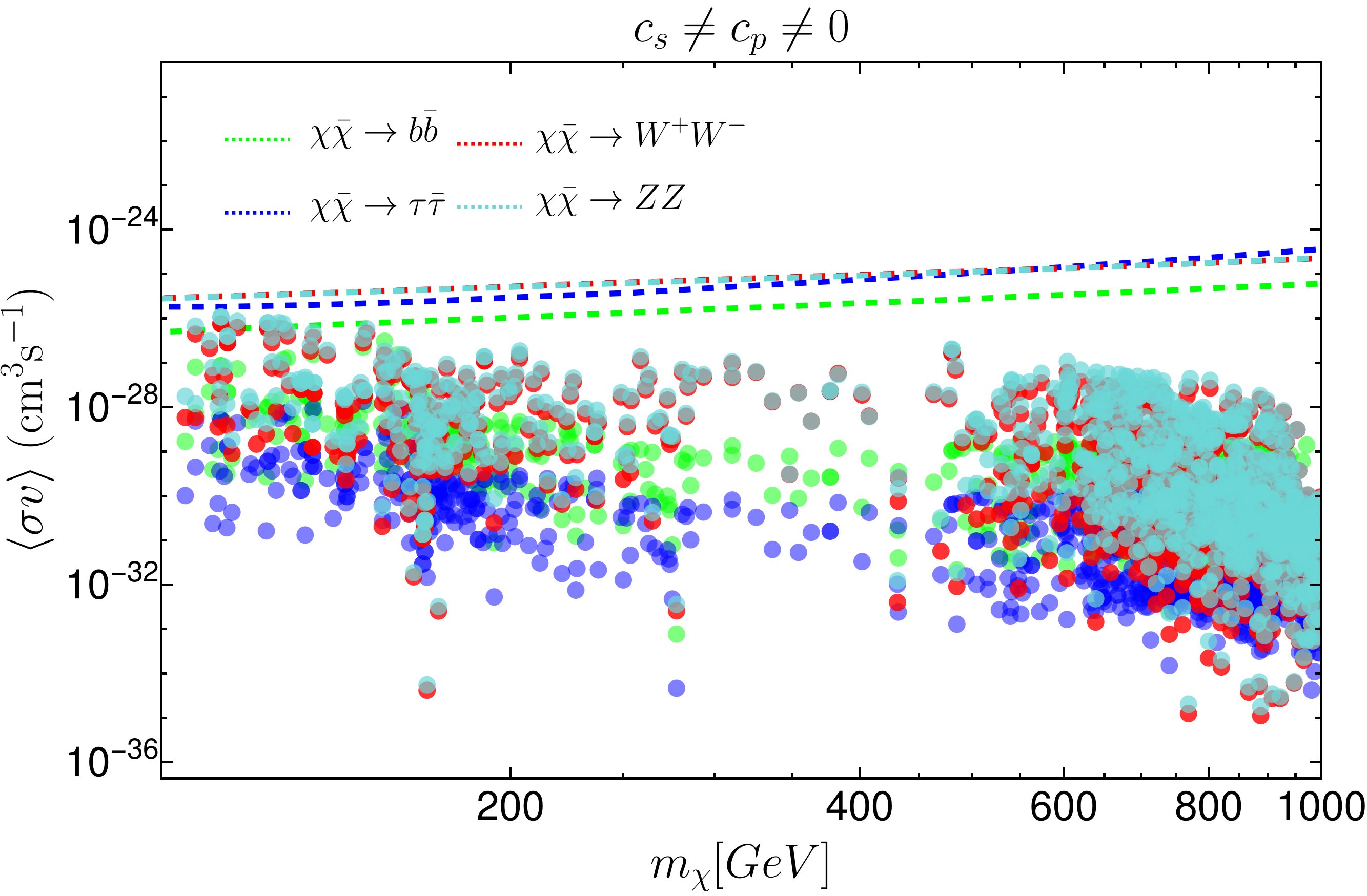}\label{fig:indirect_highmass_1}}
		\subfloat[]{\includegraphics[scale=0.1]{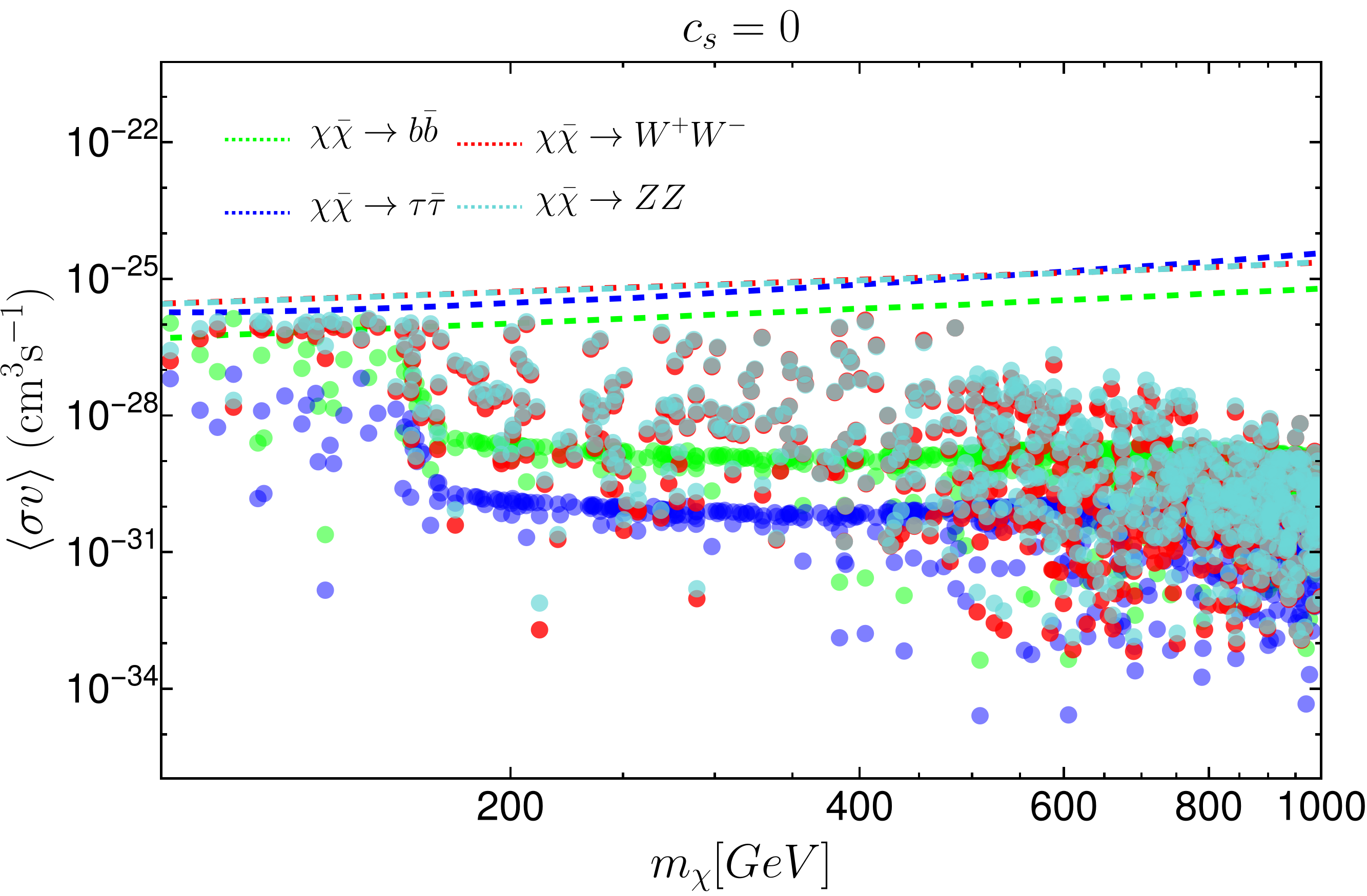}\label{fig:indirect_highmass_2}}\\
		\subfloat[]{\includegraphics[scale=0.1]{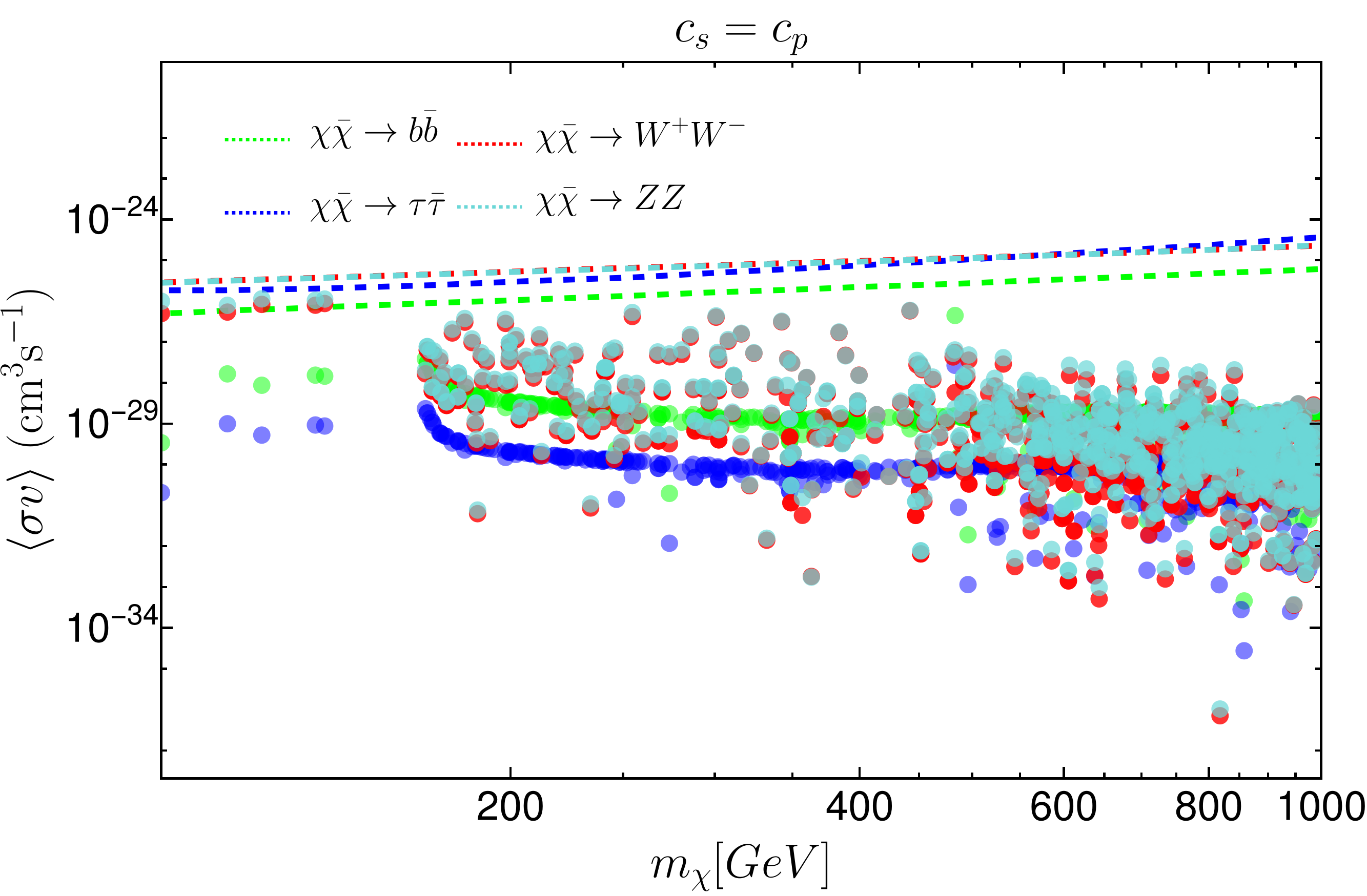}\label{fig:indirect_highmass_3}}
	\end{center}
	\caption{Variation of indirect detection cross-section of the DM with its mass $ m_{\chi}. $ Plots are shown for different combinations of scalar and pseudoscalar couplings $ c_{s} $ and $ c_{p}, $ i.e. for $ c_{s} \neq c_{p} $ (upper left), $ c_{s}=0  $ (upper right) and $ c_{s} = c_{p} $ (lower). The dashed lines are observational DM annihilation cross-sections bound to SM channels mentioned in the inset of the plot, whereas the same colour scatter plots correspond to the annihilation rate for our model to that channel. Points are allowed from both DM relic and DD cross-section as well as other observables mentioned in table \ref{tab:CKM-updated-obs}.}
	\label{fig:indirect}
\end{figure}

In fig. \ref{fig:indirect}, we have shown the parameter space in $ <\sigma v> - m_{\chi} $  plane for the annihilation rate of DM to boson and fermionic pair. The green colour corresponds to $ \chi \bar{\chi} \to b \bar{b} $, blue colour corresponds to $ \chi \bar{\chi} \to \tau \bar{\tau} $, red corressponds to $ \chi \bar{\chi} \to W^+ W^- $ and finally cyan shows $ \chi \bar{\chi} \to ZZ $ bounds. In each plot, all the scatter points correspond to one particular color, showing the annihilation cross-section rate to that particular channel for this model. Also, all the points in the plot satisfy both observed relic density value and direct detection cross-section bounds along with all other flavour and electroweak observables. Figs. \ref{fig:indirect_highmass_1} , \ref{fig:indirect_highmass_2} and \ref{fig:indirect_highmass_3}  shows the annihilation rate for different coupling combinations of $ c_{s} $ and $ c_{p}. $ The upper-left panel plot is done where $ c_{s}, ~c_{p} $ both are free and varied freely.  The plot in the upper right panel is done for the case when there is only pseudoscalar coupling between the mediator and SM, i.e., $ c_{s} =0. $ The last plot in the lower panel is done when both the couplings are there, but they are equal, i.e., $ c_{s} = c_{p}. $ In each plot, the dashed lines denote the bound of the corresponding annihilation channel given by Fermi-Lat, mentioned earlier. For the higher mass region, all the points lie well below the provided bound, but for the lower mass region, we get a comparable bound from our model. 

We can see from the figures that the parameter space, which was allowed by DM relic density, direct detection, and of the flavour, electroweak observables, are also allowed by indirect detection bound provided by Fermi-LAT. Indirect detection bound does not exclude or give better constraints in the parameter space. 

\subsection{Bounds on the dimensionless couplings}\label{subsec:boundssummary}
In our analysis, we have obtained bounds on the dimensionful couplings: $ c_s, ~c_{p} $ and $ c_{G} $ which are related to the dimensionless couplings $ g_s, ~g_p $ and $ g_V $, respectively, via the relations given in eqs. \eqref{eq:fermion_coupling} and \eqref{eq:gauge_coupling}. The bounds on the dimensionful couplings from the FCNC, FCCC and electroweak precesion observables are as follows: 
\begin{eqnarray}
|c_s| \lesssim 0.01 \rm ~GeV^{-1}, \quad |c_p| \lesssim 0.01 \rm ~GeV^{-1} ,  \quad |c_G| \lesssim 0.002 \rm ~GeV^{-1}.
\end{eqnarray}
Using these bounds we obtain the following bounds on the dimensionless couplings 
\begin{equation}\label{eq:MSgen}
|g_{s} | \lesssim 1.74, \quad |g_{p}| \lesssim 1.74, \quad |g_{V} | \lesssim 0.49.
\end{equation} 
However, while we include the data on DM relic density, direct and indirect detection bounds the above bounds will depend on $ m_{\chi}, ~c_{s\chi} $ and $ c_{p\chi} $. If we take $ m_{\chi} \geq 500  $ GeV and $ c_{s\chi} \geq 0.1,  ~c_{p\chi} \geq 0.1 $, the bounds on $ c_s, ~c_p $ will be : $ |c_{s}|, ~|c_{p}| \lesssim 0.005\rm ~GeV^{-1} $ which will lead to : 
\begin{equation}\label{eq:mX500}
|g_{s} | \lesssim 0.87, \quad |g_{p}| \lesssim 0.87.
\end{equation}
However, one should note that, $ c_s $ and $ c_p $ has a correlation which we have pointed out earlier. If we take $ c_{s\chi}, c_{p\chi} \geq 0.5  $ the corresponding bounds will be $ c_{s}, ~c_{p} \lesssim 0.001 \rm ~GeV^{-1} $ which will lead to : 
\begin{equation}
|g_{s} | \lesssim 0.17, \quad |g_{p}| \lesssim 0.17.
\end{equation}
For $ m_{\chi} < 500 $ GeV, the bounds will be little relaxed even if we take $ c_{s\chi} $ and $ c_{p\chi}  > 0.1 $ and the relevant bounds will follow eq. \eqref{eq:mX500}. The bounds on $ g_{V} $ will not depend on the data on DM searches and the relevant bound will the eq. \eqref{eq:MSgen}.

In section \ref{higher_dim_model}, we have shown an example of a toy model with dim-5 operators from which we can formulate our working model. We can see from eq. \eqref{eq:higherdimeq} that for small mixing angle, the relation between the couplings of the toy model and our working model will be as follows:
\begin{equation}\label{eq:dim5_ssf}
\mathbb{C}_{s}^{S} = m_f ~ c_s =- \frac{C ~ v ~ \alpha}{\Lambda \sqrt{2}}, \quad \mathbb{C}_{p}^{S} =m_f ~c_p = - \frac{C ~ v }{\Lambda \sqrt{2}}\,,
\end{equation} 
and
\begin{equation}\label{eq:dim5_gauge}
\mathbb{C}_{W}^{S} = \frac{C' m_{W}^2}{\Lambda}, \quad\mathbb{C}_{Z}^{S}= \frac{C' m_{Z}^2}{2 \Lambda}\,.
\end{equation}
Using eq. \eqref{eq:dim5_ssf}, we can estimate the value of $ \alpha $ as:
\begin{equation}
\alpha = \frac{\mathbb{C}_{s}^{S}}{\mathbb{C}_{p}^{S}}\,.
\end{equation}
Hence, using the bounds on $c_s, ~c_p  $ and $ c_G $ we have obtained  
\begin{eqnarray}
\left|\frac{C^{'}}{\Lambda}\right|  &~\approx~&  2 c_{G} ~\leq 0.004 \rm ~ GeV^{-1}, \\
\left|\frac{C }{\Lambda \sqrt{2}}\right| &~\approx~&   \frac{c_{p} ~m_{t}}{v} ~\leq 0.007 \rm ~GeV^{-1}\,.
\end{eqnarray}
Here, we have only used the coupling with the top quark to compare the couplings, but we can also use the other fermion couplings, which will be little relaxed.

\subsection{Comparison with the benchmark scenarios of a 2HDM with extensions}

 It is not very common in the literature to find a UV complete model with a similar operator structure to ours. However, our simplified model with a spin-0 mediator has some features which could be compared with the models of two-Higgs-doublet plus an additional pseudoscalar (2HD + P). To compare our allowed solutions, we have focused on the benchmark scenarios or the constraints in the most recent analysis of ref. \cite{Arcadi:2022lpp}. One may also look at the refs.  \cite{LHCDarkMatterWorkingGroup:2018ufk,Arcadi:2020gge}. All these analyses are based on a set-up of (2HD + P). The 2HD model under consideration could be Type-I, Type-II, Type-X and Type-Y. These different types are decided based on the interactions of the quarks and leptons with the new scalar/pseudoscalar multiplets. 
 
  We should note that apart from the SM Higgs, the (2HD+P) have one more neutral CP-even state (H), the charged Higgs states ($H^{\pm}$) and two CP-odd states $A$ and $P$, respectively, which could mix. Hence, comparing the Yukawa Lagrangian or the gauged kinetic terms of the (2HD + P) \cite{Arcadi:2022lpp} to our simplified Lagrangian will be possible only in a few limited cases. For example, one such situation could be when the masses of $H$, $A$ and $P$ are nearly equal. In such a case, one could expect that the mixing angle $\theta$ between $A$ and $P$ will be such that $\sin\theta \approx \cos\theta \approx 0.7$.  Another possible situation is when the probability of mixing between $A$ and $P$ is small, i.e. when $\sin\theta \ll 1$ \cite{Bauer:2017ota}. In such a case, it is expected that the mass splitting between $A$ and $P$ will be large. In the ref. \cite{Arcadi:2022lpp} the analyses were done considering the values of $\sin\theta$ in the range $0.1$ to $0.8$. Also, they have provided solutions for the scenario $M_H = M_A$. In such a situation, apart from the charged Higgs interaction terms, we can roughly compare our neutral scalar-fermion interaction term with the one given in the ref. \cite{Arcadi:2022lpp}. Though the comparison may not be exact, this might still help to get a rough idea about the probably allowed values of $\tan\beta$ depending on the mass of the spin-0 mediator. 
 
 \begin{table}[t]
 	\centering
 	\renewcommand{\arraystretch}{1.8}
 	\begin{tabular}{|c|c|c|c|}
 		\hline
 		\multicolumn{4}{|c|}{Couplings} \\
 		\hline
 		& up-quark & d-quark & lepton \\
 		\hline
 		\multirow{2}{*}{type-I} & $ |c_{s}| = \frac{\cot \beta}{v}  $ & $ |c_{s}| = \frac{\cot \beta}{v}  $ & $ |c_{s}| = \frac{\cot \beta}{v}  $ \\
 		&  $ |c_{p}| = \frac{\cot \beta \, \cos \theta }{v}  $ & $  |c_{p}| = \frac{\cot \beta \, \cos \theta }{v} $ & $ |c_{p}| = \frac{\cot \beta \, \cos \theta }{v} $\\
 		\hline
 		\hline
 		\multirow{2}{*}{type-II} & $ |c_{s}| = \frac{\cot \beta}{v}  $ & $ |c_{s}| = \frac{\tan \beta }{v}  $ & $ |c_{s}| = \frac{\tan \beta }{v}  $ \\
 		&  $ |c_{p}| = \frac{\cot \beta \, \cos \theta }{v}  $ & $  |c_{p}| = \frac{\tan \beta\, \cos \theta  }{v} $ & $ |c_{p}| = \frac{\tan \beta\, \cos \theta  }{v} $\\
 		\hline
 		\hline
 		\multirow{2}{*}{type-X} & $ |c_{s} | =\frac{\cot \beta}{v}  $ & $ |c_{s} | =\frac{\cot \beta}{v}  $ & $ |c_{s}| = \frac{\tan \beta }{v}  $ \\
 		& $ |c_{p}| = \frac{\cot \beta \, \cos \theta }{v}  $  & $ |c_{p}| = \frac{\cot \beta \, \cos \theta }{v}  $  & $ |c_{p}| = \frac{\tan \beta\, \cos \theta  }{v} $\\
 		\hline 
 		\hline
 		\multirow{2}{*}{type-Y} & $ |c_{s} | =\frac{\cot \beta}{v}  $ & $ |c_{s}| = \frac{\tan \beta }{v}  $ & $ |c_{s}| = \frac{\cot \beta}{v}  $ \\
 		& $ |c_{p}| = \frac{\cot \beta \, \cos \theta }{v}  $  & $  |c_{p}| = \frac{\tan \beta\, \cos \theta  }{v} $ & $ |c_{p}| = \frac{\cot \beta \, \cos \theta }{v} $\\
 		\hline 
 		All type & \multicolumn{3}{|c|}{$ |c_{G}|  = \frac{\cos (\beta - \alpha)}{v} $} \\
 		\hline
 	\end{tabular}
 	\caption{Relations of $c_s \,, $ $c_p$ and $ c_{G} $ with the equivalent parameters in the model (2HD + P). }
 	\label{tab:relations}
 \end{table}	
  
In ref. \cite{Arcadi:2022lpp}, the constraints they have obtained on $\tan\beta$ and the masses of the neutral scalar or pseudoscalar are mainly coming from the data on collider results on the Higgs coupling strengths to fermions, gauge bosons and the searches of extra Higgs states. In particular, the bounds on $|\cos(\beta-\alpha)|$ were obtained from the measurements of Higgs coupling strengths to the gauge bosons. The relevant bounds for Type-I and X are given by 
  \begin{equation}
 1 < \tan\beta \lesssim 60,\ \  |\cos(\beta-\alpha)| \lesssim 0.2
  \end{equation}
the allowed solutions for $M_A = M_H$ lie within the range $[200, 1000]$ (in GeV). While the bounds for Type-II and Y will be
\begin{equation}
1 < \tan\beta \lesssim 15, \ \ |\cos(\beta-\alpha)| \lesssim 0.1
\end{equation}
and the solutions for $M_A = M_H$ will lie within the range $[500, 1000]$ (in GeV).  Within these allowed parameter spaces they have obtained the bounds on the DM mass and it's coupling with the new spin-0 state. 
 
In the small $\sin\theta$ limit, we can relate the parameters $c_s$, $c_p$ and $c_G$ defined in our analysis with the relevant parameters in (2HD + P), and the relations are given in table \ref{tab:relations} (with $\cos\theta \approx 1$).  We note that in the model Type-I and Type-X, the coupling strengths of the up and down type quarks with the new spin-0 particles are the same. However, those in Type-II and Type-Y are different. So far, in our analyses, we have presented the results with the assumptions of universal couplings for the up and down type quarks. Therefore, for Type-I and Type-X we can use the bounds on $c_s$ and $c_p$ discussed in subsections \ref{subsec:bounds_univcoup} or \ref{subsec:boundssummary} to constraint $\cot\beta$ or $\tan\beta$. 
 
 \begin{figure}[t]
 	\begin{center}		
 		\subfloat[]{\includegraphics[scale=0.18]{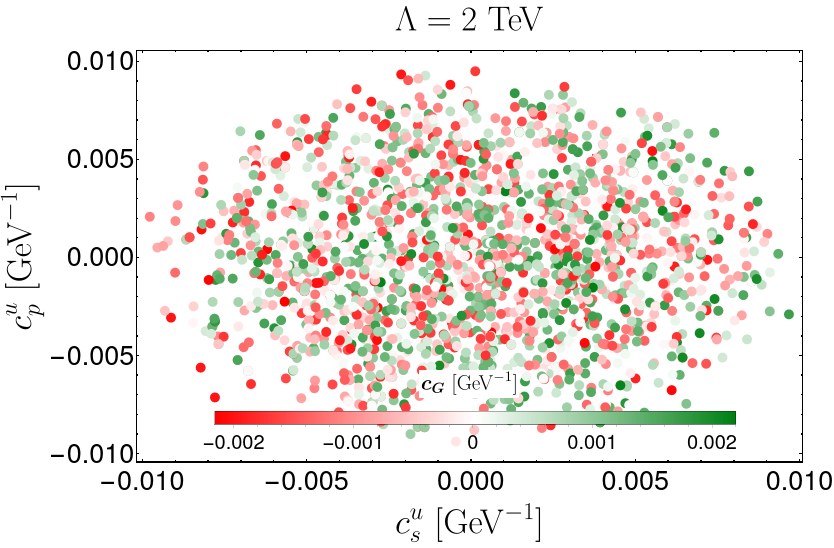}\label{fig:cst_cpt_cG_flavour}}~~~
 		\subfloat[]{\includegraphics[scale=0.17]{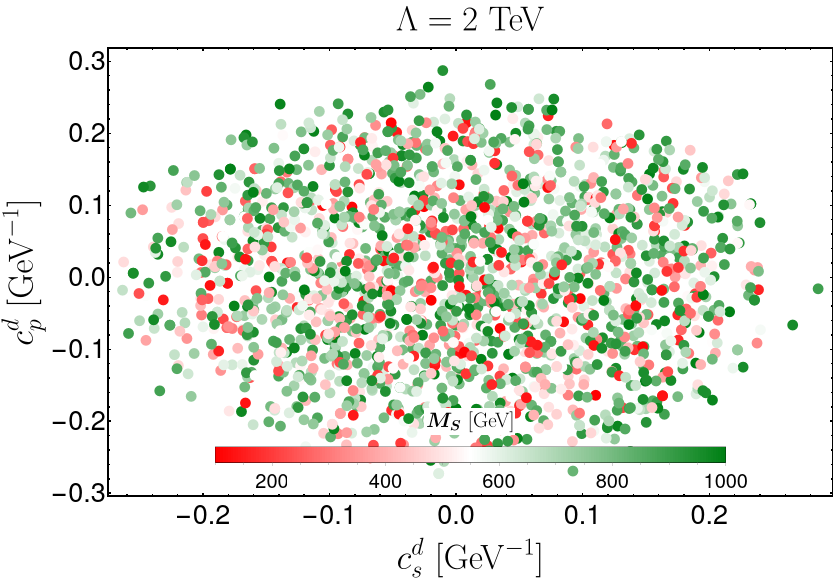}\label{fig:csb_cpb_MS_flavour}}\\
 	\end{center}
 	\caption{Allowed parameter space in $ c_{s}^{u}-c_{p}^{u} $ and $ c_{s}^{d}-c_{p}^{d} $ plane by considering all flavour and electroweak constraints. The color variation is shown with gauge coupling $c_{G}$ (left) and mass of the mediator $M_{S}$ (right).} 
 	\label{fig:flavour_cst_csb}   
 \end{figure}
 \begin{figure}[t]
 	\begin{center}		
 		\subfloat[]{\includegraphics[scale=0.18]{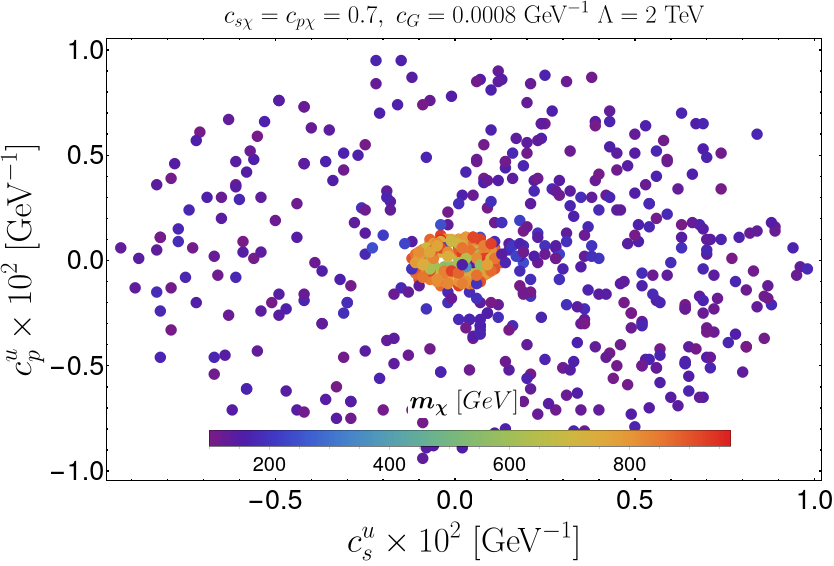}\label{fig:cst_cpt_mchi_inset_DM_flavor}}~
 		\subfloat[]{\includegraphics[scale=0.18]{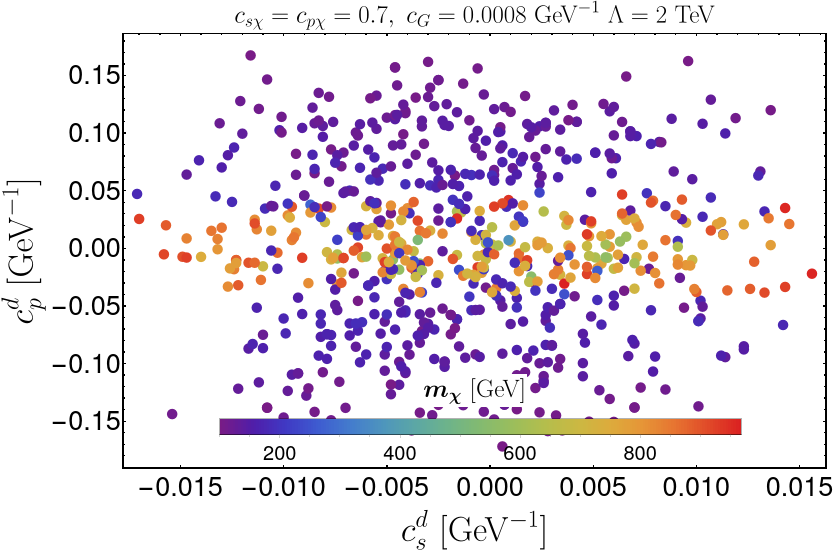}\label{fig:csb_cpb_mchi_DM_flavour}}\\
 	\end{center}
 	\caption{Allowed parameter space in $ c_{s}^{u}-c_{p}^{u} $ and $ c_{s}^{d}-c_{p}^{d} $ plane by taking into account all flavor and electroweak constraints along with DM constraints relic and direct detection bounds. The color variation is shown with mass of the DM $ m_{\chi} $ . } \label{fig:DM_flavor_cst_csb}   
 \end{figure}

For the Type-II and Type-Y models, the couplings of the up and down type quarks to the new scalar are different. Hence, in such cases, to implement the bounds from our analysis, we need to assume different coupling strengths for the up and down type quarks, like $c_{s,p}^u$ and $c_{s,p}^d$, respectively. Following the similar methodology as we have discussed earlier, we have done an analysis considering such type of couplings. We have summmarised the main results relevant to this discussion in figs. \ref{fig:flavour_cst_csb} and \ref{fig:DM_flavor_cst_csb}, respectively. The correlations in fig. \ref{fig:flavour_cst_csb} are obtained from the available data set on the FCCC, FCNC, $Z$ and $W$-pole observables we have discussed earlier. These correlations predict the following upper limits:
\begin{equation}\label{eq:flavourIIX}
|c_{s,p}^u| < 0.01  ~ \text{GeV}^{-1} ,   \ \ \ |c_{s,p}^d| \lesssim 0.25 ~ \text{GeV}^{-1}  \ \ \ \text{and}\ \ \ |c_G| \lesssim 0.002  ~ \text{GeV}^{-1},
\end{equation}
which are valid for the whole region of $ M_{S} $ that we have varied. 

However, when we include the data on the DM searches along with the above data set the above mentioned bounds will be relatively stronger for the higher values of the masses of the DM and their couplings with the mediator. In fig. \ref{fig:DM_flavor_cst_csb}, we have shown the allowed parameter space in the planes of $ c_{s}^{u}-c_{p}^{u} $ and $ c_{s}^{d}-c_{p}^{d}$, respectively. We have obtained those bounds for $c_{s\chi} = c_{p\chi} =0.7$ and $ c_{G} = 0.0008 \, \rm GeV^{-1} $. This choice of $c_G$ will lead to a value $\cos(\beta-\alpha) = 0.2$ which had been referred as a benchmark solution in ref. \cite{Arcadi:2022lpp}. For the DM mass $m_{\chi} \lesssim 500$ GeV, the allowed ranges of $ c_{s,p}^{u,d} $ are given as following
\begin{equation}\label{eq:DM_flav_cst_cpt_bound}
|c_{s,p}^{u}| < 0.01 ~ \text{GeV}^{-1} \,, \ \ |c_{s}^{d}| < 0.02 \, \text{GeV}^{-1}\ \ \text{and} \ \ |c_{p}^{d}| <   0.15 ~ \text{GeV}^{-1}\,. 
\end{equation} 
We have noted that for $|c_{s(p)\chi}| < 0.5$ the bounds on $|c_{s}^{d}|$ will be  $< 0.05$, the other bounds will remain same. This observation is due to the constraints from the direct detection cross section which is sensitive to scalar couplings.  
 
The bound will get more stronger if we go to higher mass region of DM, for example, if $m_{\chi} > 500$ GeV, with above mentioned values of $ c_{s\chi} $ and $ c_{p\chi} $, the conservative bounds on the couplings as shown in figs. \ref{fig:cst_cpt_mchi_inset_DM_flavor} and \ref{fig:csb_cpb_mchi_DM_flavour} are the following: 
\begin{equation}\label{eq:DMflavourIIX}
|c_{s,p}^u| < 0.001~ \text{GeV}^{-1} ,\ \ |c_{s}^{d}| < 0.02~ \text{GeV}^{-1} \, \ \ \text{and}\ \  |c_{p}^d| < 0.05 ~ \text{GeV}^{-1}.
\end{equation}

Furthermore, in the 2HD or in (2HD+P), the couplings $c_s$ and $c_p$ are directly related via the factor $\tan\beta$ or $\cot\beta$. In our analysis, we have treated $c_s$ and $c_p$ independently. Hence, any possible correlation between them will come only from the data. Therefore, for different bounds on $c_s$ and $c_p$ we will get different allowed ranges of $\tan\beta$. In addition, the projected bounds on $\tan\beta$ from $c_p$ will depend on the angle $\theta$. For simplicity, we will only present the projected bounds on $\tan\beta$ from the $c_s$. Interested readers can obtain the bounds on $\tan\beta$ by projecting the respective bounds on $c_p$ depending on their choices of $\theta$. 

In the limit $\sin\theta =0$, the model (2HD + P) could be considered as a general 2HD model (2HDM) without extensions. The DM particle will not couple to the 2HDM bosons in this scenario. The couplings of a 2HDM can be compared to our model with $c_s = c_p$, and the bounds $c_s =c_p \lesssim 0.01 \, \rm GeV^{-1}$ put limits on $\tan\beta$. Note that we have obtained these bounds on $c_s$ and $c_p$ from data other than the DM searches. It also applies to relatively low values of the DM masses ($m_\chi < 500$ GeV) or the couplings. For Type-I and Type-X 2HDM, the above bounds will put the limit 
\begin{equation}\label{eq:boundswoDM}
\cot\beta \lesssim 2.46 \ \ \text{or}\ \  \tan\beta \gtrsim 0.40. 
\end{equation}
While for the Type-II and Type-Y models, using the bounds on $c_{s}^u$ and $c_{s}^d$ from eq. \eqref{eq:flavourIIX} we will (roughly) obtain
\begin{equation}
0.40 \le \tan\beta \le 60.
\end{equation}

For $\sin\theta \ne 0$, the DM couples to the 2HDM pseudoscalar Higgs bosons. In such a situation, we can talk about the bounds we have obtained, including the data from the DM searches. From our analysis, we have learned that for masses $M_S \gtrsim 500$ GeV and $m_\chi \gtrsim 500$ GeV, the bounds on $c_s$ or $c_p$ will be relatively stronger as compared to those obtained for $m_{\chi} < 500$ GeV. For $m_{\chi} > 500$ and $c_{s(p)\chi} > 0.5$, for the model Type-I and X, we can apply the bounds $|c_{s}| < 0.001$ and we will obtain
 \begin{equation}
 \cot\beta \lesssim 0.25 \ \ \ \text{or}\ \ \ \tan\beta \gtrsim 4. 
 \end{equation}
For $m_{\chi} < 500$ GeV or for $c_{s(p)\chi} < 0.5$, the bounds will be as given in eq. \eqref{eq:boundswoDM}. For the model Type-II and Y, for the $m_{\chi} > 500$ GeV and $c_{s(p)\chi} \ge 0.5$ we apply the constraints $|c_{s}^u|$ and $|c_{s}^{d}|$ given in eq. \eqref{eq:DMflavourIIX} and obtain
\begin{equation}
4 < \tan\beta < 5.
\end{equation}
This bound will be little relaxed when we consider $m_{\chi} < 500$ GeV or $c_{s(p)\chi} < 0.5$. In such a case, we apply the bound $c_{s}^u < 0.001$ and $|c_s^d| < 0.05$ and obtain
\begin{equation}
4 < \tan\beta < 12.
\end{equation}
It is possible to calculate the bounds on $\tan\beta$ using the constraints on $|c_p^{u,d}|$, however, for that one need to use a particular value of $\theta$.  Also, in all types of 2HDM discussed above, the coupling $|c_G|$ is related to $\frac{\cos(\beta-\alpha)}{v}$ as defined in the table \ref{tab:relations}. The bounds on $|c_G|\lesssim 0.002 \, \rm GeV^{-1}$ which leads to a bound $|\cos(\beta-\alpha)| < 0.5$.

\section{Summary}
We have considered a simplified DM model with a fermionic dark matter and a spin-0 mediator communicating to the SM fermions and gauge bosons. We obtain the constraints on the scalar and pseudoscalar couplings of the mediator with the SM fermions ($c_s, c_p$) and the new gauge couplings ($c_G$) from a simultaneous analysis of the inputs associated with the measurements of the low energy FCCC and FCNC observables, $W$ and $Z$-pole observables and the bounds on $t\to b W$ effective couplings. Among the low-energy FCCC observables, we have considered the inputs on the CKM matrix, as well as semileptonic and leptonic rates. The FCNC observables include the available data on semileptonic, leptonic, and rare decays corresponding to $b\to s \ell^+\ell^-$, $s\to d \ell^+\ell^-$ and $c\to u\ell^+\ell^-$ transitions and the available limits on the branching fraction of the invisible decays $P \to P' \nu\bar{\nu}$. In addition, we have considered the inputs on oscillation amplitudes on the $B_{d/s}-\bar{B}_{d/s}$, $K-\bar{K}$ and $D_0-\bar{D}_0$ mixings. We have varied $M_S$ over 100 GeV to 1000 GeV. In the unit of GeV$^{-1}$, we find the following constraints: $|c_s| \lesssim 0.01 \rm ~GeV^{-1}$, $|c_p| \lesssim 0.01 \rm ~GeV^{-1}$ and $|c_G| \lesssim 0.002 \rm ~GeV^{-1}$. To get these numbers, we have not included the data on $W$ mass measurement from CDF, which shows deviation with respect to the other respective data. If we include this data, then the required value of the gauge coupling will be $|c_G| \approx 0.004$ GeV$^{-1}$.

Finally, from a simultaneous analysis of these data sets alongside the data on DM relic density and direct detection cross section we find out the allowed parameter spaces for the DM mass ($m_\chi$), mediator mass ($M_S$), the couplings ($c_s$, $c_p$) and the couplings ($c_{s\chi}$, $c_{p\chi}$) of the DM with the mediator. Also, we have shown the correlations between these parameters and noted that for the DM mass $m_\chi > 500$ GeV and $c_{p\chi} \gtrsim 0.1$, the allowed values are $|c_s| \lesssim 0.005$ GeV$^{-1}$ and $|c_p| \lesssim 0.005$ GeV$^{-1}$. Both the values independently will be even less than $0.001$ GeV$^{-1}$ if we take $c_{p\chi} \gtrsim 0.5$. We obtain these solutions within the allowed region $0.01< c_{s\chi} \lesssim 1.0$. For $m_{\chi} < 400$ GeV, the constraints on $|c_s|$ and $|c_p|$ will be little relaxed but both of them will be $\lesssim 0.005$ GeV$^{-1}$ when $c_{p\chi} \gtrsim 0.1$. Similarly, we have obtained strong correlations between $c_{s\chi}$, $c_s$ and $M_S$. We have noted that for $M_S < 500$ GeV, the bound on $|c_s|$ will be highly constrained ($\lesssim 0.005$ GeV$^{-1}$) and this bound will reduce further for $c_{s\chi} \gtrsim 0.1$. For $M_S \gtrsim 500$ GeV, we will obtain $|c_s| \lesssim 0.1$ GeV$^{-1}$ and will gradually reduce as $c_{s\chi}$ increases towards a value $\approx 1.0$. 

In addition, we have studied the case of different couplings of the down and up type quarks to the spin-0 mediator and obtained the bounds on the respective couplings. Furthermore, using the constraints on the scalar and pseudoscalar couplings, we have obtained the bounds on the parameter $\tan\beta$ of a 2HDM with an additional pseudoscalar.   

\paragraph{Note:} We have provided the final correlated data set obtained from our combined analysis, named \texttt{DM\_flavor\_2TeV.dat}, which could be useful for collider and other relevant analysis.

	\section*{Appendix}
	\appendix
	\section{Higher dimensional model}\label{sec:apphigherdimopr}
	\paragraph{Couplings with Fermions :}
	Starting from the higher dimensional Lagrangian
	\begin{equation}\label{higher_dim_ferm}
	\mathcal{L}_{ferm}=-\frac{C}{\Lambda} [\bar{\Psi}_{L} i \gamma_{5} H \psi_R P]-y_f [\bar{\psi}_L H\psi_R] + h.c,
	\end{equation}
	with\begin{equation*}
	H=  \begin{pmatrix}
	0 \\
	\frac{v+h}{\sqrt{2}}
	\end{pmatrix} \quad \text{and} \quad P= u+S_1. 
	\end{equation*}   
	$ u  $ and $ v $ are the VEV corresponding to the doublet $ H $ and singlet $ P .$  By expanding the Lagrangian (for the down quark only),
	\begin{align}\label{eqn2}
	\mathcal{L}_{ferm} = & -\frac{C}{\Lambda} \left[ \bar{d}_L i \gamma_5 \left(\frac{v+h}{\sqrt{2}}\right)d_R (u+S_1) \right] -y_f \left[\bar{d}_L\left(\frac{v+h}{\sqrt{2}}\right) d_R \right] + h.c  
	\nonumber \\
	& \Rightarrow -\bar{d} \left[\frac{y_f}{\sqrt{2}} + i \frac{Cuv}{\sqrt{2}\Lambda}\gamma_5 \right] d +  \left[..\text{interaction terms}.. \right]
	\end{align}
	We need to perform a chiral transformation to absorb the $ \gamma_5 $ part of the mass to get the mass basis.
	\begin{equation}
	\psi \to e^{\frac{i \gamma_{5} \alpha}{2}} \psi,  \quad  \psi_L \to e^{-\frac{i \alpha}{2}} \psi_L,    \quad   
	\psi_R \to e^{\frac{i  \alpha}{2}} \psi_R, 
	\end{equation}	 	 
	where  \[ \tan \alpha = \frac{C u v }{y_f \Lambda}.  \]
	With this rotation of the fermionic field, we get the correct mass term. The fermionic  bilinears will be transformed as:
	\begin{equation}\begin{split}
	\bar{\psi}i \gamma_5 \psi \to \bar{\psi}i\gamma_5\psi - \alpha \bar{\psi} \psi\,,  \\
	\bar{\psi} \psi \to \bar{\psi}\psi + \alpha \bar{\psi} i \gamma_5 \psi\,.
	\end{split}\end{equation}  
	Using this Chiral transformation, the dimension four interaction terms can  be written as: 
	\begin{equation}\label{ferm_int}
	- i\left( \frac{C u }{\Lambda \sqrt{2}} + \frac{y_f}{\sqrt{2}}  \right)  [\bar{d} \gamma_{5} d h ] - \left( \frac{y_f }{\sqrt{2}} 	- \frac{C u \alpha}{\Lambda \sqrt{2}} \right)  \left[ \bar{d} d h \right] \\ - i\left( \frac{C v }{\Lambda \sqrt{2}}\right) \left[\bar{d} d S_1 \right]\,.
	\end{equation}

	Now, another rotation of basis needs to be performed to get the scalars in mass basis by angle $ \theta $, such that: 
	\begin{equation}\begin{split}\label{scalar_mixing}
	h_{1} = \cos \theta ~ h - \sin \theta ~S_1\,, \\
	S = \sin \theta ~h + \cos \theta ~S_1 \,, 
	\end{split}
	\end{equation}
	with $ h_1 $ and $ S $ being the SM scalar and new scalar, respectively. They are in mass basis. Using this the kinetic mixing of scalars in the above eq.~\eqref{ferm_int}, we get the interaction terms of the scalars with down quark pairs as: 
	\begin{equation}\begin{split}\label{final_int}	  
	-\left[ \left( \frac{y_f}{\sqrt{2}} - \frac{C u \alpha }{\Lambda \sqrt{2}} \right) \cos \theta + \frac{C v \alpha }{\Lambda \sqrt{2}}  \sin \theta \right] \left[ \bar{d} d h_1  \right]- i \left[ \left( \frac{C u }{\Lambda \sqrt{2}} + \frac{y_f}{\sqrt{2}} \right) \cos \theta - \frac{C v }{\Lambda \sqrt{2}} \sin \theta \right]\left[\bar{d} \gamma_{5} d h_1 \right]   \\
	-\left[\left(\frac{y_f}{\sqrt{2}} - \frac{C u \alpha }{\Lambda \sqrt{2} } \right) \sin \theta - \frac{C v \alpha }{\Lambda \sqrt{2}} \cos \theta \right] \left[ \bar{d} d S \right]  - i\left[ \left( \frac{C u }{\Lambda \sqrt{2}} + \frac{y_f \alpha}{\sqrt{2}} \right) \sin \theta  - \frac{C v }{\Lambda \sqrt{2}} \cos \theta \right]  \left[ \bar{d} \gamma_5 d S\right]
	\end{split}
	\end{equation}
	
	\paragraph{Gauge boson couplings:}
	Starting from the higher dimensional Lagrangian 
	\begin{equation}
	\mathcal{L}_{gauge}=\frac{C'}{\Lambda} P |D_{\mu}H|^2 + |D_{\mu}H|^2,
	\end{equation} 
	We get interactions with the SM Higgs as well as the new scalar. After expanding the above interaction terms and giving basis rotation to the scalars to get them in mass basis as in eq.~\eqref{scalar_mixing}, the interaction terms can be written as: 
	\begin{equation}\begin{split}\label{gauge_int}
	\mathcal{L}_{gauge} =  \left( \frac{2 m_W^2}{v} \cos \theta - \frac{C' m_{W}^2}{\Lambda } \sin \theta  \right) [W_{\mu}^+ W^{\mu-} h_1] + \left(\frac{m_{Z}^2}{v} \cos \theta - \frac{C' m_{Z}^2 }{2 \Lambda} \sin \theta  \right)  [ Z_{\mu}Z^{\mu} h_1 ]  \\
	+\left( \frac{2 m_W^2}{v} \sin \theta + \frac{C' m_W^2}{\Lambda} \cos \theta  \right)  [W_{\mu}^{+}W^{\mu -} S ]  + \left( \frac{m_{Z}^2}{v} \sin \theta + \frac{C'm_{Z}^2}{2\Lambda} \cos \theta  \right)   [Z_{\mu}Z^{\mu}S]
	\end{split}
	\end{equation}

	\section{FCNC vertex correction }\label{apndxA}
	We have written in eqs. \eqref{eq:RGE} and \eqref{eq:FCNC_vertexloop_2}, the FCNC loop contribution coming from Feynman diagrams fig. \ref{fig:b_to_s} in terms of loop functions $ I $'s. The expressions for $ I_1 $ and $ I_2 $ coming from fig. \ref{fig:FCNC_loop1}, are given by as following: 
	\begin{equation}\begin{split}
	I_1 = & \frac{1}{16 \pi^2} \int_{0}^{1} dx \int_{0}^{1-x} dy \biggl \{ \frac{m_t^2}{m_W^2} \biggl ( c_s (1-2x) -ic_p \biggr ) 
	+ \frac{m_t^2}{\Delta^2} 2 \biggl ( -c_s (1+2x) + ic_p  \biggr ) 
	\\& + \frac{m_t^2}{m_W^2} \frac{1}{\Delta^2}  \biggl (  m_{d_{i}}^2 x^2 \biggl ( c_s(1-2x) + ic_p \biggr ) 
	\biggr )
	+ \frac{m_t^2}{m_W^2} \biggl ( c_s(1+6x)+ic_p   \biggr ) \log\frac{m_t^2}{\Delta^2} 
	\biggr \}\,,
	\end{split}
	\end{equation}
	\begin{align}
	I_2&   =\frac{1}{16 \pi^2} \int_{0}^{1} dx \int_{0}^{1-x} dy   \biggl \{ \frac{m_t^2}{m_W^2} \biggl ( -c_s (1+2y) + ic_p  \biggr ) 
	+ 2 \frac{m_t^2}{\Delta^2} \biggl ( c_s(1-2y) -i c_p \biggr ) \nonumber \\&
	+ \frac{m_t^2}{m_W^2 \Delta^2}  \biggl (  m_{d_{i}}^2 x^2  \biggl ( -c_s (1+2y) + ic_p   \biggr ) 
	+2 m_{d_{i}}^2 xy (c_s + ic_p) 
	\biggr )
	+   \frac{m_t^2}{m_W^2} \biggl (   c_s(1+6y) -ic_p \biggr )  \log\frac{m_t^2}{\Delta^2}     
	\biggr \} \,,
	\end{align}
	Where, $ \Delta^2 = x m_{W}^2 + (1-x) m_{t}^2,  $ for $ I_{1} $ and $ I_{2} $. Loop functions $ I_3 $ and $ I_4 $ coming from fig. \ref{fig:FCNC_loop2}, are given by:
	\begin{align}
	I_{3} = & \frac{1}{16 \pi^2} \frac{c_{G}}{2 m_{W^2}} \int_{0}^{1} dx \int_{0}^{1-x} dy \frac{1}{\Delta^2 } \nonumber \biggl\{ 2 \Delta^2 \Big[ 4 m_W^2  (3 x+3 y-1) -3 \Delta ^2 (8 x+8 y-3) \Big] \log \left(\frac{m_W^2}{\Delta ^2}\right)\\&  \nonumber 
	+4 m_{d_{i}}^2 m_W^2  \left( x(1-y+xy) -y (1+y+y^2) + \frac{y-2}{2} \right)-8 \Delta ^2 (x+y) \\ 
	&  +\Delta ^2 m_{d_i}^2 \left(x(x+y) - 4y(1+x^2-y^2)\right)-4 \Delta ^4 (5 x+5 y-3) -8 m_W^4 (x+y-1) \biggr\} \,,  
	\end{align}
	
	\begin{align}
	I_{4}= &\frac{1}{16 \pi^2} \frac{c_{G}}{4 m_{W^2}} \int_{0}^{1} dx \int_{0}^{1-x} dy \frac{1}{\Delta^2 } \biggl\{ \nonumber 4 \Delta^2 \bigg[ 4 m_{W^2} (2-3y) - 3 \Delta^2 (5-8 y) \bigg] \log \left( \frac{m_{W}^2}{\Delta^2}\right) \\&  \nonumber 
	+ 2 (1-y) \bigg[\Delta^2 m_{d_i}^2  (x-4xy + 4y^2) - 2m_{W}^2\bigg] 
	+16 y m_{W}^4 - 8 \Delta^4 (2-5y) \biggr\}\,. 
	\end{align} 
	Where, $ \Delta^2 = x m_{t}^2 + (1-x) m_{W}^2 $ for $ I_{3} $ and $ I_4. $
\section{Additional inputs for analysis}
In section \ref{all_flav_fit}, we have discussed the fit results of our model parameters $ c_s, c_p, c_G $ and $ M_S $ from the analysis of relevant FCNC, FCCC, $ W $ and $ Z $ pole observables. The observables considered are given in table \ref{tab:CKM-updated-obs}. The additional inputs that are also taken in fit are given below in table \ref{tab:theoryinputs}. 

\begin{table}[t]
	\footnotesize
	\centering
	\renewcommand{\arraystretch}{1.4}
	\begin{tabular}{|c|c|c|}
		\hline
		Input Parameters & Value & Reference \\
		\hline
		$f_+^{K \to \pi}(0)$ & $0.9706(27)$ & $N_f = 2+1+1$ \cite{FLAG:2019}\\
		$f_{K^\pm}/f_{\pi^\pm}$ & $1.1973(08)(14)$ & Average \cite{CKMFitter:2021} \\
		$f_K$ & $155.7 (3) $ & $N_f = 2+1+1$ \cite{FLAG:2021}\\
		$f_+^{DK}(0)$ & $0.7385 (44)$ & $N_f = 2+1+1$ \cite{FLAG:2021}\\
		$f_+^{D\pi}(0)$ & $0.621 (18)(12)$ & Average\cite{CKMFitter:2021}\\			
		$f_{B_s}$ & $230.3(1.3)$ MeV & $N_f = 2+1+1$ \cite{FLAG:2021}\\
		$f_{B_s}/f_B$ & $1.209(5)$ & $N_f = 2+1+1$ \cite{FLAG:2021}\\
		$B_K$ & $0.7625(97)$ & $N_f = 2+1$ \cite{FLAG:2021}\\
		$f_{D_s}$ & $249.9(5)$ MeV & $N_f = 2+1+1$ \cite{FLAG:2021}\\
		$f_{D_s}/f_D$ & $1.1783(16)$ & $N_f = 2+1+1$ \cite{FLAG:2021}\\			
		$\zeta(\Lambda_p \to p \mu^- \bar{\nu}_\mu)_{q^2 > 15}/\zeta(\Lambda_p \to \Lambda_c \mu^- \bar{\nu}_\mu)_{q^2 > 7}$ & $1.471 (96) ( 0.290)$ & \cite{CKMFitter:2021} \\
		$B_{B_s}$ & $1.327(16)(30) $ & Average \cite{CKMFitter:2021} \\
		$B_{B_s}/B_{B_d}$ & $1.007(13)(14) $ & Average \cite{CKMFitter:2021} \\
		$\bar{m}_c (m_c)$ & $1.2976(13)(120)$ GeV & Average \cite{CKMFitter:2021}\\
		$\bar{m}_t (m_t)$ & $ 165.26 (11)(30)  $ GeV & \cite{CKMFitter:2021}\\
		$\eta_{tt}$ & $0.550(0)(024)$ & \cite{CKMFitter:2021} \\
		$\eta_{ut}$ & $0.402(0)(007) $ & \cite{CKMFitter:2021} \\
		$\eta_B (\bar{\text{MS}})$ & $0.5510 (0) (22)$ & \cite{CKMFitter:2021} \\
		$ B_2 $ & $ 0.46(1)(3) $& $ N_{f} = 2+1+1 $ \cite{FLAG:2021} \\
		$ B_4 $ & $ 0.78(2)(4) $ & $ N_{f} = 2+1+1 $\cite{FLAG:2021}\\
		\hline
	\end{tabular}
	\caption{List of aditional inputs used for the CKM fit with updated values of 2021-22.}
	\label{tab:theoryinputs}
\end{table}

	\section{Fit results for $ c_s = c_p $ }
	In above section \ref{all_flav_fit}, we have shown the fit values of our model parameters $ c_s, ~ c_{p} $ and $ c_{G} $ for the general case where: $ c_{s} \neq c_{p},  $ taking into account all the relevant flavour changing changed and neutral current processes with $ W $ and $ Z $ pole observables. Here, we have added the fit results for the case: $ c_{s} = c_{p} $ in table \ref{tab:combinecseqcp_CDF}.

\begin{table}[t]
	\centering
	\rowcolors{1}{blue!5}{blue!3!red!6!green!4}
	\renewcommand{\arraystretch}{1.8}
	\begin{tabular}{*{5}{c}}
		\toprule
		\shortstack{$ \Lambda $\\ $ \text{[TeV]} $}  & \shortstack{$M_S$ \\ $ \rm [GeV] $ }   &   \shortstack{$c_s(=c_p) \times 10 $\\ $ \rm [GeV^{-1}] $}  &  \shortstack{$c_G \times 10^3 $\\ $ \rm [GeV^{-1}] $} & \shortstack{$ \Delta M_W \times 10^2$\\ $ \rm [GeV] $ }   \\
		\hline
		\hline
		\cellcolor{blue!5}  &  $250$  &  $-0.0046\pm 0.1242$  &  $2 .7703\pm 0.2585$  &  $4 .6574\pm 0.8693$  \\
		\cellcolor{blue!5}  &  $500$  &  $-0.0079\pm 0.1375$  &  $-3.4324\pm 0.3214$  &  $4 .6536\pm 0.8716$  \\
		\multirow{-3}{*}{\cellcolor{blue!5} $ 1$}  &  $800$  &  $0 .0261\pm 0.1717$  &  $4 .4872\pm 0.4196$  &  $4 .6450\pm 0.8688$  \\
		\hline
		\hline
		\cellcolor{blue!5}  &  $250$  &  $0 .0066\pm 0.1048$  &  $-2.3046\pm 0.2158$  &  $4 .6580\pm 0.8724$  \\
		\cellcolor{blue!5}  &  $500$  &  $0 .0009\pm 0.1096$  &  $-2.6456\pm 0.2478$  &  $4 .6557\pm 0.8721$  \\
		\cellcolor{blue!5} &  $800$  &  $0 .0052\pm 0.1166$  &  $3 .0478\pm 0.2846$  &  $4 .6528\pm 0.8691$  \\
		\multirow{-4}{*}{\cellcolor{blue!5} $ 2$} &  $1000$  &  $-0.0094\pm 0.1216$  &  $-3.3300\pm 0.3120$  &  $4 .6504\pm 0.8714$  \\
		\bottomrule
	\end{tabular}
	\caption{Fit results of the parameters $ c_s $ and $ c_{G} $ for different combinations of $ M_S $ and $ \Lambda $ when $ c_s = c_p $ is onsidered. All the observables, including $ \delta(\Delta r) $ (CDF, LHCb, ATLAS, D0), are taken into account. The p-value for this fit is $ \sim 3.59\% $ with 46 d.o.f. The last column shows the prediction of $ \Delta M_W $ from this fit result.}\label{tab:combinecseqcp_CDF}
\end{table}

\begin{table}[htbp]
	\centering
	\rowcolors{1}{blue!5}{blue!3!red!6!green!4}
	\renewcommand{\arraystretch}{1.8}
	\begin{tabular}{*{5}{c}}
		\toprule
		\shortstack{$ \Lambda $\\ $ \text{[TeV]} $}  & \shortstack{$M_S$ \\ $ \rm [GeV] $ }   &   \shortstack{$c_s(=c_p) \times 10 $\\ $ \rm [GeV^{-1}] $}  &  \shortstack{$c_G \times 10^3 $\\ $ \rm [GeV^{-1}] $} & \shortstack{$ \Delta M_W \times 10^2$\\ $ \rm [GeV] $ }   \\
		\hline
		\hline
		\cellcolor{blue!5}&  $250$  &  $0 .0014\pm 0.1241$  &  $-0.8202\pm 1.4140$  &  $0 .4083\pm 1.4076$  \\
		\cellcolor{blue!5} &  $500$  &  $-0.0024\pm 0.1382$  &  $-1.0125\pm 1.7567$  &  $0 .4049\pm 1.4051$  \\
		\multirow{-3}{*}{\cellcolor{blue!5} $ 1$}  &  $800$  &  $0 .0085\pm 0.1548$  &  $1 .3217\pm 2.2803$  &  $0 .4030\pm 1.3905$  \\
		\hline
		\hline
		\cellcolor{blue!5}&  $250$  &  $-0.0020\pm 0.1039$  &  $0 .6834\pm 1.1601$  &  $0 .4096\pm 1.3907$  \\
		\cellcolor{blue!5} &  $500$  &  $-0.0003\pm 0.1105$  &  $0 .7819\pm 1.3376$  &  $0 .4066\pm 1.3913$  \\
		\cellcolor{blue!5} &  $800$  &  $0 .0016\pm 0.1174$  &  $0 .8982\pm 1.5465$  &  $0 .4041\pm 1.3915$  \\
		\multirow{-4}{*}{\cellcolor{blue!5} $ 2$}  &  $1000$  &  $0 .0029\pm 0.1215$  &  $0 .9799\pm 1.6930$  &  $0 .4026\pm 1.3914$  \\
		\bottomrule
	\end{tabular}
	\caption{Fit results of the couplings for different sets of $ \Lambda $ and $ M_S $ in the presence of both scalar and pseudoscalar coupling are equal, i.e., $ c_s = c_p $. All the observables of table \ref{tab:CKM-updated-obs} are taken into account along with the weighted mean data of $ \delta(\Delta r) $  observable from LHCb, ATLAS and D0 experiment \cite{LHCb:2021bjt,ATLAS:2017rzl,ATLAS:2023fsi,D0:2012kms}. The p-value of this fit is $ \sim 34.36 \%.  $ with 43 d.o.f. The last column shows the prediction of $ \Delta M_W $ from this fit result.  }\label{tab:combinecseqcpwithout_CDF}
\end{table}

\section{Dark Matter }\label{apndxC}
The expression for the cross-section of DM annihilating to SM fermion pair $ ( \bar{\chi}\chi \to \bar{f}f ) $ is given by \cite{Berlin:2014tja}: 
\begin{eqnarray}
\sigma v  &\approx& \frac{N_c c_{p \chi}^2 m_{f}^2\sqrt{1-m_f^2/m_\chi^2} \Big[m_{\chi }^2 (c_{p}^2+c_{s}^2)-m_f^2 c_{ s}^2\Big]}{2 \pi   \left(m_A^2-4 m_{\chi }^2\right)^2} \nonumber \\
&+&\frac{N_c v^2}{16 \pi  m_{\chi }^2 \left(4 m_{\chi }^2-m_A^2\right)^3 \sqrt{1-m_f^2/m_\chi^2}} \Bigg[ c_{p \chi}^2 \bigg\{c_{p}^2 m_f^2 m_{\chi }^2 \Big(m_A^2 (m_f^2-2 m_{\chi }^2)+12 m_f^2 m_{\chi }^2-8 m_{\chi }^4\Big)  \nonumber \\
&+&c_{s}^2 m_f^2 (m_f^2-m_{\chi }^2) \Big(m_A^2 (m_f^2+2 m_{\chi }^2)-20 m_f^2 m_{\chi }^2+8 m_{\chi }^4\Big)\bigg\}  \label{sigVsdirS} \\
&-&2 c_{s \chi}^2 m_{f}^2 (m_A^2-4 m_{\chi }^2) (m_{\chi }^2-m_f^2) \Big(m_{\chi }^2 (c_{p}^2+c_{s}^2)-m_f^2 c_{s}^2\Big)\Bigg]. \nonumber
\end{eqnarray}
Where, $ N_c =3, $ for quarks and $ N_c =1, $ for annihilation to lepton pairs.

The effective spin-independent nucleon-WIMP $ (N-\chi) $ inelastic scattering cross-section in zero transfer momentum limit $ (q^2 \to 0) $, can be given by: 
\begin{equation}
\sigma^{\rm SI}_{\chi N} \simeq \frac{\mu_{\chi N}^2 c_{s\chi}^2}{\pi M_S^4} \left(   Z \widetilde f_p +  (A-Z) \widetilde  f_n \right)^2, 
\end{equation}
where, $ \mu_{\chi N} $ is the reduced mass of DM-nuclei and $ \widetilde  f_n $ are defined below. 
The nucleon form factor is defined as : 
\begin{align}
\frac{\widetilde f_n}{m_n} =& \sum_{q=u,d,s} f_{T_q}^n\,\frac{\widetilde f_q}{m_q} +\frac{2}{27} \, f_{TG} \sum_{q=c,b,t} \frac{\widetilde f_q}{m_q}\,,
\\ & \to \frac{m_{q} c_{s}}{GeV} \left( \frac79 \sum_{q=u,d,s} f_{T_q}^n +\frac29 \right),
\end{align}
where in the second line we take $\widetilde f_q = c_s m_q^2  /{\rm GeV}$ and $f_{TG}=1-f^n_{T_u}-f^n_{T_d}-f^n_{T_s}$, with $f_{T_u}^{p(n)}=0.018(0.013)$ , ~ $f_{T_d}^{p(n)}=0.027(0.040)$ and $f_{T_s}^{p(n)}=0.037(0.037)$ \cite{Bhattacharya:2017fid}.

\bibliographystyle{JHEP} 
\bibliography{FDM_highmass_lipika}		


\end{document}